\documentclass[12pt,letterpaper]{article}
\pdfoutput=1

\usepackage{graphicx,array}
\usepackage{url}
\usepackage{color}
\usepackage{latexsym}
\usepackage{amsthm}
\usepackage{amsmath}
\usepackage{amssymb}
\usepackage{amsfonts}
\usepackage[numbers,sort&compress]{natbib}
\usepackage{bm}
\usepackage{slashed}
\usepackage{mathrsfs}
\usepackage{enumerate}
\usepackage{tikz}
\usepackage{siunitx}
\usepackage{mdframed}
\usepackage{lipsum}
\usepackage{setspace}

\usepackage[utf8x]{inputenc}%
\usepackage{tcolorbox}%

\usepackage{hyperref} 
\hypersetup{
    colorlinks=true,       
    linkcolor=red,          
    citecolor=blue,        
    filecolor=magenta,      
    urlcolor=blue           
}
\usepackage[all]{hypcap} 

\usepackage{natbib}
\newcommand{\nc}{\newcommand}

\nc{\beq}{\begin{equation}}  
\nc{\eeq}{\end{equation}}  
\nc{\beqa}{\begin{eqnarray}}  
\nc{\eeqa}{\end{eqnarray}}  

\def\be{\begin{equation}}
\def\ee{\end{equation}}
\def\bea{\begin{eqnarray}}
\def\eea{\end{eqnarray}}

\nc{\bit}{\begin{itemize}}  
\nc{\eit}{\end{itemize}}

\def\GeV{\mathrm{GeV}}     

\newcommand{\avg}[1]{\left\langle#1\right\rangle}
\newcommand{\Mphi}{m_{\phi,0}}
\newcommand{\Veff}{U_{\rm eff}}

\def\sol/{\textrm{DM soliton}}
\def\sols/{\textrm{DM solitons}}
\def\Tittlesol/{\textrm{DM Soliton}}
\def\Tittlesols/{\textrm{DM Solitons}}
\def\EDB/{\textrm{EWS-DMB}}
\def\EDBs/{\textrm{EWS-DMBs}}
\def\DMB/{\textrm{DMB}}
\def\DMBs/{\textrm{DMBs}}
\def\DMN/{\textrm{DM number}}
\def\DMNs/{\textrm{DM numbers}}

\DeclareRobustCommand\encircle[1]{%
  \tikz[baseline=(X.base)] 
    \node (X) [draw, shape=circle, inner sep=-1] {\strut \raisebox{-0.5pt}[0pt]{#1}};}
\newcommand{\Pcirc}{\encircle{$\Phi$}}
\newcommand{\Msol}{M_{\tiny \Pcirc}}
\newcommand{\Rsol}{R_{\tiny \Pcirc}}

\usepackage{floatrow}
\newfloatcommand{capbtabbox}{table}[][\FBwidth]

\usepackage{blindtext}

\title{ \bf
Electroweak Symmetric Dark Matter Balls
\author{\large Eduardo Pont\'on$\,^\star$,  Yang Bai$^{\,\star,\diamond}$ and Bithika Jain$\,^\star$}
\date{\small \it 
$^\star$ICTP South American Institute for Fundamental Research \& Instituto de F\'isica Te\'orica \\ 
  Universidade Estadual  Paulista, S\~ao Paulo, Brazil \\
$^\diamond$Department of Physics, University of Wisconsin-Madison, Madison, WI 53706, USA  \\
}
}

\begin{document}

\maketitle

\setlength{\parskip}{0.2ex}

\begin{abstract}
In the simple Higgs-portal dark matter model with a conserved dark matter number, we show that there exists a non-topological soliton state of dark matter. This state has smaller energy per dark matter number than a free particle state and has its interior in the electroweak symmetric vacuum. It could be produced in the early universe from first-order electroweak phase transition and contribute most of dark matter. This electroweak symmetric dark matter ball is a novel macroscopic dark matter candidate with an energy density of the electroweak scale and a mass of 1 gram or above. Because of its electroweak-symmetric interior, the dark matter ball has a large geometric scattering cross section off a nucleon or a nucleus. Dark matter and neutrino experiments with a large-size detector like Xenon1T, BOREXINO and JUNO have great potential to discover electroweak symmetric dark matter balls. We also discuss the formation of bound states of a dark matter ball and ordinary matter. 
\end{abstract}

\thispagestyle{empty}  
\newpage  

\setcounter{page}{2}

\begin{minipage}{0.925\textwidth}
\vspace{7cm}
\definecolor{lightgray}{gray}{0.85}
\setstretch{1.3} 
\begin{tcolorbox}[arc=0mm,auto outer arc, colback=lightgray,leftrule=3pt,rightrule=3pt,toprule=3pt,bottomrule=3pt]
\begin{center}
{\fontfamily{ariel}\selectfont
\large{
We are deeply saddened by the passing away of Eduardo Pont\'on (4 April 1971 -- 13 June 2019). Eduardo has been an excellent mentor and collaborator. He has provided invaluable guidance for us from his passion and rigorous scientific attitude. We wish that he has a peaceful life in the other, maybe electroweak symmetric, world. 
}
}
\end{center}%
\end{tcolorbox}%
\end{minipage}

\newpage

\hypersetup{linktocpage} 
\tableofcontents 
\hypersetup{linkcolor=red} 

\newpage

\section{Introduction} 
\label{sec:intro}
Dark matter (DM) is one of the remaining mysteries in particle physics after the discovery of Higgs boson in 2012. After a few decades of searching for electroweak-sector-related dark matter particles with a mass around 100 GeV and with a null result~\cite{Aprile:2018dbl}, we have started to enlarge the scope of dark matter masses from both the theoretic model and the experimental search sides. For our visible sector, we have many interesting states of ordinary matter ranging from diluted gas to a dense neutron star. Analogously, it will not be surprising that there are many types of states for dark matter.  Under certain circumstance, the majority of dark matter could be in a form of macroscopic state instead of free particle states. The well-known example is the primordial black hole dark matter~\cite{Hawking:1971ei}, which has the Schwarzschild radius as its macroscopic size. Another established example the so-called ``quark nugget"~\cite{Witten:1984rs} with around nuclear energy density, which has the constituents of dark matter to be fermionic quarks and the geometrical size of $0.01$--$10$~cm. In this paper, we will focus on another type of macroscopic dark matter with a bosonic constituent or ``non-topological soliton" as named in the literature. 

For a scalar field with non-linear interactions, it has long been pointed out that there exists a spacially-localized state that can be a solution to the scalar classical equation~\cite{RosenSoliton}.  The existence and properties of the non-topological soliton as a field-theory object have been studied extensively by T. D. Lee~\cite{Friedberg:1976me} and S. R. Coleman~\cite{Coleman:1985ki} and their collaborators (see Ref.~\cite{Lee:1991ax} for a review), while its primordial production from early universe physics has also been worked out in Refs.~\cite{Frieman:1988ut,Griest:1989bq,Frieman:1989bx}. In supersymmetrical models, Q-balls (the soliton states and named by Coleman), built of squarks and sleptons have also been proposed as a potential dark matter candidate~\cite{Kusenko:1997si,Dvali:1997qv,Kusenko:1997vp}. With a conserved global internal symmetry, the non-topological soliton is an extended object with the lowest value of the energy for a fixed conserved charge, and therefore is stable at quantum level. The non-topological soliton is simply different from topological solitons, which has a quantized charge related to the algebraic geometry. For instance, a nucleon can be regarded as a topological soliton state of pions or Skyrmion because of $\pi_3[SU(2)]=\mathcal{Z}$~\cite{Witten:1983tx}. 

After some preparation of soliton basics, we want to point out the main observation of this paper: in the simple Higgs-portal complex scalar dark matter model, a non-topological soliton state exists for dark matter and could be the lowest energy state per dark matter number. For such a simple dark matter model, the dark matter could be in the macroscopic soliton state with a very large dark matter number, which we will refer as dark matter balls (DMBs). One possible mechanism to produce dark matter soliton states from early-universe dynamics could come from the first-order phase transition of electroweak (EW) symmetry, which can be naturally realized based on the quantum-corrected Higgs potential from the complex dark matter particle loop. Below the EW phase transition temperature, the EW symmetry-breaking bubbles grow and push the dark matter number to be in front of the bubble wall. After a few bubbles meet each other and coalesce, the dark matter number is enclosed in a small region and is still in the high-temperature EW-unbroken phase. Based on our later estimation and assuming some initial dark matter-antimatter asymmetry, we have found that the dark matter number is mainly stored in the soliton or \DMB/ state instead of free dark matter particle state. 

One interesting feature of \DMBs/ based on the Higgs-portal dark matter model is the interplay between the Higgs potential and dark matter field strength. For a positive quartic interaction of the two fields, a large complex scalar dark matter field inside the \DMB/ provides an effective positive mass for the Higgs field and prefers the Higgs field to sit at zero or a negligible value. So, the dark matter state could be a {\it Electroweak Symmetric DM Ball} (EWS-DMB) in sense that the electroweak symmetry is unbroken inside \DMBs/. This particular feature of \DMB/ also means that the interaction of \DMB/ with ordinary matter is relative large. From  our later detailed calculations, we have found that the scattering cross section of \DMBs/ off a nucleon or nucleus saturates to the geometrical cross section, when a \DMB/ has a large radius. 

Our observation could dramatically change the experimental search strategy for dark matter: instead of searching for the single-hit scattering event with a small recoil energy in a location deep underground, one could search for multi-hit scattering events at a location not necessarily underground. In consequence, neutrino-oriented experiments with a large size detector become suitable for this type of macroscopic dark matter. Or, searching for tracks in an ancient mineral like Mica may also discover this type of \DMBs/ because of its very long, billion-year, exposure time. We will also discuss various search strategies for EWS-DMB. 

The paper is organized as follows. We first work out the properties of soliton states with and without the dark matter bare mass and self-quartic interaction in Section~\ref{sec:soliton}. In Section~\ref{sec:early-universe}, we study the early-universe productions of \DMBs/ based on the first-order electroweak phase transition and obtain the characteristic charge, mass and radius for \DMBs/. We then calculate the scattering cross sections of \DMBs/ with Standard Model (SM) particles in Section~\ref{sec:scattering}. The detection of \DMBs/ in various experiments will be discussed in Section~\ref{sec:pheno}. We summarize our results in Section~\ref{sec:conclusion}. Furthermore, we have also included four Appendixes: the calculation of the number of \DMB/ nucleation sites in Appendix~\ref{sec:nucleation-sites}, the calculation of the binding energy of bound states of EWS-DMBs and ordinary matter in Appendix~\ref{sec:Backreaction}, the calculation of the bound states in a Higgs potential well in Appendix~\ref{sec:HiggsWell} and a simple example of scattering against a heavy object in Appendix~\ref{sec:heavysource}.

\section{Soliton States in a Higgs-Portal Dark Matter Scenario} 
\label{sec:soliton}

In the Higgs-portal dark matter scenario with a complex scalar particle $\Phi$,\footnote{Although we will not do so here, one could consider the fermionic case. Due to the Pauli exclusion principle, it is qualitatively different from the bosonic example that is the focus of this work.}  the most general renormalizable Lagrangian preserving a $U(1)_\Phi$ symmetry is
\beqa
\mathcal{L} = \partial_\mu\Phi^\dagger {\partial^\mu} \Phi + \partial_\mu H^\dagger {\partial^\mu} H - \lambda_h \left( H^\dagger H - \frac{v^2}{2} \right)^2 - \lambda_{\phi h}\,\Phi^\dagger \Phi H^\dagger H  - \Mphi^2\, \Phi^\dagger \Phi - \lambda_\phi \, (\Phi^\dagger \Phi)^2~.
\label{MasterL}
\eeqa
The $U(1)_\Phi$ symmetry ensures that the elementary $\Phi$ quanta are stable, and therefore a DM candidate.
This is one of the simplest extension of the SM to include dark matter. For reasons that will become clear in the following, we will focus on the region of parameter space with $\lambda_{\phi h} > 0$ and $\Mphi^2 \geq 0$, so that the physical $\Phi$ mass squared is never negative, even in the absence of a vacuum expectation value (VEV) for $H$. We will also take $\lambda_\phi > 0$.\footnote{Furthermore, we restrict ourselves to the perturbative regime $\lambda_{\phi h} \ll 4\pi$ and $\lambda_\phi \ll 16\pi^2$.} In this case, the global minimum of the tree-level potential breaks the EW symmetry spontaneously: $\langle H \rangle^T  = (0, v/\sqrt{2})$ with $v=246$~GeV, and $\langle \Phi \rangle = 0$. The quartic coupling $\lambda_h$ is related to the Higgs boson mass $m_h \approx 125$~GeV~\cite{Aad:2015zhl} by $\lambda_h = (m_h/v)^2/2 \approx 0.13$. After electroweak symmetry breaking (EWSB), the free $\Phi$ particle mass is 
\beqa
m_{\phi}^2 = \frac{\lambda_{\phi h}}{2} v^2 + \Mphi^2~.
\label{mphi}
\eeqa
When the bare dark matter mass $\Mphi = 0$, the $\Phi$ particle obtains all of its mass from EWSB and $m_{\phi}  = \sqrt{\lambda_{\phi h}} \, v/\sqrt{2}$. 

We are interested here in non-vacuum field configurations that are nevertheless stable due to the conservation of the charge associated with the global $U(1)_\Phi$ symmetry. In the theory given by Eq.~(\ref{MasterL}), the existence and properties of such solutions were worked out in~\cite{Friedberg:1976me} (assuming $\Mphi=0$ and $\lambda_\phi=0$), thus  providing an example of a ``non-topological soliton" (for a review, see~\cite{Lee:1991ax}). We will briefly review how these solutions arise and their salient features. We start with the case $\Mphi = 0$ and $\lambda_\phi = 0$, to establish that such \sols/ exist even in this minimal case, which depends on a single free parameter, $\lambda_{\phi h}$. This will also highlight the crucial role played by this coupling. In a second stage we will include the effects of the remaining two free parameters, $\Mphi$ and $\lambda_\phi$, which can affect the qualitative properties of the soliton solutions. We will describe the relevant features in Section~\ref{OtherTwoParameters}. 

The \sols/ are characterized by a non-vanishing charge
\bea
Q &=& i \int \! d^3x \, \left( \Phi^\dagger \partial_t \Phi - \Phi \partial_t \Phi^\dagger \right) ~=~ \omega \int \! d^3x \, \phi^2~,
\label{QCharge}
\eea
which is obtained from the time-dependence $\Phi(x) = e^{-i \omega t } \phi(\vec{x}) /\sqrt{2}$, with $\phi(\vec{x}) $ real. We will focus on spherically symmetric solitons (that have the lowest energy) with $\phi(\vec{x}) = \phi(r)$ and $H^T  = \big(0, h(r)/\sqrt{2}\big)$, obeying the classical equations of motion
\beqa
&&\phi^{\prime \prime}(r) + \frac{2}{r}\, \phi^\prime(r) + \left[ \omega^2 - \frac{1}{2}\lambda_{\phi h}\, h(r)^2 \right] \phi(r) =0~, 
\label{phiEOM} \\ [0.5em]
&& h^{\prime \prime}(r) + \frac{2}{r}\,h^\prime(r)  + \left[ \frac{m_h^2}{2} - \lambda_h \, h(r)^2 - \frac{1}{2} \lambda_{\phi h} \, \phi(r)^2 \right]\,h(r) =0~,
\label{hEOM}
\eeqa
and subject to the boundary conditions $\phi^\prime(0)=h^\prime(0)=0$, $\phi(\infty)=0$ and $h(\infty) = v$.  

In order to develop an intuition it is useful to write down an approximate description by neglecting the Higgs derivatives in Eq.~(\ref{hEOM}). The motivation is that often the Higgs profile is nearly vanishing inside the \sol/ and takes the (almost) constant value $v$ outside, approximating a step function. Thus, apart from the relatively small transition region, the neglect of the spatial derivatives can be justified a posteriori, thus  permitting an effective description in terms of a single degree of freedom.\footnote{When the transition region is not small, the approximation can deviate by order one from the exact solution, but the qualitative features remain the same. We will also show numerical solutions that solve the full system of Eqs.~(\ref{phiEOM}) and (\ref{hEOM}).} Eq.~(\ref{hEOM}) then shows that one can have configurations obeying
\bea
h^2 &\approx& 
\left\{ 
\begin{array}{ccc}
 \dfrac{m_h^2}{2\lambda_h} - \dfrac{ \lambda_{\phi h}}{{2\lambda_h}} \, \phi^2 & \hspace{5mm} \textrm{for} & 
 \lambda_{\phi h} \, \phi^2 < m_h^2~,  \\ [1.0em]
 0  &  \hspace{5mm} \textrm{for}  &  \lambda_{\phi h} \, \phi^2 > m_h^2~.
\end{array}
\right.
\label{happrox}
\eea
Inserting Eq.~(\ref{happrox}) into Eq.~(\ref{phiEOM}) one gets
\bea
\phi^{\prime \prime} + \frac{2}{r}\, \phi^\prime + \Veff'(\phi) &\approx& 0~,
\label{phiEffEOM}
\eea
where the effective potential is obtained by using Eq.~(\ref{happrox}) in (minus) the potential terms of Eq.~(\ref{MasterL}), but including the terms coming from the time derivatives:
\bea
-V_\omega(h, \phi) &\equiv& \frac{1}{2} \omega^2 \phi^2 - \frac{1}{4} \lambda_h \left( h^2 - v^2 \right)^2 - \frac{1}{4} \lambda_{\phi h}\,\phi^2 h^2  - V_{\Phi}(\phi)~,
\label{Vomega}
\eea
giving
\bea
\Veff(\phi) &=& - V_{\Phi}(\phi) + 
\left\{ 
\begin{array}{ccc}
\dfrac{1}{2} \left( \omega^2 - \dfrac{\lambda_{\phi h}\,m_h^2}{4\,\lambda_h} \right) \phi^2 + \dfrac{ \lambda^2_{\phi h}}{{16\lambda_h}} \, \phi^4~ & \hspace{5mm} \textrm{for} & 
 \lambda_{\phi h} \, \phi^2 < m_h^2~,  \\ [0.7em]
 \dfrac{1}{2} \omega^2 \phi^2 - \dfrac{m_h^4}{16\,\lambda_h}  &  \hspace{5mm} \textrm{for}  &  \lambda_{\phi h} \, \phi^2 > m_h^2~.
\end{array}
\right.
\label{Veff}
\eea
For later use, we reintroduced here the pure $\phi$-dependent terms
\bea
V_{\Phi}(\phi) &=& \frac{1}{2} \Mphi^2 \, \phi^2 + \frac{1}{4} \lambda_\phi  \phi^4~,
\label{Vphi}
\eea
although for the time being we are setting them to zero. We then see that at large $\phi$ values, $\Veff$ increases quadratically with $\phi$. Importantly, the origin is unstable provided 
\bea
\omega^2 &<& \frac{\lambda_{\phi h}\,m_h^2}{4\,\lambda_h} +  \Mphi^2~=~ m_{\phi}^2~,
\label{HillCond}
\eea
where we again wrote the $\Mphi^2$ dependence for later reference. As we will see, the small $\omega$ limit corresponds to large charge $Q$. Assuming Eq.~(\ref{HillCond}), one can see that there is a solution that starting ``at rest" (using an effective 1-particle mechanics in 1D language, with time evolution parametrized by $r$) at $\phi(r=0) = \phi_0$, rolls down the effective potential towards the hill at $\phi = 0$, loosing in the process energy due to the effective friction term in Eq.~(\ref{phiEffEOM}). It is clear that by adjusting $\phi_0$, it is always possible to arrange for this motion to stop at $\phi(r=\infty) = 0$. One can also see that since $\Veff(\phi=0) = 0$, one must have $\Veff(\phi_0) > 0$. At the saturation point of Eq.~(\ref{HillCond}), the term in braces in $\Veff(\phi)$ evaluated at the matching point $\lambda_{\phi h} \, \phi^2 = m_h^2$ takes the positive value $m_h^4/(16\lambda_h^2)$. Thus, for $\omega^2 \lesssim \lambda_{\phi h}m_h^2/(4\lambda_h)$ it is possible to find solutions fully contained in the region $\lambda_{\phi h} \, \phi^2 < m_h^2$. Such solutions can have a non-negligible $h$ inside the core of the \sol/, and typically require a more careful analysis that takes into account the $h$ derivatives that have been neglected in the effective description. However, when $\omega$ is very small, $\phi_0$ must be such that $\lambda_{\phi h} \, \phi^2_0 > m_h^2$ to satisfy $\Veff(\phi_0) > 0$. Translated into the behavior for $h$ this corresponds to situations with (nearly) vanishing $h$ inside the core of the \sol/. We will therefore sometimes refer to such solutions as \textit{Electroweak Symmetric DM Balls} (\EDBs/), or \DMBs/ for short. The associated $h$-profile typically resembles the step-like profile captured by the effective description. 

\begin{figure}[t]
\centering
\includegraphics[width=0.6\textwidth]{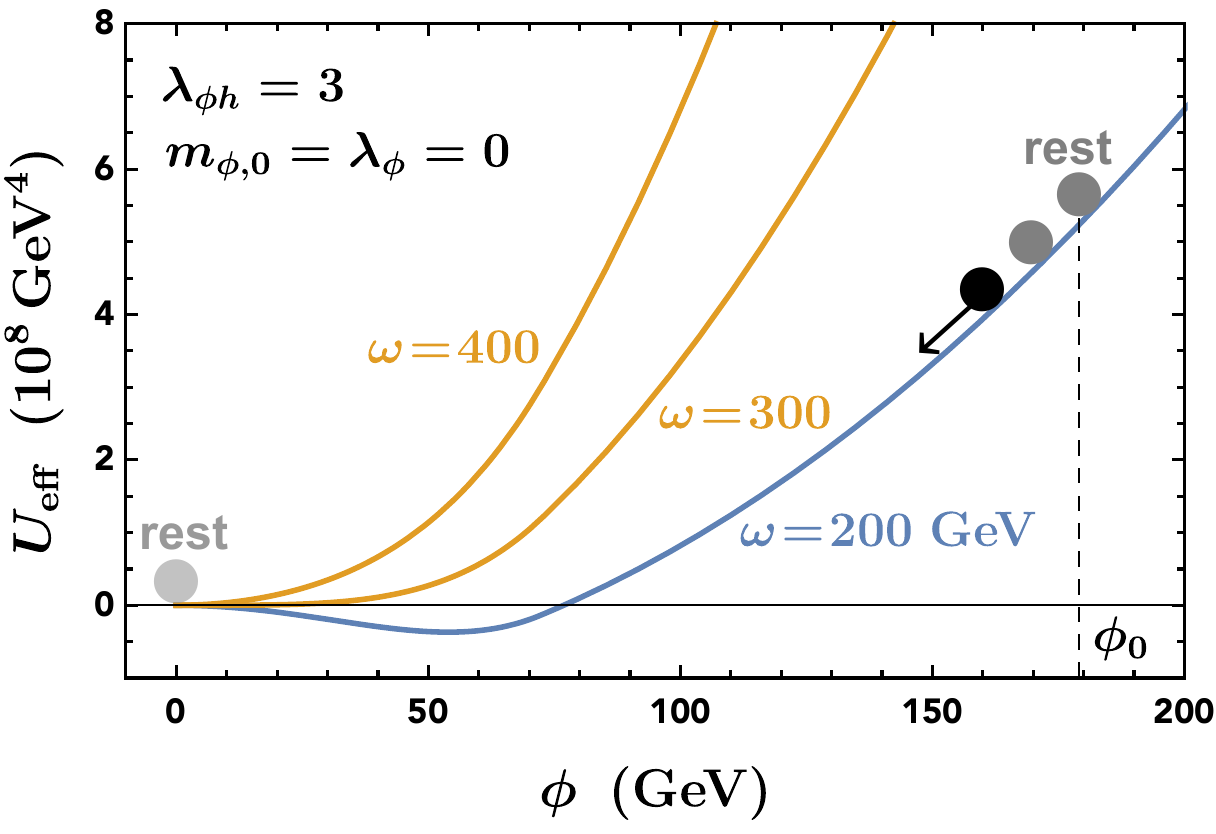}
\caption{Examples of the effective potential defined in Eq.~\eqref{Veff} as a function of $\phi$ for different values of $\omega$. It describes the 1-particle mechanics in 1D analogue, with a friction term, as given in Eq.~\eqref{phiEffEOM}. The particle starts at rest at $\phi = \phi_0$  (for $r=0$) and comes to rest at $\phi=0$ (for $r \to \infty$). For the $\omega=200$~GeV case one has $ \phi_0 \approx 179$~GeV.}
\label{fig:Veff1}
\end{figure}
We show in Fig.~\ref{fig:Veff1} the effective potential, $\Veff$, taking $\lambda_{\phi h} = 3$ and $\Mphi = \lambda_\phi = 0$, for several values of $\omega$. The threshold value defined by the saturation of the inequality (\ref{HillCond}) is about $301$~GeV, and we show an example in its vicinity. Together with the $\omega = 200$~GeV case, it gives rise to \sols/ with a sizable Higgs VEV inside the core. (As we will see, for $\omega = 100$~GeV one obtains solutions displaying a core with a small Higgs VEV, i.e.~an \EDB/.) Finally, the $\omega = 400$~GeV case does not satisfy Eq.~(\ref{HillCond})  and leads to oscillating solutions that tend slowly to zero as $r\to \infty$, and are therefore not localized. These do not belong in the class of solitonic solutions. 

From the previous discussion, we also see that for \DMB/ solutions one must have the scaling
\bea
\phi_0 &\sim& \frac{1}{\omega}~.
\label{phi0Scaling}
\eea
The size of the \DMBs/ can be easily estimated as follows: setting $h=0$ in Eq.~(\ref{phiEOM}), as is appropriate inside the soliton, leads to
\bea
\phi(r) &\approx& \phi_0 \,  \frac{\sin(\omega\,r)}{\omega\,r}~.
\label{phiApprox}
\eea
This function has an infinite number of zeros, each of which corresponds to a solution. We will focus on the solution associated with the first zero, which has the lowest energy. Near this first zero, $h$ turns on leading quickly to the asymptotic value $\phi \to 0$ as $r\to \infty$ (the excited solutions arise in a similar manner, but with additional nodes). The size of the transition, i.e.~the thickness of the surface boundary separating the EW breaking and EW preserving phases is of order $\pi/v$. We therefore see that the size of the lightest \DMB/, denoted by $\Rsol$, is about
\bea
\Rsol &\approx& \frac{\pi}{\omega}~.
\label{EWSBSize}
\eea
Inserting the approximate solution (\ref{phiApprox}) in Eq.~(\ref{QCharge}), together with Eq.~(\ref{phi0Scaling}), one finds
\bea
Q &\approx& 4\pi \omega \phi_0^2 \int_0^{\Rsol} \! r^2 dr \, \frac{\sin^2(\omega r)}{(\omega r)^2}
\nonumber \\[0.5em]
&\approx& \frac{2\pi^2 \phi_0^2}{\omega^2} ~\sim~ \frac{1}{\omega^4}~.
\label{QomegaScaling}
\eea
As stated earlier, small $\omega$ maps into large $Q$. We also see that in this limit, we have the scaling
\bea
Q &\sim& \Rsol^4~.
\eea
With this qualitative understanding, let us now consider some examples of the full solutions to Eqs.~(\ref{phiEOM}) and (\ref{hEOM}).

\subsection{Solutions to the Classical Equations of Motion}
\label{ClassicalEOM}

It is possible to obtain numerical solutions to the system (\ref{phiEOM}) and (\ref{hEOM}) and the specified boundary conditions by the ``shooting method". This depends on two variables: $\phi(0) = \phi_0$ and $h(0) = h_0$. The first derivatives vanish, which provides the four initial conditions to uniquely specify the solution. In practice, one starts at a small $r_0$ to avoid the singular point at the origin. One can then adjust $\phi_0$ and $h_0$ to obtain the solution that obeys $\phi(\infty)=0$ and $h(\infty) = v$. In practice one takes $r = \infty$ to mean an $r_{\rm max}$ large enough that the neglected part can be seen to be numerically close to the desired solution. This procedure can be followed for any fixed set of Lagrangian parameters, and fixed $\omega$. For a given model, one is interested in scanning over $\omega$, i.e.~in obtaining soliton solutions of different charge $Q$.

\begin{figure}[t]
\centering
\includegraphics[width=0.475\textwidth]{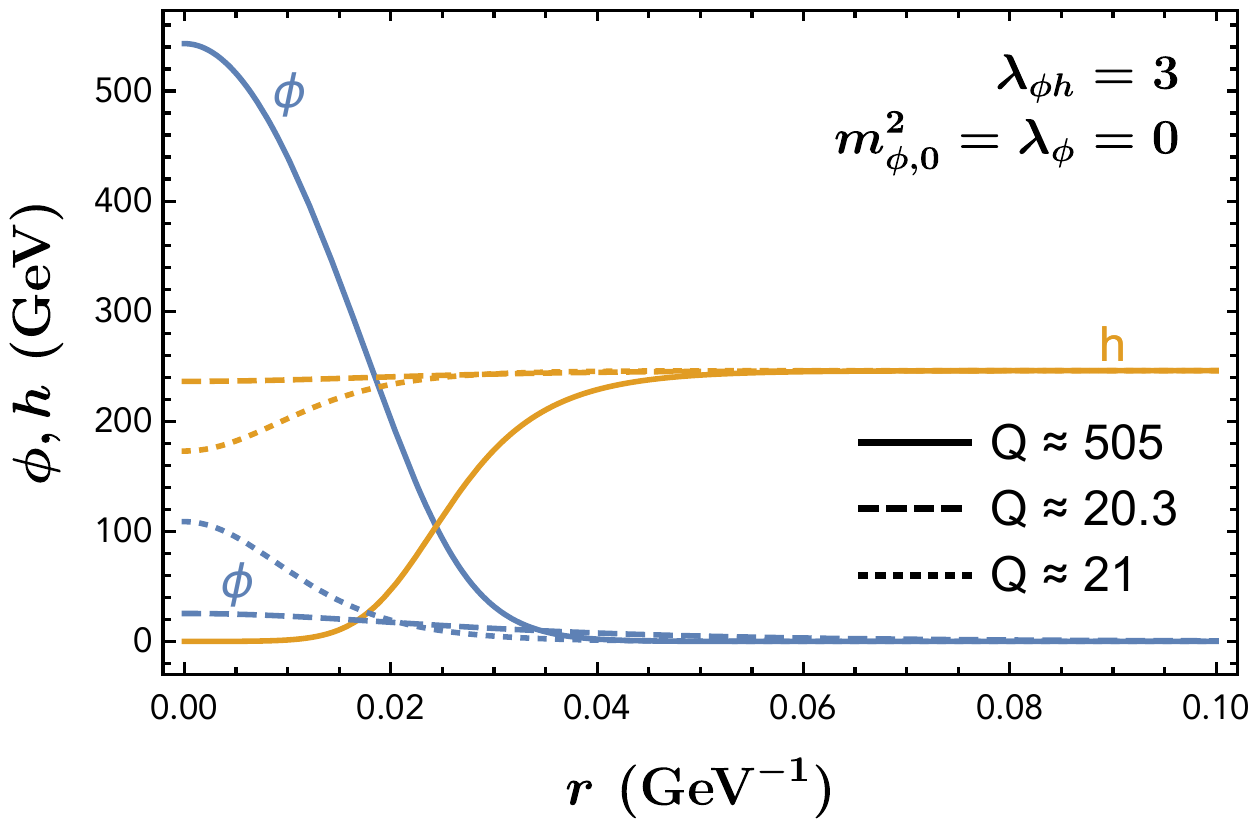}
\hspace{3mm}
\includegraphics[width=0.485\textwidth]{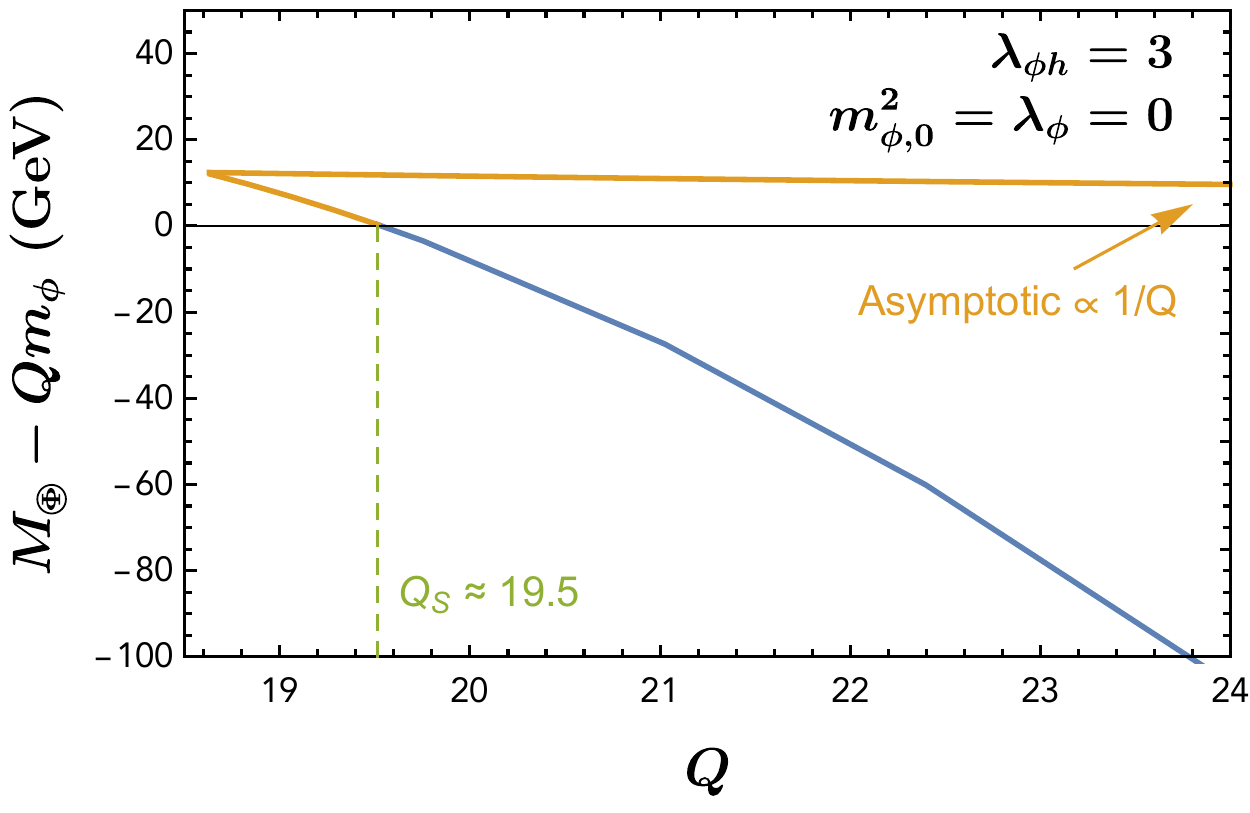}
\caption{Left panel: profiles for \sols/ for three different charges, $Q\approx 505$, $Q\approx 20.3$ and $Q \approx 21$. The $Q\approx20.3$ case (dashed lines) corresponds to $\omega = 300$~GeV, close to the threshold value $\omega_{\rm th} = m_\phi \approx 301$~GeV that allows such types of configurations, as determined from Eq.~(\ref{HillCond}). This solution is quantum mechanically unstable against decay into $Q$ free particle states. The $Q = 21$ case ($\omega = 280$~GeV), although having a similar charge, is stable. The $Q \approx 505$ case (solid lines) corresponds to $\omega = 110$~GeV. The size of this \DMB/, as estimated from Eq.~(\ref{EWSBSize}), is $\Rsol \approx 0.03~{\rm GeV}^{-1}$, which is reasonable from the figure. Right panel:  difference between the \sol/ mass, $\Msol$, and the energy of $Q$ free $\Phi$ particles of mass $m_\phi \approx 301$~GeV, for $\lambda_{\phi h}=3$, as a function of the charge $Q$ (low $Q$ region). 
The orange branch corresponds to soliton solutions that are unstable against decay into such non-bound free $Q$-particle states. The blue branch is stable. For the model shown, the boundary between the two branches is at $Q_{S} \approx 19.5$, corresponding to $\omega_{S} \approx 286$~GeV. This is the smallest charge for a stable \sol/.}
\label{fig:ProfilesLowQ}
\end{figure}
We show in the left panel of Fig.~\ref{fig:ProfilesLowQ} the $\phi$ (blue) and $h$ (orange) profiles for three different charges, in the model defined by $\lambda_{\phi h} = 3$ and $\Mphi = \lambda_\phi = 0$. The choice of $\lambda_{\phi h} = 3$ is motivated by the mechanism of formation of such \sols/, to be described later, but the features discussed in this section are similar for any $\lambda_{\phi h}$ of order one. The $Q = 20.3$ case (flatter, dashed profiles) corresponds to a choice of $\omega$ close to the threshold value $\omega_{\rm th} = m_\phi$ defined by the saturation of the inequality (\ref{HillCond}). One can see that it is very close to the vacuum solution. There is a second solution with a similar charge with $Q=21$ that displays a better defined core. As we will explain next, the former solution is unstable against decay into $Q$ free elementary $\Phi$ quanta, while the latter is a stable \sol/. The third example has a larger charge $Q \approx 505$, corresponding to a smaller $\omega = 110$~GeV. It shows a clear core with a small Higgs VEV and a large value for $\phi$. This \sol/ would fall in the category of \EDBs/ defined above. Although the transition in the Higgs profile from zero to $v$ is comparable to the core, one can see that the $\phi$ profile is reasonably well described by the approximate solution (\ref{phiApprox}) (for $r \lesssim \Rsol$). Indeed, for $\omega = 110$~GeV, Eq.~({\ref{EWSBSize}}) gives $\Rsol \approx 0.03~{\rm GeV}^{-1}$, in good agreement with what is seen in the figure. Obtaining full numerical solutions with larger cores is challenging, as the solutions are sensitive to an exponentially small $h_0$. Such solutions can nevertheless be easily obtained in the framework of the effective description. We will also use the effective description to discuss the effects of the two additional parameters, $\Mphi$ and $\lambda_\phi$.

Before turning to the general case, we consider the mass of the \sol/. In the mean field approximation we are using, this can be obtained by computing the classical energy of the configuration:
\bea
\Msol &=& 4\pi \int_0^\infty \! dr \, r^2 \left\{ \frac{1}{2} \,\omega^2 \phi^2 + \frac{1}{2} (\phi')^2 + \frac{1}{2} (h')^2 + V_H(h) + \frac{1}{4} \lambda_{\phi h} h^2 \phi^2 + V_\Phi(\phi) \right\}~,
\label{MPhi}
\eea
where $V_H(h)$ is the SM Higgs potential and $V_\Phi(\phi)$ was defined in Eq.~(\ref{Vphi}). Here, $\phi'$ and $h'$ are derivatives with respect to $r$. Note that localized field configurations cannot have a well-defined energy: although the mean energy in the rest frame is given by $\Msol$, there are quantum mechanical fluctuations in the 3-momentum. Such effects can be taken into account by a proper separation between the collective center of mass coordinates and the vibrational modes. This can be achieved by defining appropriate coherent states followed by a projection onto zero-momentum eigenstates~\cite{Lubeck:1986if}. We will ignore such corrections and use $\Msol$ above as a proxy for the soliton mass, since the above precision is sufficient for our purposes. 

We show in the right panel of Fig.~\ref{fig:ProfilesLowQ} the mass of the lightest \sol/ as a function of~$Q$.\footnote{We show here the low $Q$ region of a scan over $\omega$, obtained by solving the EOM numerically, as described above. However, for the (almost) horizontal (orange) part of the curve we use instead the first order approximation (for  $\omega \approx \omega_c$) derived in~\cite{Friedberg:1976me}:
%
$\Msol - Q m_\phi =
(2 \pi^2 M_2^2 m_h^8)/(\lambda_h^2 m_\phi ^7 Q) + {\cal O}(Q^{-3})$
%
where the moment $M_2$ is calculable and gives $M_2 \approx 0.75$. The reason is that in this region the excited states are split by small energy differences (that tend to zero as $Q \to \infty$ on this branch), and it is difficult to isolate the ground state numerically.
}
One can distinguish two branches as one increases $\omega$ from small values up to $\omega_{\rm th} = m_\phi$ where the \sol/ solutions cease to exist. The charge decreases monotonically down to a minimum value ($\sim 17.9$ in the figure), then increases rapidly again and diverges as $\omega \to m_\phi$. It can be shown that all such \sol/ solutions are stable against small classical fluctuations~\cite{Friedberg:1976me}. It is, however, important to compare the \sol/ mass against the energy of $Q$ free elementary $\Phi$ quanta. We plot the difference $\Msol - Q m_\phi$, which shows that there are two branches to be distinguished. The first branch (orange) is unstable against quantum mechanical decay into $Q$ free particle states. The second branch (blue) is forbidden from decaying by a combination of energetic considerations and the conservation of the $Q$ charge. In fact, they correspond to stable quantum mechanical states. This defines a $Q_{S}$ that separates the two types of solutions. For the model parameters used in the figure, one finds $Q_{S} \approx 19.5$ (corresponding to $\omega_{S} = 286$~GeV). Thus, the $Q \approx 20.3$ profiles in the left panel of the figure correspond to an unstable soliton, while the $Q \approx 21$ and $Q \approx 505$ cases correspond to stable \sols/. 

In order to establish the scaling of $\Msol$ with $Q$, let us focus on \DMBs/ by assuming that the $h$ field vanishes inside the core. Then $\phi$ is given by Eq.~(\ref{phiApprox}) for $r < \Rsol \approx \pi/\omega$ (and zero for $r > \Rsol$). Neglecting the surface tension contributions from $h$ (i.e.~setting $h' = 0$ everywhere), one has from Eq.~(\ref{MPhi}):
\bea
\Msol &\approx& \frac{\pi^4\,m_h^4}{12\, \lambda_h\, \omega ^3} + Q\,\omega~,
\label{SimpleDMBQuadraticModel}
\eea
where we exchanged $\phi_0$ for $Q$ using Eq.~(\ref{QomegaScaling}). We are interested in the lowest energy solution for fixed $Q$. Minimizing against the dynamical variable $\omega$, we find that the minimum is at
\bea
\Msol &\approx& \frac{2 \sqrt{2}\,\pi}{3\,\lambda_h^{1/4}} \, Q^{3/4}\,m_h~.
\label{MPhiLargeQ}
\eea
Thus, together with the results of the previous section, we have that for \DMBs/ in the case that $\lambda_\phi = 0$, the following scaling laws between the charge, the size, and the mass of the \DMB/ apply~\footnote{We will see in the next section that $\Mphi^2$ by itself does not change these scaling relations.}
\bea
Q &\sim& \Rsol^4~,
\hspace{1cm}
\Msol ~\sim~ Q^{3/4} ~\sim~ \Rsol^3~.
\label{ScalingLawsQuadratic}
\eea
These scaling laws hold for \sols/ with a large $Q$. In the low $Q$ region displayed in the right panel of Fig.~\ref{fig:ProfilesLowQ} somewhat different relations are obeyed, that can only be found by a more detailed numerical analysis.

\begin{figure}
\centering
\includegraphics[width=0.47\textwidth]{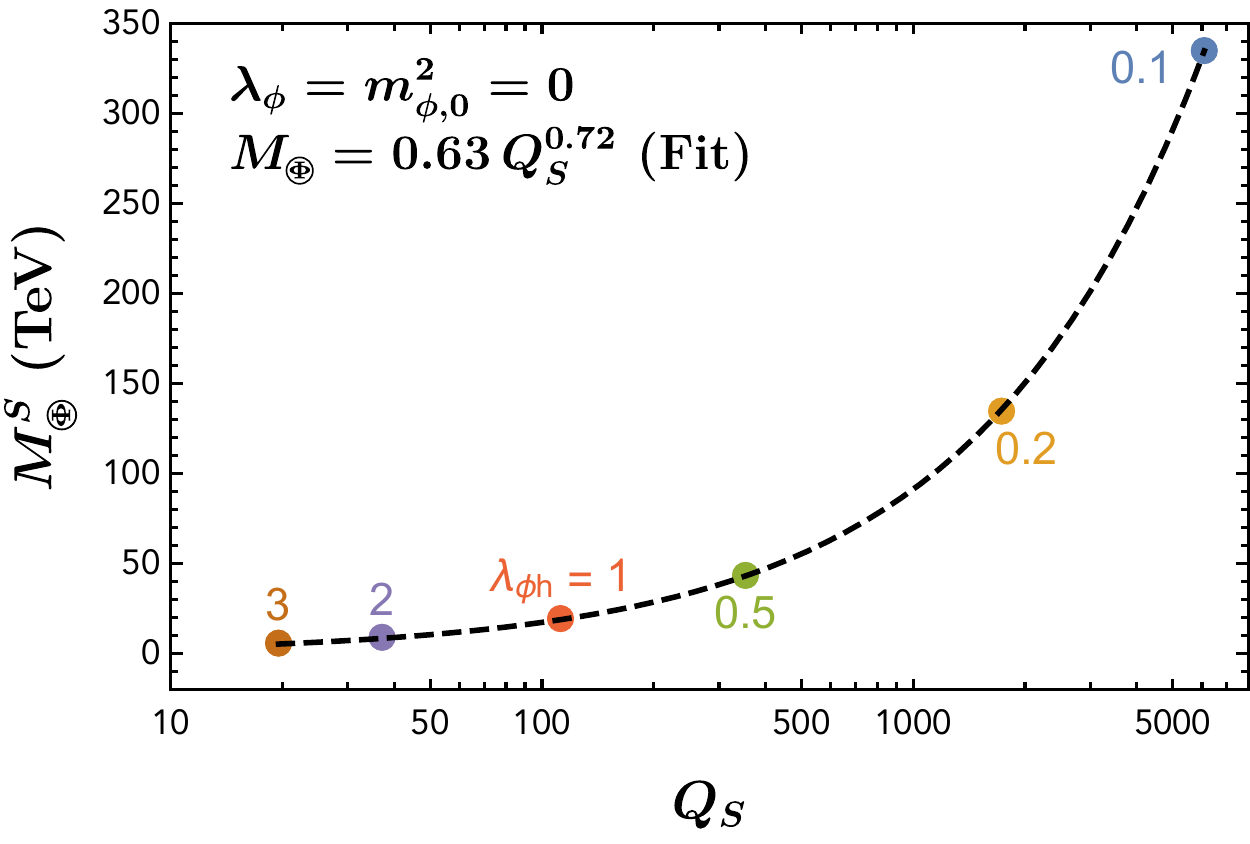}
\hspace{3mm}
\includegraphics[width=0.48\textwidth]{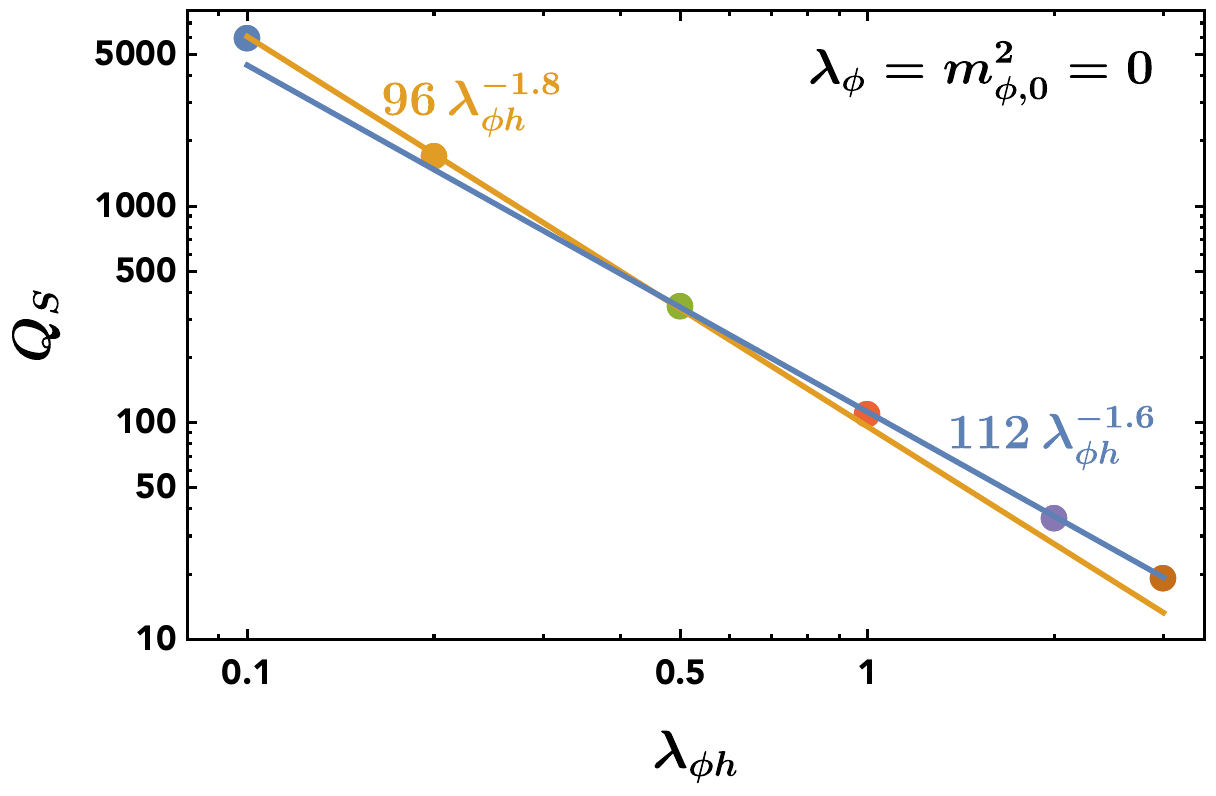}
\caption{Left panel: the \DMB/ mass with the critical charge $Q_S$ to have a stable \DMB/ for different values of $\lambda_{\phi h}$. Right panel: the critical charge $Q_S$ for several values of $\lambda_{\phi h}$ and a simple power-law fit. It is assumed that $ \lambda_\phi=\Mphi = 0$.
}
\label{fig:RMvsQ}
\end{figure} 

We focus on the stability point associated with the charge $Q_S$ that delimits the stable/unstable soliton configurations. In the left panel of Fig.~\ref{fig:RMvsQ} we show the corresponding \sol/ mass, $\Msol^S$, as a function of $Q_S$, as we vary $\lambda_{\phi h}$ in the range $[0.1,3]$. The simulated models are well fitted by~\footnote{If one uses Eq.~(\ref{MPhiLargeQ}) to determine $Q_S$, even though it is not meant to hold for the lowest charges corresponding to \sols/ that are not \DMBs/, one can estimate $Q_S \sim 512 \pi^4 m_h^2 / (81 \lambda_{\phi h}^2 v^2)$. This gives a reasonable order of magnitude estimate for $Q_S$, working better for larger $\lambda_{\phi h}$.}
\bea
\Msol^S &=& 0.63 \times Q_S^{0.72}~{\rm TeV}~.
\eea
This is a parametric relation across models as we vary $\lambda_{\phi h}$. The parametric dependence for $Q_S(\lambda_{\phi h})$ is shown in the right panel of Fig.~\ref{fig:RMvsQ}, and can be reproduced by a broken  power law in this range, as shown in the figure. From the information in both panels one can get $\Msol^S(\lambda_{\phi h})$. One can similarly consider the radius for such a minimum charge \sol/, which is well described by
\bea
\Rsol^S &=& 0.004 \times Q_S^{0.25}~{\rm GeV}^{-1}~.
\eea
Thus, the typical radii for such charges are of order $R \sim \textrm{few}~10^{-2}~{\rm GeV}^{-1} \sim 10^{-3}~{\rm fm}$, while the masses are in the tens of TeV and above region. These scales arise from the weak scale due to charge enhancements, qualitatively similar to the scaling laws discussed above, but not as simple. Based on these plots we can estimate the energy density associated with the \sol/ configuration to be of order
\bea
\rho &=& \frac{\Msol}{(4\pi/3) \Rsol^3} ~\sim~ (100~{\rm GeV})^4~.
\eea
To the extent that the scaling laws given in Eq.~(\ref{ScalingLawsQuadratic}) connect the low $Q$ and high $Q$ cases, we expect the same estimate to hold for very large $Q$ \DMBs/.

\subsection{Effects of the Dark Matter Bare Mass and Self-Quartic Interaction}
\label{OtherTwoParameters}

\begin{figure}[t]
\centering
\includegraphics[width=0.475\textwidth]{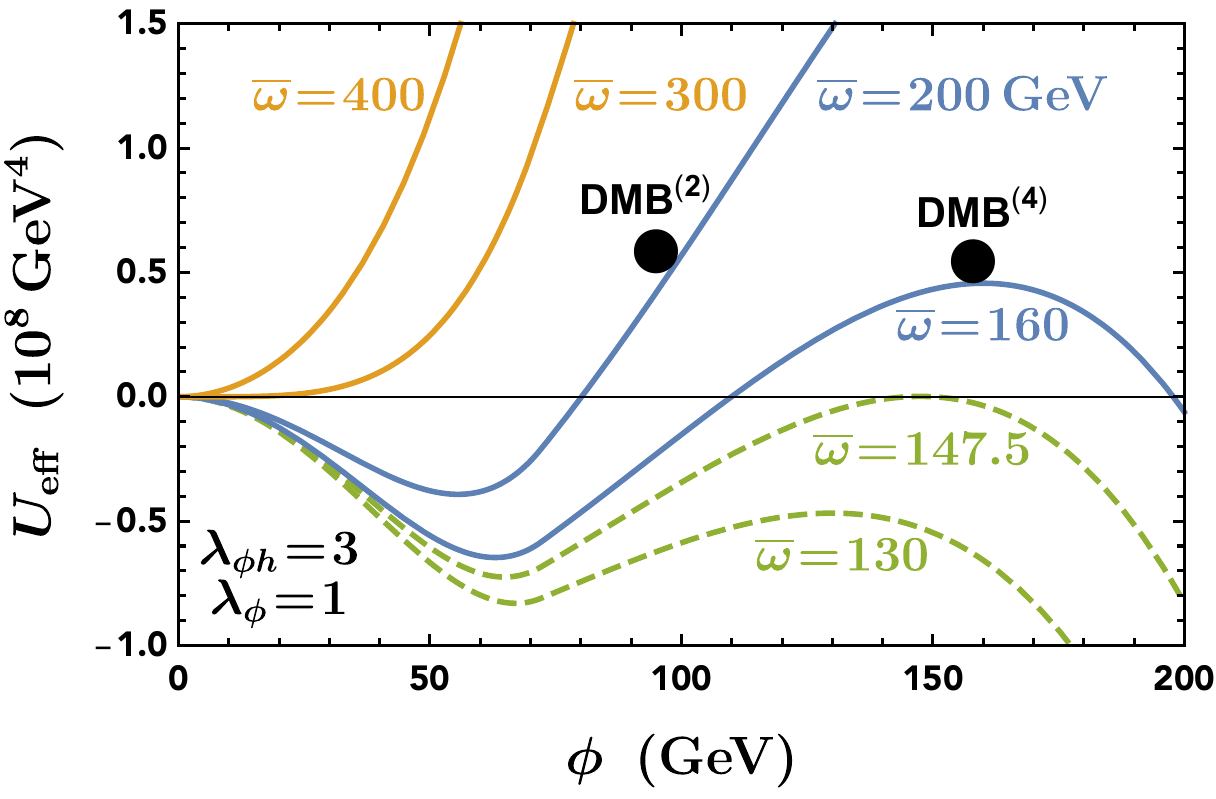}
\hspace{3mm}
\includegraphics[width=0.475\textwidth]{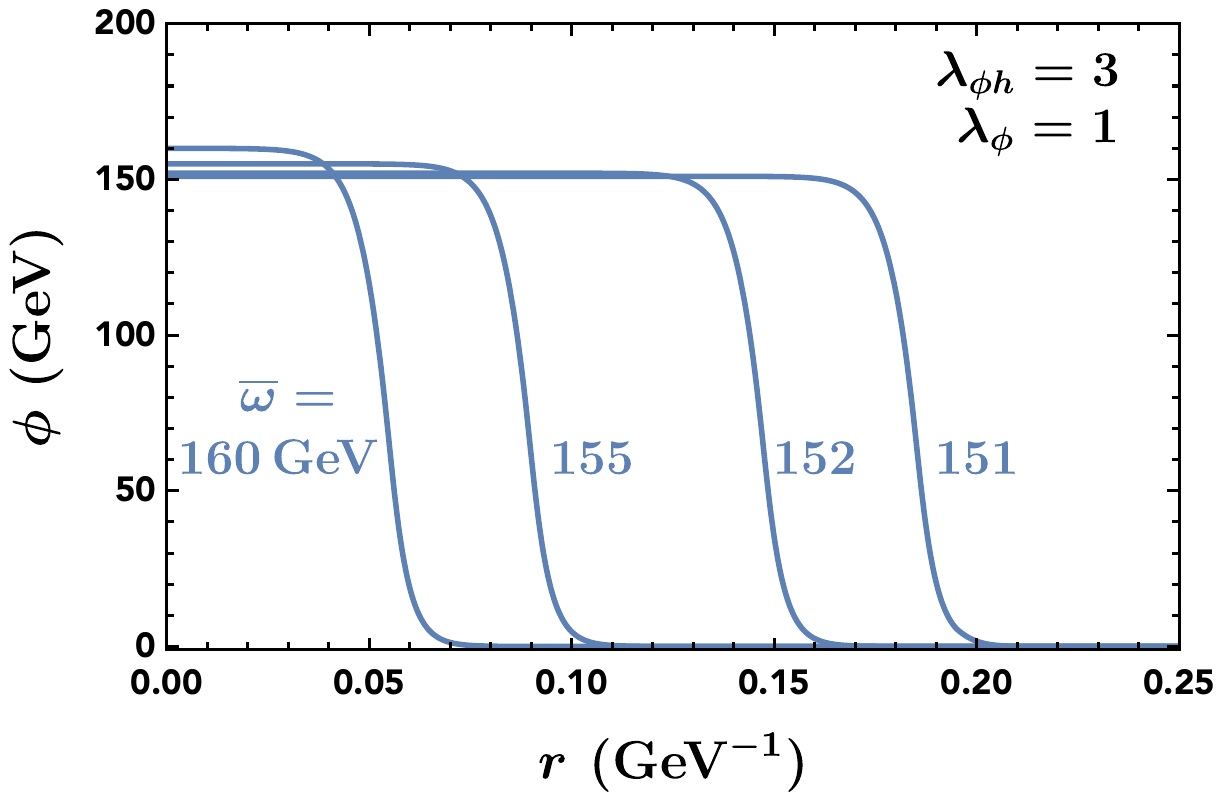}
\caption{Left panel: similar to Fig.~\ref{fig:Veff1} for the effective potential as a function of $\phi$, but including the  $\Phi$ bare mass and quartic self-interactions. For $\lambda_{\phi h}=3$ and $\lambda_\phi =1$, only the range $147.5\,\mbox{GeV} < \overline{\omega} < 301\,\mbox{GeV}$ allows for \sols/. The particles marked as $\DMB/^{(2)}$ and $\DMB/^{(4)}$ are meant to \textit{illustrate} the two classes of \DMBs/ that can appear in this system (see footnote~\ref{illustrationOnly}).
Right panel: \sol/ profiles for several values of $\overline{\omega}$ near $\overline{\omega}_c \approx 147.5$~GeV, taking $\lambda_{\phi h} =3$, $\lambda_{\phi} = 1$. The plateau approaches $\phi \approx \phi_{\rm max} \approx 147.5$~GeV as $\overline{\omega}$ approaches $\overline{\omega}_c$. The transition in the approximate Higgs profile, Eq.~(\ref{happrox}) is expected to also occur near $r = \Rsol$, where $\Rsol$ is the size of the \sol/, as read from the figure.}
\label{fig:ProfilesQuartic}
\end{figure}

We now comment on the effects of the remaining two parameters of the model, $\Mphi^2$ and $\lambda_\phi$. Within the context of the effective potential description defined in Eq.~(\ref{Veff}), one can see that
\begin{enumerate}

\item The bare mass $\Mphi^2$ can be easily included by defining an effective $\overline{\omega}^2 \equiv \omega^2 - \Mphi^2$ in the effective potential and associated EOM. One must only remember that when computing observables such as the mass of the \sol/ via Eq.~(\ref{MPhi}), it is the orthogonal combination $\omega^2 + \Mphi^2$ that appears. Similarly, the charge $Q$ is proportional to $\omega$, and the combination $\overline{\omega}$ enters only through $\phi$. With the solutions for the $\Mphi^2 = 0$ case at hand this can be easily taken into account.

\item The quartic self-interaction $\lambda_\phi$ has a more significant effect: it changes the large $\phi$ behavior of the effective potential from the quadratic one used in the previous section, turning it down to reach an asymptotic behavior $-\frac{1}{4} \lambda_\phi \phi^4$ (for $\lambda_\phi > 0$) (see the left panel of Fig.~\ref{fig:ProfilesQuartic}). This creates a maximum in the potential at some $\phi_{\rm max}$. The soliton solutions must therefore satisfy $\phi_0 < \phi_{\rm max}$, since for $\phi_0 > \phi_{\rm max}$ the solutions would run down the hill in the wrong direction and not be bounded. This is the scenario considered for $Q$-balls in Ref.~\cite{Coleman:1985ki}, and we know that stable solitonic configurations exist in this case.

\end{enumerate}

The first point could mean that even for ultraheavy elementary $\Phi$ particles that receive only a small contribution to their mass from EWSB, it could be possible to have solitonic configurations related to the weak scale, i.e.~sustaining an EW symmetric ``vacuum" in a finite region of space, inside the normal EW breaking vacuum.

Let us now describe some of the consequences of the quartic coupling $\lambda_\phi$, assuming for simplicity that we are interested in \sols/ with a large charge $Q$, such that they fall in the class of \EDBs/. In this case, the maximum of the effective potential described in point 2 above lies in the region $\lambda_{\phi h} \, \phi^2 > m_h^2$, where according to Eq.~(\ref{Veff}),
\bea
\Veff &=& \frac{1}{2} \, \overline{\omega}^2 \phi^2 - \frac{1}{4} \lambda_\phi  \phi^4 - \frac{m_h^4}{16\lambda_h}~.
\eea
This determines $\phi_{\rm max} = \overline{\omega}/\sqrt{\lambda_\phi}$ and $\Veff^{\rm max} = \overline{\omega}^4/(4\lambda_\phi) - m_h^4/(16\lambda_h)$. Since $\Veff(\phi = 0) = 0$, one must have $\Veff^{\rm max} > 0$, which defines a critical frequency
\bea
\overline{\omega}_c&=& \left( \frac{\lambda_\phi}{4\lambda_h} \right)^{1/4} m_h~,
\label{omegac}
\eea
such that soliton solutions must obey $\overline{\omega} > \overline{\omega}_c$. The conditon (\ref{HillCond}) must also be imposed, so that the origin be a maximum as opposed to a minimum, as discussed in the previous section. Thus, in the presence of $\lambda_\phi$, $\omega$ is bounded by non-zero values both from below and above. In the left panel of Fig.~\ref{fig:ProfilesQuartic}, we show the effective potential as a function of $\phi$ for several choices of $\overline{\omega}$ and fixed $\lambda_{\phi h}=3$ and $\lambda_\phi=1$. Only when $\overline{\omega}\in(147.5, 301)$~GeV one finds trajectories where the effective particle, starting at an appropriate $\phi_0$, comes to rest at $\phi=0$ at $r=\infty$, with an exponential approach, so that it effectively reaches the second hill in a finite $r$. These are the finite energy, finite $Q$, \sols/. We also indicate a categorization of two distinct classes of \sols/ in terms of the initial conditions in the particle mechanics analogue. The ``quadratic \sols/", discussed in the previous section, are denoted by $\DMB/^{(2)}$ in the figure. Those for which the quartic $\Phi$ self-interaction plays an essential role, as we are discussing in this section, are denoted by $\DMB/^{(4)}$, i.e.~we will refer to them sometimes as ``quartic \DMBs/".\footnote{\label{illustrationOnly}Unlike in Fig.~\ref{fig:Veff1}, this is only to illustrate the concept. In particular, as was shown in that figure, the $\DMB/^{(2)}$ should start higher on the $\overline{\omega} = 200$~GeV curve than the region shown in Fig.~\ref{fig:ProfilesQuartic}, where the quartic effects become more important. For sufficiently small $\lambda_{\phi}$ both types of \DMBs/ can coexist.} 

Although the allowed range of $\overline{\omega}$ is limited, soliton solutions with arbitrarily large charges exist. These occur for $\overline{\omega}$ close to $\overline{\omega}_c$, and are obtained by making the volume large, as they display a uniform charge density. Thus, such balls behave like aggregates of $\Phi$ matter~\cite{Coleman:1985ki}. We show in the right panel of Fig.~\ref{fig:ProfilesQuartic} the numerical solutions obtained within the effective potential approach, for $\lambda_{\phi h} =3$, $\lambda_{\phi} = 1$, and for several values of $\overline{\omega}$ near $\overline{\omega}_c \approx 147.5$~GeV, as obtained from Eq.~(\ref{omegac}). One can see that as $\overline{\omega}$ approaches $\overline{\omega}_c$, the size of the soliton increases, and the profiles resemble a step-function, much more than when the quartic coupling is absent or negligible. This can be easily understood from the particle mechanics analogy: one starts at rest near the top of the potential maximum at $\phi_{\rm max}$, and slowly picks up speed for a long ``time"~$r$, generating a nearly constant inner core. At some point enough speed is attained and the particle falls down the potential in a short time, decelerates rapidly as it approaches the local maximum at the origin, and eventually comes to rest there. We can therefore use a simple step function profile for $\phi$ to estimate quantities of interest, where for concreteness we can take the size of the $\phi$ profile as the $\Rsol$ such that $\phi(\Rsol) = \phi_{\rm max}/2$. In the limit of $\overline{\omega} - \overline{\omega}_c \ll \overline{\omega}_c$, the \DMB/ radius shows a simple scaling as $\Rsol \propto 1/(\overline{\omega} - \overline{\omega}_c)$, which can be seen in the right panel of Fig.~\ref{fig:ProfilesQuartic}. For $\lambda_{\phi h} =3$ and $\lambda_{\phi} = 1$, the overall coefficient can be fitted from the numerical results: $\Rsol \approx 0.66/ (\overline{\omega} - \overline{\omega}_c)$.

The Higgs profile also takes a step-like form, with 
\bea
h &\approx& 
\left\{ 
\begin{array}{ccc}
 v & \hspace{5mm} \textrm{for} & 
 \lambda_{\phi h} \, \phi^2 < m_h^2~,  \\ [0.5em]
 0  &  \hspace{5mm} \textrm{for}  &  \lambda_{\phi h} \, \phi^2 > m_h^2~.
\end{array}
\right.
\label{happroxQBall}
\eea
The Higgs profile ``size" is determined by the $R_h$ such that $\phi(R_h) = m_h/\sqrt{\lambda_{\phi h}}$, which we have assumed is smaller than $\phi_{\rm max}$. Taking the ratio of the two $\phi$ values that define these radii, we have
\bea
\frac{\phi(\Rsol)}{\phi(R_h)} &=& \frac{1}{2} \left( \frac{\lambda_{\phi h}}{\lambda_{\phi}} \right)^{1/2} \frac{\overline{\omega}}{m_h}
~\approx~ \left( \frac{\lambda_{\phi h}^2}{64 \lambda_h \lambda_{\phi}} \right)^{1/4}~,
\eea
where in the second relation we assumed $\overline{\omega} \approx \overline{\omega}_c$. For order one couplings $\lambda_{\phi h}$ and $\lambda_\phi$, the ratio is of order one, so that the two radii can be identified: $R \equiv R_h \sim \Rsol$. When $\lambda_{\phi}$ is small, there can be some difference between the two radii. However, one expects at least a 1-loop size of order $\lambda_\phi^{\rm 1-loop} \sim \lambda_{\phi h}^2/(16\pi^2)$, so that the above ratio of VEVs is not expected to be greater than $(2\pi^2)^{1/4} \sim 2$, and therefore the two radii should be close enough to be identified as in the case of order one couplings.

The charge of the \sol/, Eq.~(\ref{QCharge}), in the case $\overline{\omega} \approx \overline{\omega}_c$, is then approximately given by
\bea
Q &\approx& \left( \frac{4\pi}{3} \, \Rsol^3 \right) \omega_c \, \phi_{\rm max}^2 ~=~ \frac{4\pi}{3} \, \frac{1}{\lambda_\phi} \, \omega_c \, \overline{\omega}_c^2\,\Rsol^3~.
\label{QQBall}
\eea
The soliton mass, Eq.~(\ref{MPhi}), neglecting the $h$ surface tension contributions, is given here by
\bea
\Msol &\approx& 4\pi \int_0^{\Rsol} \! dr \, r^2 \left\{ \frac{1}{2} \omega^2_c \phi^2_{\rm max} + V_H(0) + V_\Phi(\phi_{\rm max}) \right\}
\nonumber \\[0.5em]
&=& \left( \frac{4\pi}{3} \, \Rsol^3 \right) \left\{ \frac{1}{2} (\omega^2_c + \Mphi^2) \phi^2_{\rm max} + \frac{\overline{\omega}^4_c}{4\lambda_\phi} + \frac{1}{4} \lambda_\phi  \phi^4_{\rm max} \right\}
\nonumber \\[0.5em]
&=& Q\,\omega_c~,
\label{MphiQBall}
\eea
where we used $V_H(0) = m_h^4/(16\lambda_h) = \overline{\omega}_c^4/(4\lambda_\phi)$ due to the condition $\Veff^{\rm max} = 0$ at $\overline{\omega} = \overline{\omega}_c$. Compared to the energy, $Q\,m_\phi$, of $Q$ free quanta, each of mass $m_\phi$ as given in Eq.~(\ref{mphi}), we have
\bea
\frac{\Msol}{Q\,m_\phi} &=& \left( \frac{\Mphi^2 + \sqrt{\lambda_\phi}\,m_h^2/\sqrt{4\lambda_h}} {\Mphi^2 +  \lambda_{\phi h}\,m_h^2/(4\lambda_h) } \right)^{1/2}~.
\label{eq:DMB-free-particle-mass-ratio}
\eea
We see that the above ratio is less than one when
\bea
\frac{\sqrt{\lambda_\phi \lambda_h}}{\lambda_{\phi h}} &<& \frac{1}{2}~\, \qquad \mbox{i.e.} \qquad \frac{\sqrt{\lambda_\phi}}{\lambda_{\phi h}} < 1.4 ~.
\label{stabilityQuarticDMB}
\eea
 For instance, if $\lambda_\phi \sim \lambda_\phi^{\rm 1-loop} \sim \lambda_{\phi h}^2/(16\pi^2)$, this is always satisfied. The \sol/ is then the \textit{lowest energy per dark number} state, and stable. 

We also see that for large $Q$, large $\Rsol$ \DMBs/ in the presence of a $\lambda_\phi \neq 0$, which have $\omega \approx \omega_c$, one has the following scaling laws between the \DMB/'s charge, size and  mass:
\bea
Q &\sim& \Rsol^3~,
\hspace{1cm}
\Msol ~\sim~ Q ~\sim~ \Rsol^3~.
\label{ScalingLawsQuartic}
\eea
Thus while in both types of \DMBs/ we have discussed, $\Msol$ scales with volume, they can carry very different charges. The formation mechanism for such aggregates of charges will determine the type of \DMBs/ one would expect. We discuss these issues next.

\section{Early Universe Production of \DMBs/} 
\label{sec:early-universe}
Depending on the early-universe history, there could be several possible ways to produce \DMB/s, in this section we concentrate a simple mechanism based on first-order phase transition. Especially, with the extension of the singlet scalar of the SM Higgs potential, the electroweak symmetry breaking is naturally a first-order one. Furthermore, the typical dark matter number in one \DMB/ depends whether there is an asymmetry in dark matter and antimatter or not. For simplicity, we just assume that the dark matter asymmetry is given by some ultra-violent physics and have the production mechanism similar to the ``quark nugget" in Ref.~\cite{Witten:1984rs}.

\subsection{First-Order Electroweak Phase Transition}
\label{StrongEWPT}
The tree level scalar potential is given by $V_{\omega=0}(h,\phi)$ in Eq.~(\ref{Vomega}), setting $\omega=0$. The form of this potential is equivalent to one obtained by addition to the SM Higgs potential of two real scalar singlets, corresponding to the real and imaginary parts of $\Phi$. In the early universe, at very high temperature, the global minimum occurs at the electroweak symmetry preserving point $(\avg{h},\, \avg{\phi}) = (0,\, 0)$. As the universe cools down, the global minimum happens at an EWSB vacua with $(\avg{h},\, \avg{\phi}) = (v,\, 0)$. Depending on the coupling $\lambda_{\phi h}$, one can have a ``one-step" phase transition where the phase transition occurs purely along the Higgs direction. 

To study the electroweak phase transition (EWPT), we consider the effective finite-temperature potential $V_{\rm eff}(h, T)$ where $h$ is the real component of the SM Higgs doublet, $H^T = (0,h/\sqrt{2})$ and $T$ is temperature~\cite{Gross:1980br,Parwani:1991gq,Arnold:1992rz,Carrington:1991hz,Delaunay:2007wb,Carena:2008vj,Katz:2014bha,Curtin:2016urg,Jain:2017sqm}
\beq
\label{eq:TFD}
V_{\rm eff}(h, T) \equiv V_{\rm tree}(h) + \sum_i V_{\rm CW}\left[m^2_i(h)\right] + \sum_i  V_{\rm T}\left[m^2_i(h) + \Pi_i,T\right]~,
\eeq
where $\Pi_i$ is thermal masses (or Debye masses) (see its formulas in \cite{Curtin:2016urg} for instance). The first term $V_{\rm tree}(h) = \lambda_h(h^2 - v^2)^2/4$ is the tree-level SM Higgs potential. The second term $V_{\rm CW}$ is the one-loop contribution to the zero-temperature effective potential, also known as Coleman-Weinberg potential~\cite{Coleman:1973jx}. Using the on-shell renormalization scheme in the Landau gauge, it is given by~\cite{Delaunay:2007wb}
\bea
\sum_i  V_{\rm CW}\left[m^2_i(h) \right]&=& \sum_i  (-1)^{F_i}\frac{g_i}{64 \pi^2} \bigg[ m_i^4(h) \left ( \log \frac{m_i^2(h) }{m_i^2(v) } - \frac{3}{2} \right ) \,+ \,2\, m_i^2(h)\, m_i^2(v) \bigg] \label{eq:A_VCW}~,
\eea
where $g_i$ is the degree of freedom for each particle, $F_i = 1(0)$ for fermions(bosons), $m_i(h)$ are masses in the presence of a background Higgs field with $i= {t,W,Z,h, \Phi}$ and ignoring lighter fermions. The finite-temperature correction term has
\beq
\sum_i V_{\rm T}\left[m^2_i(h) + \Pi_i,T\right] = \sum_i (-1)^{F_i} \, \frac{g_i\,T^4}{2 \pi^2} \int_0^{\infty} dx ~x^2 \log \bigg[1 \mp e^{\big(-\sqrt{x^2+ \left ( m_i^2(\phi)+ \Pi_i \right )/T^2} \big)}    \bigg] \label{eq:A_VT}~,
\eeq
where the integral with ``$-/+$" sign denotes the thermal bosonic/fermionic function. 

Before we provide the parameter space for first-order phase transition, we want to note that requiring the ordinary electroweak vacuum with $\langle h \rangle=v=246$~GeV as the global vacuum at $T=0$ or $V_{\rm eff}(v,0) < V_{\rm eff}(0, 0)$ sets a constraint on the coupling $\lambda_{\phi h}$ and the bare mass $\Mphi$~\cite{Chung:2012vg}. When $\Mphi=0$, this requires
\beqa
\lambda_{\phi h} \lesssim \frac{4\sqrt{2}\,\pi\,m_h}{v} \approx 9.0  ~.
\label{eq:lambdaphih-upper-bound}
\eeqa
The two-loop effective potential could slightly change this numerical number. For the range of $0 \leq \Mphi \leq 200$\,GeV, the upper bound on $\lambda_{\rm \phi h}$ varies from 9.0 to 10.0. Therefore, in our numerical calculation for the parameter space of phase transition, we will restrict ourselves to this allowed range. 

\begin{figure}[thp] 
	\centering
		\includegraphics[width=0.46\linewidth]{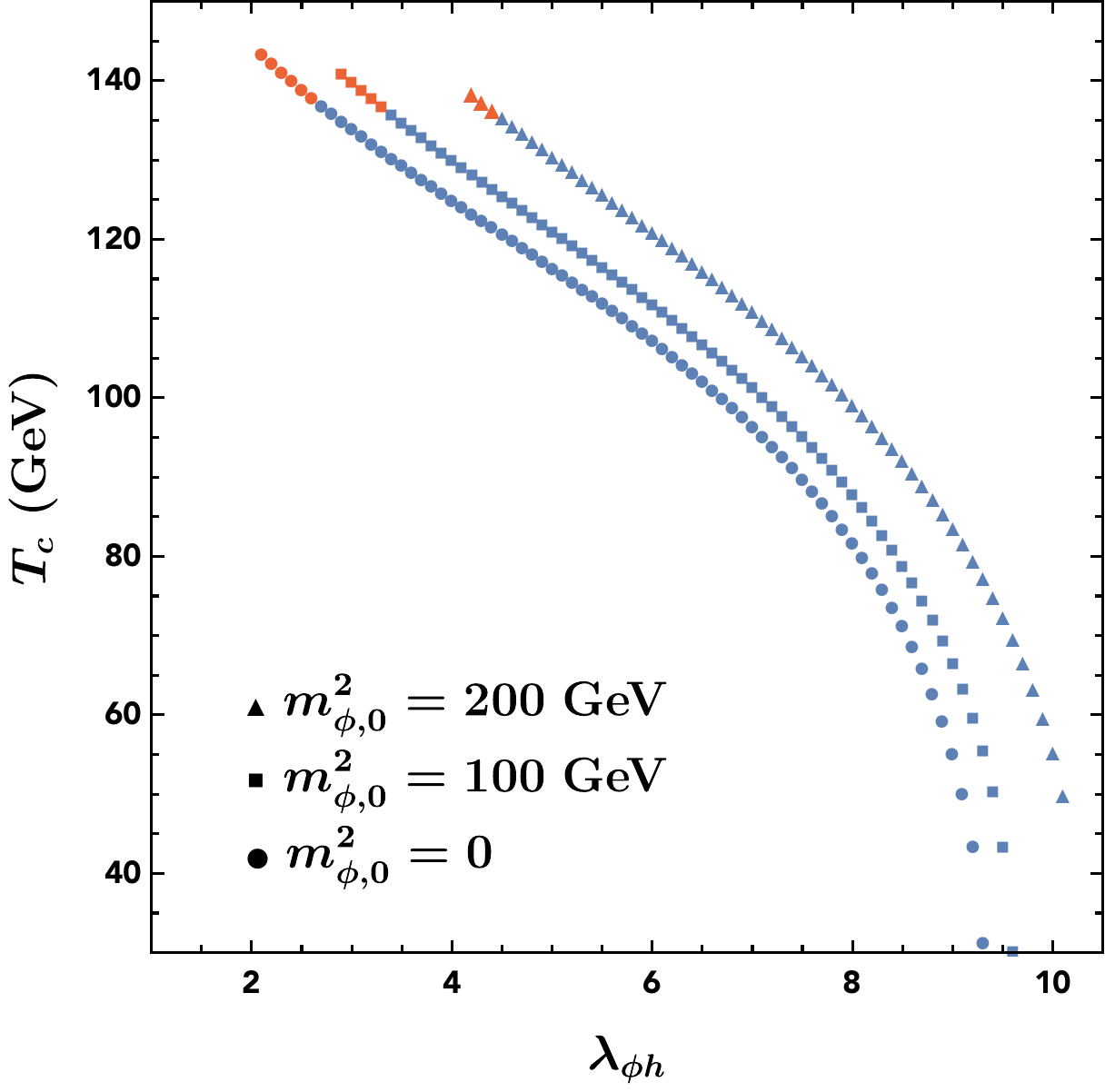} 
\quad \quad \hspace{3mm}
	\includegraphics[width=0.445\linewidth]{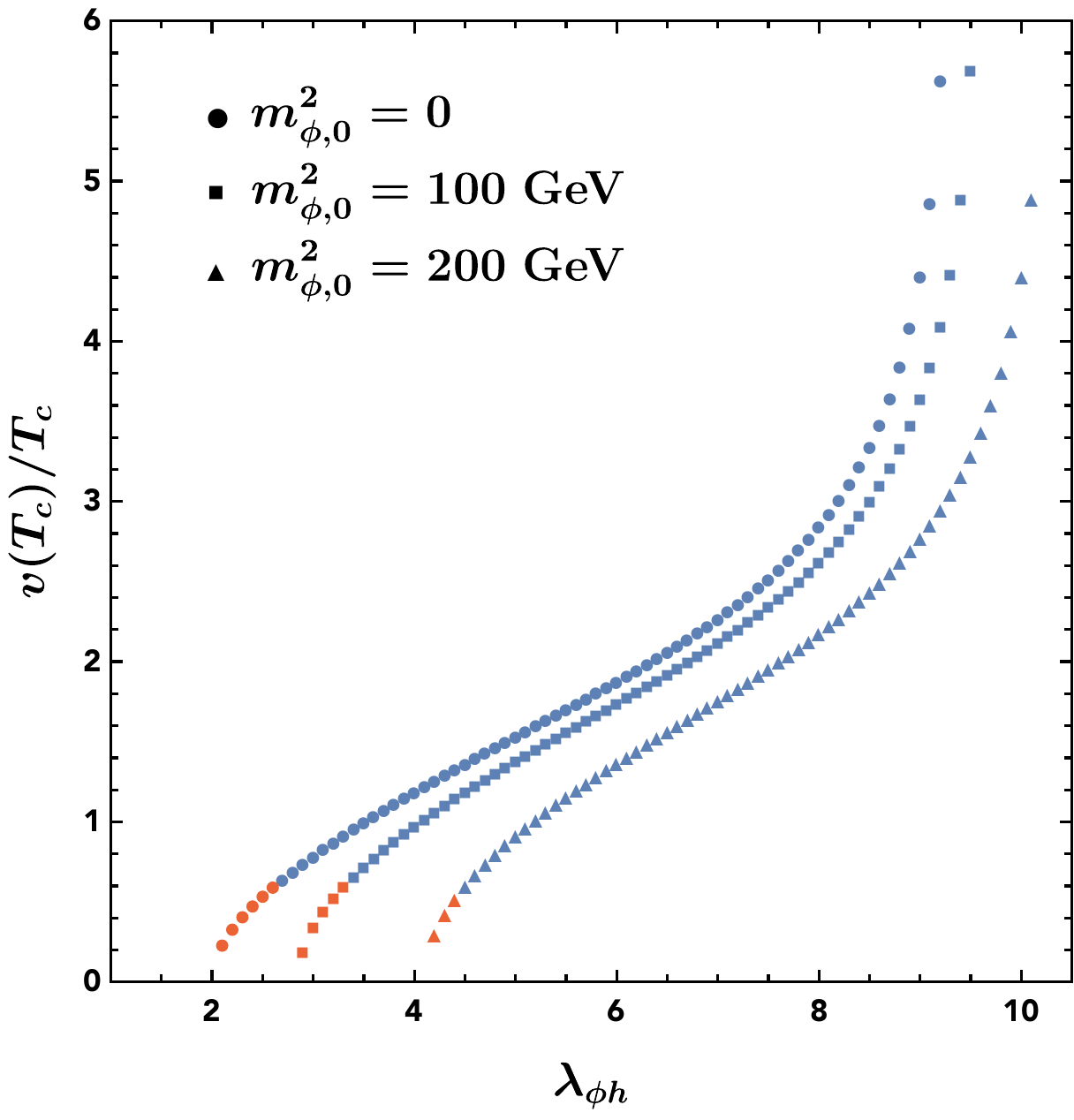} 
	\caption{Left panel: $T_c$ as a function of $\lambda_{\phi h}$ for different values of $\Mphi$.  Right panel: Strength of EWPT, $v(T_c)/T_c$ as a function of $\lambda_{\phi h}$. In both plots the red points correspond to $v(T_c)/T_c < 0.6$ or a weak first-order phase transition, while the blue points are for $v(T_c)/T_c > 0.6$ or a strong first-order phase transition, which we call strong EWPT. The ratio $v(T_c)/T_c$ grows very rapidly as $\lambda_{\phi h}$ approaches around 9 for $\Mphi=0$. }
	\label{fig:vcTclam}
\end{figure}

In the left panel of Fig.~\ref{fig:vcTclam}, we show the first-order phase transition temperature as a function of $\lambda_{\phi h}$ for different bare mass $\Mphi = \{0, 100, 200\}~\GeV$. In each curve, we also separate it into two regions with the strong first-order phase transition region in blue with $v(T_c)/T_c \geq 0.6$~\cite{Patel:2011th} and the weak first-order phase transition region in red with  $v(T_c)/T_c < 0.6$. For $\Mphi = 0$~GeV, the first-order phase transition happens for $\lambda_{\phi h} \gtrsim 2$, while the strong first-order phase transition happens for $\lambda_{\phi h} \gtrsim 2.6$. As $\lambda_{\phi h}$ increases but below the upper bound in \eqref{eq:lambdaphih-upper-bound}, the phase transition temperature decreases. For the benchmark point with $\lambda_{\phi h} = 3$, the phase transition temperature has $T_c \approx 134$~GeV. In the right panel of Fig.~\ref{fig:vcTclam}, we show the ratio of the $T_c$-dependent EWSB VEV $v(T_c)$ over $T_c$ as a function of $\lambda_{\phi h}$. Again, the strong(weak) first-order phase transition region is denoted in blue(red) color. As the coupling $\lambda_{\phi h}$ increases, the ratio of $v(T_c)/T_c$ increases. We also note that for both plots, the $\Phi$ self-interaction quartic coupling $\lambda_\phi$ does not play a role for the one-step phase transition evaluated at the one-loop level. 

\subsection{Formation of \DMBs/ from First-Order Phase Transition}
\label{BallProduction}

We discuss now how \DMBs/ might be formed during the EWPT in the early universe. As we will see, the formation of the \DMBs/ requires the transition to be a strong first-order. We will also discuss their expected average properties such as charge, mass and size. 

For the purpose of this section, we assume that some high-scale physics, analogous to leptogenesis, has already generated a $Q$ asymmetry, that we will call \DMN/,\footnote{To emphasize its connection to DM, we will sometimes refer to the $Q$ charge of a state as \textit{\DMN/}. When dealing with free fundamental $\Phi$ quanta, this is the difference between $\Phi$-particles and $\Phi$-antiparticles. It applies more generally to extended classical field configurations with no well-defined number of particles.} with a yield $Y_{\Phi} \equiv n_{\Phi}/s$, which we treat as a UV-dependent free parameter. Here,  the entropy density is $s = (2\pi^2/45)g_{*S}\,T^3$, with $g_{*S}$ being the effective number of relativistic degrees of freedom. As a reference point, the SM baryon number asymmetry is measured to be $Y_{\rm B} \simeq 10^{-10}$~\cite{Aghanim:2018eyx}. It would be interesting if there was a common origin for $Y_\Phi$ and $Y_{\rm B}$, in which case one would expect $Y_\Phi \sim Y_{\rm B}$, at least if the generation occurs at the same  time. Realizing such a scenario would require additional model assumptions. However, one should note that the presence of the complex scalar can already lead a strong first-order EWPT, which is one of the conditions for EW baryogenesis. Thus, one may be able to build a model to also generate the \DMN/ asymmetry within the framework of EW baryogenesis, which we will not explore in this paper. 

We organize the analysis in three stages for conceptual clarity:
\begin{enumerate}

\item The ``snowplow" stage, taking place around $T_c$, when the EWSB nucleation process happens. We will argue that a large fraction of the \DMN/ ends up in the unbroken phase, as opposed to the true vacuum (broken) phase.

\item The second stage is delimited by the formation of \DMBs/ from the \DMN/ stored inside regions of unbroken phase. 

\item Subsequent to the \DMB/ formation, the free \DMN/ gets rid of its symmetric component, leaving behind  the asymmetric yield $Y_\Phi$.

\end{enumerate}
We will argue that up until the freeze-out temperature $T_F$ of the free $\Phi$ particles in the broken phase, \DMN/ continues being accumulated inside the \DMBs/. The end result is that the amount of \DMN/ stored in elementary $\Phi$ quanta is exponentially suppressed.

We start with the snowplow stage. Just below the EWPT temperature $T_c \sim 130$~GeV, the EWSB (true vacuum) bubbles start to pop up, and grow when they surpass a critical size. During the bubble nucleation process, one immediate question is whether the \DMN/ stays mainly in the unbroken or broken phases. To address this, we first give a simple kinetic argument, assuming that $\Mphi^2=0$ (or that it can be neglected).\footnote{\label{smallAsymmetry}For large $\Mphi^2$, such that the $\Phi$ particles are non-relativistic already at $T_c$ in the unbroken phase, taking into account the conserved \DMN/ is more involved. Considerations analogous to the ones detailed in this section would allow to determine how much of the  \DMN/ ends up in the broken versus unbroken phases. However, this would be  relevant only in the presence of additional physics that would account for the first-order phase transition, since the small abundance of $\Phi$ particles would have a negligible effect on the finite-temperature Higgs effective potential. Thus, we do not consider this case, and focus on $0 \leq \Mphi^2 \leq (200~{\rm GeV})^2$, as discussed in Section~\ref{StrongEWPT}. Note however, that Eqs.~(\ref{energyconservation}) and (\ref{minlambdaphih}) remain unchanged in the presence of an arbitrary $\Mphi^2$.} At leading order, the answer involves the $\Phi$ particle mass $m_\phi(T) \approx \sqrt{\lambda_{\phi h}/2}\,v(T)$, the phase transition temperature, $T_c$, and the bubble wall speed $\beta_{\rm w}$ (or the corresponding boost factor $\gamma_{\rm w}=1/\sqrt{1-\beta_{\rm w}^2}$). It is convenient to work in the bubble wall's rest frame, which sees a stream of $\Phi$ particles moving in the $\hat{z}$ direction (this is just the direction of expansion of  the bubble wall in the plasma frame). From energy conservation, the condition for a $\Phi$ particle to remain in the unbroken phase, where it is massless, is $\hat{p}_z^2 \leq m_{\phi,c}^2$, where $m_{\phi,c} \equiv m_\phi(T_c)$. Here the hat denotes that $\hat{p}_z$ is the momentum of the particle in the wall's rest frame. Boosting this condition back to the plasma frame, we arrive at
\beqa
(\beta_{\rm w}\gamma_{\rm w} E + \gamma_{\rm w} \,p_z)^2 \leq m_{\phi,c}^2 ~.
\label{energyconservation}
\eeqa
For a non-relativistic wall speed, $\beta_{\rm w} \ll 1$, this condition simplifies to $p_z^2 \leq m_{\phi,c}^2$. Using the Bose-Einstein statistics distribution, the average momentum is $\langle p_z^2 \rangle = \langle p^2\rangle /3 = [4 \zeta(5)/\zeta(3)] \, T_c^2 \approx 3.5\, T_c^2$. So for the bubble wall to ``snowplow" the \DMN/ into the unbroken phase one needs 
\bea
\lambda_{\phi h} &\geq& \frac{8 \zeta(5)}{\zeta(3)} \, \frac{T_c^2}{v(T_c)^2} ~\approx~ 7.0 \times \frac{T_c^2}{v(T_c)^2}~.
\label{minlambdaphih}
\eea 
From the relation between $\lambda_{\phi h}$ and $v(T_c)/T_c$ shown in the right panel of Fig.~\ref{fig:vcTclam}, one can infer that one needs a modestly large value of $\lambda_{\phi h} \gtrsim 4$ so that most of the \DMN/ stays in the unbroken phase. 

Instead of kinematic arguments, one can also provide an estimation based on chemical equilibrium considerations. Here the condition of chemical equilibrium, $\mu_\Phi^{\rm (h)}=\mu_\Phi^{\rm (l)}$, allows to estimate the ratio of \DMN/ in the high-temperature, ``h", and  low-temperature, ``l", phases. In both phases and not far below $T_c$, one has $\mu_\Phi/T \ll 1$ (small asymmetry, see footnote~\ref{smallAsymmetry}). For a relativistic gas of elementary $\Phi$ particles, one has at $T$
\bea
n_\Phi^{\rm (h)} &\approx& \frac{1}{3} \, \mu_\Phi^{\rm (h)}\,T^2~,
\eea
where $n_\Phi = n_{\Phi} - n_{\Phi^\dagger} $, while for non-relativistic $\Phi$ particles~\footnote{This is the case, in particular around $T_c$, inside the true EWSB vacuum during a sufficiently strong first-order EWPT induced by the $\lambda_{\phi h}$ coupling,. Referring to Fig.~\ref{fig:vcTclam}, the non-relativistic limit should hold approximately for $v(T_c)/T_c \gtrsim 1$, and to good accuracy for $v(T_c)/T_c \gtrsim1.5$.}
\bea
n_\Phi^{\rm (l)} &\approx& \left(\frac{2\,\mu_\Phi^{\rm (l)}}{T} \right) \left( \frac{T\,m_{\phi}(T)}{2\pi} \right)^{3/2}\, e^{- m_{\phi}(T) / T}~.
\eea
Thus, when in chemical equilibrium,
\beqa
r &\equiv& \frac{n_\Phi^{\rm (l)}}{n_\Phi^{\rm (h)}} ~\approx~ 6 \, \left( \frac{m_{\phi}(T) }{2\pi\,T} \right)^{3/2}\, e^{- m_{\phi}(T) / T}~.
\label{eq:r-equilibrium}
\eeqa
For a heavy elementary $\Phi$ particle, $r$ is suppressed. In the case that $\Mphi^2 = 0$, and assuming that the inequality (\ref{minlambdaphih}) is saturated, one has $m_{\phi,c}/T_c \approx 1.86$, and $r\approx 0.15$. However, the chemical equilibrium between inside and outside of the \DMB/ could be kept until a lower temperature, $T_F$. This is because the free $\Phi$ and $\Phi^\dagger$ can be absorbed by the \DMBs/ or a large binding energy can be released when free $\Phi$ and $\Phi^\dagger$ particles enter \DMBs/. 

The relevant process is $\footnotesize{ \Pcirc}_{Q} + \Phi \leftrightarrow \footnotesize{ \Pcirc}_{Q+1} + X$ with $X$ denoting SM particles. We first note that when the temperature is above the ``binding energy", $E_{\rm bind}(T) \equiv m_{\phi}(T) - \omega(T)$ [with $\omega(T)$ as the temperature-dependent energy per charge for the soliton state], both forward and backward processes are efficient. The chemical equilibrium between \DMB/  and free $\Phi$ state is reached. As $T <  E_{\rm bind}(T)$, the free $\Phi$ can be absorbed by the \DMBs/, but not the other way. The freeze-out temperature, $T_F$, for $\footnotesize{ \Pcirc}_{Q} + \Phi \rightarrow \footnotesize{ \Pcirc}_{Q+1} + X$, is anticipated to be satisfy $T_F < E_{\rm bind}(T_F)$. The free $\Phi$ particle absorbing rate by \DMBs/ is estimated to be
\beqa
\Gamma_{Q + \Phi \rightarrow Q+1} =  \langle \sigma v \rangle\,n_{\tiny{\Pcirc}} \,  \simeq  4\,\pi\, R^2_{\tiny{\Pcirc}}(T) \, \frac{Y_{\Phi} \,s}{Q} = 4\,\pi\, R^2_{\tiny{\Pcirc}}(T) \, \frac{Y_{\Phi}}{Q}\,  \frac{2\pi^2}{45}\,g_{*s} \, T^3  \,,
\label{eq:absorb-rate}
\eeqa
with the radius of \DMB/ as a function of $T$.  Just below the temperature of the ending of nucleations $T_f$, the number of \DMB/, $N^{\rm Hubble}_{\DMB/}$, within one Hubble patch is estimated in Eq.~(\ref{Nnucleations}). Using it, the averaged radius of \DMB/ is around $R_{\tiny{\Pcirc}}(T_f)  \sim (N^{\rm Hubble}_{\DMB/})^{1/3}\,d_H$. The inverse Hubble distance is $1/d_H = H(T_f) = \sqrt{ \pi^2 g_*/90}\,T_f^2/M_{\rm P}$, where the reduced Planck mass $M_{\rm P} = 2.43 \times 10^{18}$~GeV. As the Universe cools, the radius of \DMB/ also reduces, which requires a more detailed understanding of how \DMB/ evolves at a non-zero temperature and non-zero vacuum pressure. As a simplistic estimation, we assume its radius shrinking velocity is $\beta_{\rm w}$ from $T_f$ to the freeze-out temperature $T_F$. Using the relation $t = 1/(2 H)$, we have the radius as a function of temperature as
\beqa
R_{\tiny{\Pcirc}}(T) \simeq  R_{\tiny{\Pcirc}}(T_f)  - \frac{\beta_{\rm w}}{2} \left[ H(T)^{-1} - H(T_f)^{-1} \right] \,.
\eeqa
Substituting $R_{\tiny{\Pcirc}}(T)$ into Eq.~\eqref{eq:absorb-rate} and requiring $\Gamma_{Q + \Phi \rightarrow Q+1}  \simeq H(T)$, we have the approximate freeze-out temperature as
\beqa
T_F \approx \frac{\beta_{\rm w}^{1/2}}{\sqrt{2}} \, \frac{T_f}{(N^{\rm Hubble}_{\DMB/})^{1/6}}  ~, 
\label{eq:freeze-out-temperature}
\eeqa
For $\beta_{\rm w} =1/\sqrt{3}$, $T_f \approx 133.4$~GeV and $N^{\rm Hubble}_{\DMB/} = 1.0 \times 10^{13}$ for $\lambda_{\phi h}=3$ from Eq.~(\ref{Nnucleations}), one has $T_F \approx 0.49$~GeV. One can then substitute $T_F$ into \eqref{eq:r-equilibrium} to obtain $r\approx 4.3\times 10^{-265}$ for $\lambda_{\phi h}=3$. For sure, our estimation of the freeze-out temperature and the ratio $r$ is a naive one, but the results do suggest that  the fraction of dark number in the free $\Phi$ particle state can be neglected for the phenomenological purpose.

As the universe cools down, $\Phi$ and $\Phi^\dagger$ states annihilate into SM particles and leave behind the asymmetric component. If one is allowed to ignore additional \DMN/ shuffling processes from one phase to the other or below the freeze-out temperature $T_f$, DM in our Universe could be composed of both macroscopic \DMBs/ and microscopic $\Phi$-particle states, if both states are stable. Focusing on quartic \DMBs/, we saw in Eq.~(\ref{MphiQBall}) and the subsequent discussion, that the energy per charge of the \DMB/ is given by $\omega_c$, as given in Eq.~(\ref{omegac}), which is less than the free particle mass, $\omega_c < m_\phi$, for the generic parameter space defined by $\sqrt{\lambda_\phi}/\lambda_{\phi h} < 1.4$ [see Eq.~(\ref{stabilityQuarticDMB})]. Given an asymmetric yield $Y_\Phi$, one can calculate the ratio of the DM and ordinary matter energy densities as
\beqa
\frac{\Omega_{\Phi} } {\Omega_{\rm B} } &=& \frac{[(1-r)\,\omega_c + r\,m_\phi]\,Y_\Phi}{m_p \, Y_{\rm B}}  ~\approx~ \frac{\omega_c\, Y_\Phi}{m_p \,Y_{\rm B}} \,.
\label{YPhiOverYB}
\eeqa
To fit the measured value of $\Omega_{\rm DM}/\Omega_{\rm B}  \approx 5.4$~\cite{Aghanim:2018eyx}, one needs $\omega_c\, Y_\Phi \approx 5.4\,\times\,10^{-10}$~GeV.  If the yield $Y_\Phi$ is comparable to the ordinary baryon one, the model parameters are then required to satisfy $\omega_c \sim 5.4$~GeV, which is the well-known situation of asymmetric DM models. We note that, to achieve such a small value of $\omega_c$, we need $\lambda_\phi \approx 1.8 \times 10^{-6}$, which is much smaller than its natural lower-limit value of $\mathcal{O}(\lambda_{\phi h}^2/16\pi^2)$ for $\lambda_{\phi h} = {\cal O}(1)$. We therefore take $\lambda_\phi \sim 10^{-2}$ and choose $\omega_c \sim 50$~GeV as a benchmark model point, for which one has $Y_\Phi \sim 10^{-11}$. 

Having discussed the DM abundance and its rough composition in terms of free elementary $\Phi$ particles versus such trapped inside \DMBs/, we can now estimate the average \DMN/ in a \DMB/. Since, as we have argued, we expect most of the \DMN/ to stay in the false EWS vacuum, we will neglect the small contribution inside the EWSB bubbles in the following estimates.  Given the \DMN/ density around $T_c$, the average \DMN/ inside a \DMB/ can be estimated as the ratio of the total number within one Hubble patch over the number of \DMBs/ in one Hubble patch. The total \DMN/ within one Hubble patch has
\beqa
N_\Phi^{\rm Hubble} &\approx& Y_\Phi\, s\, d_H^3  ~\simeq~ \left(7.8 \times 10^{37} \right)\, \left( \frac{Y_\Phi}{10^{-11}} \right)\left( \frac{134\,\mbox{GeV}}{T_c} \right)^3 ~.
\eeqa
Here, we also took $g_{*S}\approx g_* \approx108.75$. The number of \DMBs/ in one Hubble patch is of the same order of magnitude as the number of EWSB nucleation sites, which is sensitive to the detailed properties of the EW bubbles or the model parameter $\lambda_{\phi h}$. In Appendix~\ref{sec:nucleation-sites}, we have estimated the number of nucleation sites in terms of the model parameter $\lambda_{\phi h}$ [see Eq.~(\ref{Nnucleations})].
After some numerical fit, the number of \DMBs/ within one Hubble patch reads
\beqa
 N^{\rm Hubble}_{\DMB/} \sim 1.0\times 10^{13}\,\times \,\left(\frac{\lambda_{\phi h}}{3}\right)^{-14}  ~,
\eeqa
which captures the dominant dependence on $\lambda_{\phi h}$ (see also Table~\ref{table:S3-over-T} for numbers for $\lambda_{\phi h}$ from 3 to 7). When $\lambda_{\phi h}$ varies from 3 to 7, we find that $N^{\rm Hubble}_{\DMB/}$ decreases from $1.1 \times 10^{13}$ to $4.0 \times 10^{7}$. 

Finally, the average \DMN/ in one \DMB/ is estimated to be
\beqa
Q \sim \left(7.8 \times 10^{24} \right)\, \left( \frac{Y_\Phi}{10^{-11}} \right)\left( \frac{134\,\mbox{GeV}}{T_c} \right)^3\,  \left(\frac{\lambda_{\phi h}}{3}\right)^{14}  ~,
\eeqa
with a fitted $T_c$ as a function of $\lambda_{\phi h}$ as $T_c \approx 134.5\,{\rm GeV}  - 9.3\,{\rm GeV} \,\times \, \left(\lambda_{\phi h}- 3\right)$ [see Eq.~\eqref{eq:Tc-fit}]. 
Multiplying by the energy per charge, $\omega_c$, the average \DMB/ mass is
\beqa
\Msol \sim  \left(3.9\times 10^{26} \,\mbox{GeV} \right)\, \left( \frac{\omega_c\,Y_\Phi}{5\times10^{-10}\,\mbox{GeV}} \right)\left( \frac{134\,\mbox{GeV}}{T_c} \right)^3\,\left(\frac{\lambda_{\phi h}}{3}\right)^{14} ~.
\label{eq:DMB-mass-bench}
\eeqa
In the range $\lambda_{\phi h} \in [2,9]$, the average \DMB/ mass ranges from $1.1\times 10^{24}$~GeV to $9.2\times 10^{33}$~GeV or from $1.9$~g to $1.6\times 10^{10}$~g. In our subsequent  phenomenological considerations we will allow for a wider range of \DMB/ masses, and use the first-order phase transition values as guidance. From Eq.~(\ref{QQBall}), which applies to quartic \DMBs/, we see that  the \DMB/ radius scales like $\omega_c^{-1}\,\lambda_{\phi}^{1/3}\,Q^{1/3}$, in the limit that $\Mphi^2=0$ (i.e.~$\overline{\omega}_c = \omega_c$). Using also Eq.~(\ref{omegac}), we therefore have 
\beqa
\Rsol &\approx& \left(0.004~\mbox{GeV}^{-1}\right) \, \lambda_\phi^{1/12}\,Q^{1/3}  \nonumber \\
&\approx&  \left(5.8\times 10^5~\mbox{GeV}^{-1}\right) \left( \frac{\lambda_\phi}{0.013} \right)^{1/12} \, \left( \frac{Y_\Phi}{10^{-11}} \right)^{1/3}\left( \frac{134\,\mbox{GeV}}{T_c} \right)\,  \left(\frac{\lambda_{\phi h}}{3} \right)^{4.7} \,.
\label{eq:DMB-radius}
\eeqa
For the range of $\lambda_{\phi h} \in [2,9]$, the \DMB/ radius varies from $8.1\times 10^{4}~\mbox{GeV}^{-1}$ to $1.7\times 10^{8}~\mbox{GeV}^{-1}$ or from $1.6\times 10^{-9}~\mbox{cm}$ to $3.3\times 10^{-6}~\mbox{cm}$. 

\section{Scattering of \DMBs/ with SM Particles} 
\label{sec:scattering}

In this section we discuss a number of issues related to the scattering of \DMB/s. We start with a discussion of bound states of SM particles inside the \DMB/, as this bears on its scattering cross section, and other possible effects.

\subsection{Bound States} 
\label{sec:BoundStates} 

We have seen how \DMBs/ sustain a core where the EW symmetry is essentially unbroken, while outside the soliton the usual EWSB vacuum is quickly reached. Such Higgs profiles display a sharp transition of size $\sim 1/v$, separating a much wider EW preserving region from the symmtry-breaking vacuum outside. It acts as a potential well, seen by any SM particle or bound state with a strength dictated by its coupling strength to the Higgs boson. Typically, the Higgs well is invisible only to massless particles like photons (at tree level), as well as neutrinos due to their lightness. We can model this physics as a 3D potential well (i.e.~with a sharp transition), and use the intuition from the quantum mechanical treatment of such a problem. However, since elementary particles trapped inside the \DMB/ see essentially no Higgs VEV, they behave like trapped massless states. We will therefore include the kinematic relativistic effects. For the SM fermions and gauge bosons, spin effects are also expected to be important in determining the spectrum of bound states. Although our methods can be generalized in a straightforward manner to include such effects, we will neglect them for simplicity, and aim at getting only a qualitative understanding. Hence, we will be thinking of appropriate scalar particles as proxies for the SM fermions and gauge bosons. Similarly, we will neglect corrections from the possible creation of particle-antiparticle pairs, which would require a significantly more complex quantum field theory treatment. In summary, we are interested here in the Klein-Gordon equation in the presence of a 3D well, in its one-particle interpretation.

We will also be interested in particles, such as hadrons or nuclei, that get most of their mass from sources (QCD dynamics) other than EWSB. Such bound states can be described by a non-relativistic analysis, which can be obtained by taking the non-relativistic limit of the case above.

We describe the appropriate treatment in Appendix~\ref{sec:HiggsWell}, and invoke here only the main results, which are sufficiently intuitive. In Appendix~\ref{sec:Backreaction} we discuss the backreaction of such bound states on the \DMB/, which is expected to be small due to the large dark matter or $\Phi$ number composing a \sol/ with a large $Q$. This is in spite of the large number of particles that can get bound by the \DMB/, as we shall discuss.

The structure of the Klein-Gordon equation is identical to the Schr\"odinger equation in a 3D potential well, with the replacement $E \rightarrow E^2/2m_\chi$ (where $m_\chi$ is the mass of the particle in the normal EWSB vacuum), together with an appropriate mapping of the potential (by a constant rescaling). Thus, the solutions are given by spherical Bessel functions inside and outside the \DMB/, whose matching at the boundary lead to the condition for the spectrum. For example, for particles that get their mass solely from EWSB, one gets for the $s$-wave states for a \DMB/ with a radius $R$
\bea
- \cot\left(E R \right) &=& \sqrt{\frac{m^2_\chi - E^2}{E^2}}~,
\label{relativisticspectrumSection}
\eea
which determines the bound state spectrum $E_n$. The threshold radius to have a bound state can be estimated to be 
\beqa
R_{\rm th}  = \frac{\pi}{2}\,m^{-1}_\chi ~. 
\eeqa
Numerically, one has $R_{\rm th}^{t} \approx 0.009\,\mbox{GeV}^{-1}$,  $R_{\rm th}^{e} \approx 3142\,\mbox{GeV}^{-1}$, $R_{\rm th}^{\nu}  \gtrsim 1.6\times10^{10}\,\mbox{GeV}^{-1}$ for $m_\nu < 0.1$~eV. So, for the \DMB/ radius from first-order phase transition in Eq.~\eqref{eq:DMB-radius}, neutrinos can not be bounded, but other fermions do have bound states. 

For particles that get only part of their mass from EWSB, the $s$-wave spectrum is determined by 
\bea
- \cot\left(\sqrt{E^2 - m_N^2 + y_{hNN}^2 v^2} \, R \right) = \sqrt{\frac{m_N^2 - E^2}{E^2 - m_N^2 + y_{hNN}^2 v^2}}~,
\label{NucleonspectrumSection}
\eea
where we denote the mass in the EWSB vacuum by $m_N$ (as for nucleons), and the coupling of the $N$ particle to a single Higgs boson by $y_{hNN}$ (e.g.~the nucleon Yukawa coupling). Note that Eq.~(\ref{relativisticspectrumSection}) can be obtained from Eq.~(\ref{NucleonspectrumSection}) by setting $m_N = y_{hNN} v$. Also, if $E \equiv m_N + \Delta E$ with $|\Delta E| \ll m_N$, one can check that Eq.~(\ref{NucleonspectrumSection}) reduces to the non-relativistic result determining the binding energies $|\Delta E_n|$ in the presence of a potential well of depth $V_0 = (y_{hNN} v)^2/2m_N$. The threshold radius to have a bound sate for nucleons is
\beqa
R^N_{\rm th} = \frac{1}{y_{h NN} \, v}\, \frac{\pi}{2}  \,,
\eeqa
or numerically $R^{N}_{\rm th} \approx 5.8\,\mbox{GeV}^{-1}$. The corresponding threshold radius for a nucleus is even smaller by a factor of $1/A$.

As a first application, consider Eq.~(\ref{relativisticspectrumSection}) in the case of a large \DMB/ with $m_\chi R \gg 1$. The eigenvalues are then close to the poles of the cotangent. In particular, keeping the first order correction, the lowest energy solution is at 
\bea
E^{\chi}_0 \approx \frac{\pi}{R} \left( 1 - \frac{1}{m_\chi R} \right) ~. \qquad 
\label{E0Approximate}
\eea
As shown in Appendix~\ref{sec:Backreaction}, when backreaction effects can be neglected, the mass of the \DMB/ with a bound state $\chi$ as above is given simply by $\Msol^{(0)} + E^\chi_n$, where $\Msol^{(0)}$ is the \DMB/ mass in the absence of $\chi$, and $E^\chi_n$ is the bound spectrum as described above. So, for the particles satisfying $m_\chi R \gg 1$, the bound state masses for different species are still ordered in their SM relations or $E^t_0 > E^b_0 > E^c_0$ for instance. On the other hand, the mass spectrum is very degenerate, which may kinematically forbid some SM decaying channels. The energy balance for a process $A \to B + C + \cdots$ can be described simply in terms of the bound state spectrum. For example, for the SM process $t \to W^+ b$, assuming all the particles are bound to the \DMB/ in the ground state, and using Eq.~(\ref{E0Approximate}), one can easily see that this decay channel is kinematically forbidden in the limit of $m_b R \gg 1$. However, the top quark in \DMB/ could still decay into light fermions via three-body processes like $t \to e^+ \nu_e\, d$ for $m_e R, m_d R \ll 1$ and no bound states for electron, neutrinos and down quark.

Consider now a nucleon. Here, for $y_{h NN} v R\gg 1$, the binding energy for the lowest energy state is determined by the vanishing of the denominator on the r.h.s.~of Eq.~(\ref{NucleonspectrumSection}),  $E^{\rm bind}_{N, 0} \approx \sqrt{m_N^2 - y_{hNN}^2 v^2} - m_N \approx -0.04 m_N$ and around 38~MeV numerically.~\footnote{Although the expression for $E^{\rm bind}_{N, 0}$ we have quoted is only approximate, one can check that it reproduces the correct difference in binding energies between the proton and neutron bound states by solving Eq.~(\ref{NucleonspectrumSection}) for the ground states numerically.} Hence the nuclei can be safely treated in the nonrelativistic limit. Compared to the typical electron binding energies which are of order $m_e$ (at least if $R$ is large enough for the potential well to sustain several bound states), the typical nucleon binding energies are larger, but only by a factor of order $0.04 m_N/m_e \approx 80$. Let us compare the proton versus neutron cases. Using $m_p = 938.27$~MeV and $m_n = 939.56$~MeV~\cite{Patrignani:2016xqp}, together with $y_{hpp} = 1.12 \times 10^{-3}$ and $y_{hnn} = 1.14 \times 10^{-3}$~\cite{Cheng:2012qr}, we get
\beqa
\sqrt{m_p^2 - y_{hpp}^2 v^2}  - \sqrt{m_n^2 - y_{hnn}^2 v^2} \approx 0.17~{\rm MeV}~.
\label{eq:proton-neutron-difference}
\eeqa
Thus, the neutron is slightly more deeply bound and ``lighter" than the proton inside a \DMB/. This reflects the fact that inside the \DMB/ the quarks are massless, and the fact that $m_d > m_u$ in the normal vacuum gives a positive contribution to $m_n - m_p$ (that dominates over the negative electromagnetic contribution) is absent inside the soliton. Hence, the proton is ``heavier" than the neutron when bound to a \DMB/. This means, in particular, that for such bound states, we can have the process
\bea
p &\to& n + e^+ + \nu_e~,
\eea
with the neutrino escaping the \DMB/ and the positron bounded inside \DMB/.\footnote{We are assuming here that QCD is in its standard chirally broken phase, an issue that requires analysis. Notice also that in this case the $W$ gauge boson receives a contribution to its mass, unrelated to the Higgs, which for simplicity we have ignored in Eq.~(\ref{eq:proton-neutron-difference}).} Thus, trapped protons tend to decay into neutrons, in analogy to a neutron star. 

One can reason along similar lines to address other processes involving particles bound to a \DMB/. One should keep in mind that, in general, a \DMB/ can be expected to be a complicated object carrying with it a cloud of different types of particles. Finally, we also note that there are many bound states in a \DMB/ with a large $R$. The maximum number of angular momentum for the bound states has $l_{\rm max} \sim R/R_{\rm th}$, i.e. roughly when the order of the spherical Bessel function passes the number of cycles that the $l = 0$ Bessel function can fit in the size $R$. Thus, we can put an upper bound on the number of orbital states as $(R/R_{\rm th})^3$ or more explicitly 
\beqa
N_{\rm orbital} \sim \frac{1}{4}\, \left( \frac{R}{R_{\rm th}}\right)^3 ~.
\label{eq:number-orbital}
\eeqa
One should keep in mind that the above estimates do not take into account the strong and electromagnetic interactions between these particles. Using a benchmark $R=2\times 10^4\,\mbox{GeV}^{-1}$, the number of states for various states have $N_{\rm orbital}^t \sim 10^{18}$, $N_{\rm orbital}^{N} \sim 10^{10}$ and $N_{\rm orbital}^{e} \sim 10^2$. 

If all the nucleon states are occupied, we estimate a nucleon density of about $10^{10}/(4\pi R^3/3) \sim (100~\mbox{MeV})^3$, which corresponds to an inter-nucleon distance of about 2 fm, a density of the order of the nuclear density [as seen from Eq.~\eqref{eq:number-orbital}, this is independent of $R$]. The total contribution to the \DMB/ mass due to these nucleons is about $10^{10}\,m_N\approx 10^{10}$~GeV, where the small binding energy per nucleon is neglected. The captured nuclear matter gives a negligible contribution to the \DMB/ mass, which is dominated by the $\phi$ and $h$ contributions.

\subsection{Scattering Off a Nucleon or Nucleus} 
\label{sec:scattering-nucleon} 

After the previous cursory description of bound states of SM particles in a \DMB/, we turn to the question of \DMB/ scattering from normal matter such as nucleons or nuclei.

The first observation is that \DMB/s are expected to be heavy compared to the target particles. For instance, Figs.~\ref{fig:ProfilesLowQ} and \ref{fig:RMvsQ} show examples with masses in the multi-TeV range and above,\footnote{In particular, the \sols/ discussed in this work cannot be lighter than the weak scale.} and we will actually discuss much heavier objects like in \eqref{eq:DMB-mass-bench}, as expected from our discussion on how they can be produced. This means that one can analyze the scattering problem treating the \DMB/ as an infinitely heavy object generating a fixed Higgs background field, while treating the nucleon or nucleus as a light particle scattering against such a fixed potential---{\it a quantum mechanics scattering problem}. We describe the procedure in Appendix~\ref{sec:heavysource}, in the context of a simple toy model. Since the typical velocities involved if the \DMBs/ are gravitationally bound to our galaxy are of order $10^{-3}\,c$, one is safely in the non-relativistic regime. As discussed previously, we can further model the Higgs background as a 3D potential well. 

As is familiar from the non-relativistic quantum mechanical treatment of scattering processes, the scattering cross section can be affected significantly when bound states are available. In the present context, the presence of bound states depends on the size of the \DMB/, with a threshold radius $R_{\rm th} \sim (y_{hNN}\,v)^{-1}$ that depends on the scattering particle (normal vacuum) mass $m_N$ (as discussed in Appendix~\ref{sec:HiggsWell}). When the \DMB/ is small enough so as to not allow for bound states, one can treat the problem in the Born approximation (and safely in the $q=0$ limit). For \DMB/s with a large charge, on the other hand, they are large enough to contain a very large number of bound states. In such a situation a partial wave analysis is more appropriate. We will describe the partial wave analysis in some generality first, as the Born approximation case can be understood as an appropriate limit of the partial wave result.

\subsubsection{Partial Wave Analysis} 
\label{sec:PartialWaves} 

As explained above and in Appendix~\ref{sec:HiggsWell} our problem is mapped into the corresponding non-relativistic problem by a simple reinterpretation of certain quantities. In particular, for a partial wave $l$ in the 3D potential well we are using, the spectrum is determined by matching the logarithmic derivatives at $r = R$, and takes exactly the form of the non-relativistic result:
\bea
K \, \frac{j_l'(K R)}{j_l(K  R)} &=& k \, \frac{j'_l(k R) \cos \delta_l + n'_l(k R) \sin \delta_l}{j_l(k R) \cos \delta_l + n_l(k R) \sin \delta_l}~,
\label{LogDer}
\eea
where $j_l$ and $n_l$ are spherical Bessel functions of the first and second kind respectively, and $\delta_l$ is the scattering phase shift for the $l$-th partial wave. The only difference is in the interpretation of the wavevectors $k$ (outer region) and $K$ (inner region). For particles that get their mass only from EWSB, such as an electron, they are given by 
\bea
K &=& |E|~,
\label{KelectronScatt}
\\ [0.5em]
k &=& \sqrt{E^2 - m^2}~,
\label{kelectron}
\eea
where $m$ is the mass of the particle in the normal EWSB vacuum.
For particles like nucleons that  get contributions to their mass from sources other than EWSB, they are given by 
\bea
K &=& \sqrt{E^2 - m_N^2 + y_{hNN}^2 v^2}~,
\label{KNucleonScatt}
\\ [0.5em]
k &=& \sqrt{E^2 - m_N^2}~,
\label{kNucleon}
\eea
where $m_N$ is the mass in the normal vacuum, and $y_{hNN}$ is their coupling to a single Higgs (e.g.~a Yukawa interaction). 

The cross section is given by
\bea
\sigma &=& \sum_{l=0}^{\infty} \, \frac{4\pi(2l+1)}{k^2} \, \sin^2\delta_l~.
\label{sigmaPartialWave}
\eea
For given parameters, it is straightforward to obtain numerically $\sin\delta_l$ from Eq.~(\ref{LogDer}), and do it up to  large enough $l$ that the sum in Eq.~(\ref{sigmaPartialWave}) is observed to converge.

\begin{figure}
\centering
\includegraphics[width=0.48\textwidth]{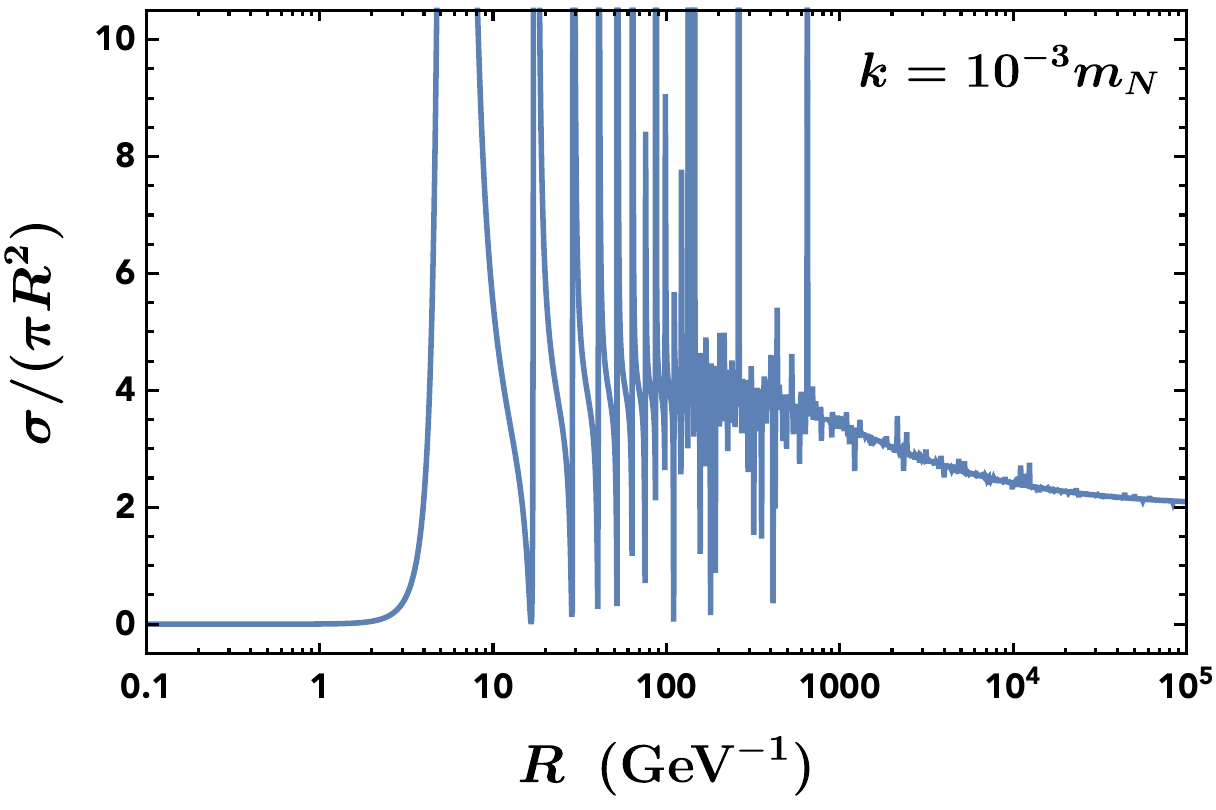}
\hspace{3mm}
\includegraphics[width=0.48\textwidth]{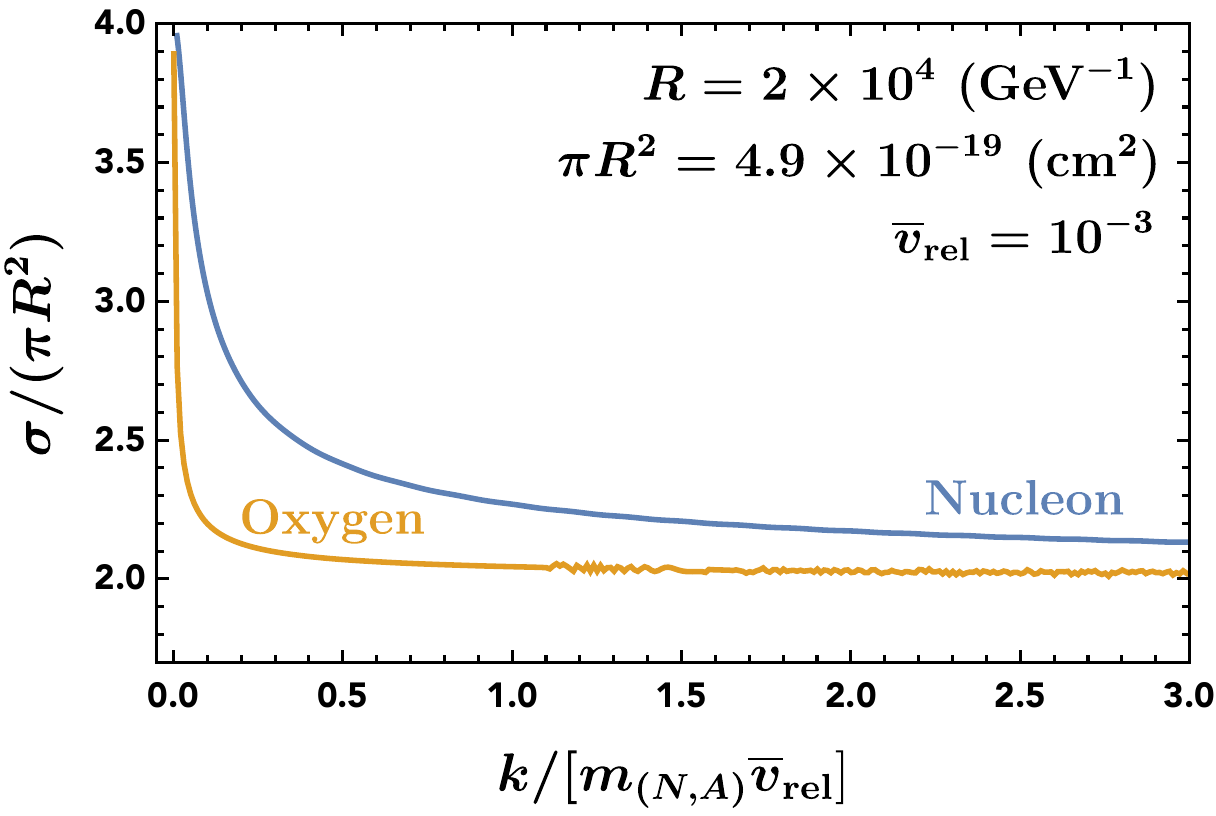}
\caption{Left panel: \DMB/ scattering against nucleons, as a function of the \DMB/ radius and for a typical momentum at the Earth's position. Right panel: \DMB/ scattering cross section against a nucleon or an oxygen nucleus, as a function of the scattering momentum. We take a benchmark \DMB/ with radius $R = 2\times 10^4~{\rm GeV}^{-1} \approx 0.04\,\si{\angstrom}$ and average relative velocity $\overline{v}_{\rm rel} = 10^{-3}$.
}
\label{fig:sigmavsRk}
\end{figure} 
In the left panel of Fig.~\ref{fig:sigmavsRk} we show the result for scattering against a nucleon, as a function of the \DMB/ radius $R$. We plot the cross section, as computed from Eqs.~(\ref{KNucleonScatt}), (\ref{kNucleon}) and (\ref{sigmaPartialWave}), summing up to $l = 100$, where we have checked that the sum has been saturated in the range of $R$ shown. We use $m_N = 938.9$~MeV and  $y_{hNN} = 1.1 \times 10^{-3}$, and assume a typical \DMB/ scattering momentum of $k = 10^{-3} m_N$. We see that, as a function of $R$, the cross section displays a complicated resonant structure. For small \DMBs/ with $R < 1~{\rm GeV}^{-1}$, as those shown in Figs.~\ref{fig:ProfilesLowQ} and \ref{fig:ProfilesQuartic}, the cross section is suppressed. This is the regime where it can be computed in the Born approximation, as will be discussed in the next subsection. Well above the most prominent resonances, it displays a ``hard ball" behavior that drops from $\sigma = 4\pi R^2$ to $\sigma = 2\pi R^2$, although with some additional small scale resonant structures. These are large \DMBs/ that are of phenomenological interest given our previous considerations on their production.

In the right panel of Fig.~\ref{fig:sigmavsRk} and summing up to $l = 100(900)$ for nucleon(oxygen), we show the cross section for \DMB/ scattering against a nucleon or an oxygen nucleus,\footnote{Here we use $m_O \approx A\,m_N$ and $y_{hOO} \approx A\,y_{hNN}$, with the atomic number $A=16$.} for a benchmark \DMB/ of radius $R = 2\times 10^4~{\rm GeV}^{-1} \approx 0.04~\si{\angstrom}$, which corresponds to a geometrical cross section of $\pi R^2 \approx 4.9 \times 10^{-19}~{\rm cm}^2$. One can see that the cross sections follow roughly a ``hard ball" behavior. For a smaller radius (for instance $R = 6000~{\rm GeV}^{-1}$), there exists also a superimposed resonant structure. We note, however, that the detailed structure is rather sensitive to the coupling of nucleons to the Higgs. 
One should keep this in mind, as we have made approximations and neglected physical effects such as spin or other relativistic effects that can affect the spectrum of bound states. Nevertheless, we learn that for such large \DMBs/, the cross section is expected to be between 2 and 4 times the geometric cross section. This holds, in particular, whether the scattering is off a nucleon or a much heavier nucleus.

Our considerations in this section depend only on the \DMB/ radius. However, the charge and mass of the \DMB/ for this radius depend on the type of \DMB/. For instance, for \DMBs/ sustained with $\lambda_\phi = 0$ (quadratic \DMBs/), which obey the scaling laws of Eq.~(\ref{ScalingLawsQuadratic}), one has for $R = 2\times 10^4~{\rm GeV}^{-1}$
\bea
Q &\approx& 1.2 \times 10^{26}~,
\hspace{1cm}
\Msol ~\approx~ 2.6 \times 10^{22}~{\rm GeV} ~\approx~ 0.05~{\rm g}~.
\label{QuadraticBenchmark}
\eea
For a \DMB/ sustained by the quartic self interaction $\lambda_\phi = 10^{-2}$ [see paragraph after Eq.~(\ref{YPhiOverYB})], one finds from Eqs.~(\ref{omegac}) and (\ref{QQBall}),
\bea
Q &\approx& 3.2 \times 10^{20}~,
\hspace{1cm}
\Msol ~\approx~ 1.6 \times 10^{22}~{\rm GeV} ~\approx~ 0.03~{\rm g}~.
\label{QuarticBenchmark}
\eea
We see that their masses are comparable, as expected from the fact that in both cases the mass scales with volume, and the underlying scales involved are just the weak scale times functions of couplings taken to be of order one. Only the associated charges are significantly different. From our discussion in Section~\ref{OtherTwoParameters} we expect that the second case is more likely (i.e.~the quartic instead of the quadratic \DMBs/).

\subsubsection{Born Limit} 
\label{sec:Born} 

Let us now turn to the Born limit, applicable for small enough \DMBs/ that do not sustain any bound states (for the given target particle). We will see that in this case one can compute in more detail the scattering cross section, without assuming a 3D well model. It also addresses directly the constraints on \DMBs/ with masses not exceedingly above the weak scale.

In the presence of the nontrivial Higgs background induced by the \DMB/, a particle approaching this region sees a potential
\bea
V(r) &=& y\, [h(r) - v]~,
\eea
where we choose that $V(r) \to 0$ as $r \to \infty$, and $y$ is the coupling of the scattering particle to the Higgs (e.g.~$y_{hNN}$ for nucleons).
In the first Born approximation the scattering amplitude is given by
\bea
f(E,\theta) &=& - \frac{2m}{q} \, \int_0^\infty \! dr \, r \sin(qr) V(r)~,
\label{BornApprox}
\eea
where $m$ is the reduced mass of the system and $q = |\vec{q}|$ is the momentum transfer. For the case at hand, the range of the Higgs potential well is much shorter than the length scales that can be probed by the typical $q \sim 10^{-3} m$. We can therefore set $q=0$, and compute Eq.~(\ref{BornApprox}) numerically for the Higgs profiles found as described in Section~\ref{sec:soliton}.

We have computed the \DMB/-nucleon scattering cross sections for a number of models with different $\lambda_{\phi h}$ and different \DMB/ charges (see Section~\ref{ClassicalEOM}). Depending on the value of $\lambda_{\phi h}$, these span \DMB/ masses from about $5-500$~TeV.\footnote{The precise range covered in our scan of models depends on $\lambda_{\phi h}$. For instance, the lightest stable solutions are shown in Fig.~\ref{fig:RMvsQ}.} We find that the cross sections for this range of masses are well described by the relation
\bea
\sigma_{\tiny \Pcirc N}(M) &=& 8.2 \times 10^{-42} \left( \frac{M}{{\rm TeV}} \right)^{2.1}~{\rm cm^2}~,
\eea
where we have considered all the models together as they all fall reasonably close to the above parametrization. Comparing to the Xenon1T bound~\cite{Aprile:2018dbl}, which in this region can be parametrized as
\bea
\sigma^{\rm SI}(M) &\lesssim& 1.2 \times 10^{-45} \left( \frac{M}{{\rm TeV}} \right)~{\rm cm^2}~,
\label{DirDetConstrXENON}
\eea
we see $i)$ that at the lowest masses of order several TeV, $\sigma_{\tiny \Pcirc N}$ is already excluded by direct detection searches, and $ii)$ that $\sigma_{\tiny \Pcirc N}$ increases with mass faster than the linear growth in Eq.~(\ref{DirDetConstrXENON}). We note that the models considered above all have a vanishing bare mass $\Mphi = 0$ (and $\lambda_{\phi} = 0$). Increasing the value of $\Mphi$ does not qualitatively relax the constraint. For quartic \DMBs/, where $\lambda_{\phi}$ is important, the \DMBs/ have a large radius and a large geometric nucleon scattering cross section, much larger than those captured by the Born approximation above.

Hence, all the models at low masses are excluded by direct detection constraints up to the experimental reach (about $2.8 \times 10^{18}$~GeV for Xenon1T and $1.4 \times 10^{21}$~GeV for BOREXINO), and we do not consider them in further detail. Our focus is instead in the large \DMB/ mass limit.

\section{\DMB/ Detection} 
\label{sec:pheno}
Before we discuss the detection of \DMB/, we want to briefly discuss the collider search for Higgs-portal dark matter. For the dark matter particle mass is below one half of the Higgs boson, or $m_\phi < m_h/2$, the Higgs boson can decay into two dark matter particles and has additional invisible decay branching ratio~\cite{Sirunyan:2018owy}. For the \DMB/ formation scenario from first-order phase transition, the coupling $\lambda_{\phi h}$ is needed to be above around 2, such that the dark matter particle mass $m_\phi$ is heavier than the Higgs boson mass. For this case, the collider constraints from the LHC are dramatically reduced because of the three-body production phase space for producing off-shell Higgs-mediated two dark matter particles and one additional jet. For instance, the parton-level production cross section for $\lambda_{\phi h}=2$ or $m_\phi = 246$~GeV with $\Mphi=0$ is around 0.1 fb with a missing transverse energy cut above 250 GeV at the 13 TeV LHC. With 36.1 fb$^{-1}$, the number of signal events is three orders of magnitude smaller than the uncertainty of measurement~\cite{Aaboud:2017phn}. So, there is no collider constraint on the model parameter space with $\lambda_{\phi h} \gtrsim 2$ considered in this paper. 

Another constraints on the model parameter space come from direct detection of the free $\Phi$ particle. As discussed around Eq.~\eqref{eq:freeze-out-temperature}, the free dark matter particle is subdominant of the total dark matter energy density. However, given the stringent direct detection constraints on a dark matter with a mass around 100 GeV, a non-trivial constraint on the coupling $\lambda_{\phi h}$ may apply here. Using the coupling $\lambda_{\phi h}\,v\, h\,\Phi \Phi^\dagger$ and $y_{h NN}\,h\,\overline{N}N$ with $N=n,p$ and $y_{hNN}\approx 1.1\times 10^{-3}$, the spin-independent scattering cross section has the formula of
\beqa
\sigma^{\rm SI}_{\Phi-N} = \frac{\lambda^2_{\phi h}\,y^2_{hNN}}{4\pi} \, \frac{v^2\,m_p^2} {m_h^4\,m_\phi^2}  \approx \frac{\lambda_{\phi h}\,y^2_{hNN}}{2\pi} \, \frac{m_p^2} {m_h^4}
 ~,
\eeqa
where we have used the reduced mass to be $\mu_{\Phi-N} \approx m_p$ in the limit of $m_p \ll m_\phi$ and the small bare mass limit with $\Mphi \ll m_\phi$. The latest constraints from Xenon1T have set an upper bound for dark matter scattering  cross sections in Eq.~\eqref{DirDetConstrXENON}, which can be translated into a constraint on our model parameter space as
\beqa
r \lesssim 7.8 \times 10^{-4}\times \lambda_{\phi h}^{-1/2} \,,
\eeqa
which is well satisfied for a freeze-out temperature below 1 GeV in Eq.~\eqref{eq:r-equilibrium}. 

\subsection{Multiple Scattering Signals for a \DMB/ with a Large $Q$} 
\label{sec:multiple-scattering} 
For \DMBs/ with a small $Q$ and as we discussed around Eq.~\eqref{DirDetConstrXENON}, the scattering cross section is small for \DMB/ to reach the underground detectors, but not small enough to satisfy the direct detection constraints. On the other hand, the \DMB/ could have a very heavy mass and a large scattering cross section (see Fig.~\ref{fig:sigmavsRk}). Although we did discuss the potential early universe productions of \DMB/ from a first-order phase transition and obtained some benchmark radii for \DMB/ in Eq.~\eqref{eq:DMB-radius}, we will keep the \DMB/ radius and mass as free parameters to discuss its detection potential. As a simple recap what we have learned for the properties of \DMB/ in Section~\ref{sec:soliton}: for $\Mphi=0$ with $\overline{\omega}_c = \omega_c$, the mass $\DMB/$ has $\Msol = Q\, \omega_c$, while the radius has $\Rsol = (3\lambda_\phi/4\pi)^{1/3}\,\omega_c^{-1}\,Q^{1/3}$. For a large value of $\Rsol$, the scattering cross section of \DMB/ with a nucleus reaches a geometrical one with (see Fig.~\ref{fig:sigmavsRk})
\beqa
\sigma_{{\tiny\Pcirc}A} &\approx& 2 \pi \Rsol^2  = \left(\frac{9\pi}{2}\right)^{1/3}\,\lambda_\phi^{2/3}\, \omega_c^{-8/3}\, \Msol^{2/3}    \nonumber \\
&=& \left( 9.8 \times 10^{-19}\,\mbox{cm}^2 \right) \left( \frac{\lambda_\phi}{0.013} \right)^{2/3}\, \left( \frac{50\,\mbox{GeV}}{\omega_c} \right)^{8/3}\, \left( \frac{\Msol}{1.6\times 10^{22}\,\mbox{GeV}} \right)^{2/3}
~,
\label{eq:cross-section-nucleus}
\eeqa
which will be used as a prediction based on \DMB/ properties. We emphasize here that once the radius of \DMB/ is large enough, the scattering cross section with a nucleus is approximately independent of the nucleus mass number. 

For the heavy \DMB/ with a large scattering cross section, the ordinary underground dark matter direct detection experiments become less sensitive. The ideal experiment would be the one with a large product of the exposure time and the effective detector area. For a long exposure time, one could consider Mica type experiment~\cite{Price:1986ky} that looks for tracks generated by \DMBs/ passing by. For a large effective detector area, one could adopt the neutrino-oriented detectors with a small enough energy-trigger threshold. In the following, we discuss the search potential for a few experiments.  

\subsubsection*{Mica Constraints} 
\label{sec:mica}

For our \DMB/ with or without quark or neutron matter inside, the scattering cross section is in the range of $2$--$4$ times the geometric cross section (see Fig.~\ref{fig:sigmavsRk}). Since the scattering off a nucleon and a nucleus have similar cross sections, we can ignore the detailed chemical components of Mica for constraining the scattering cross section or the geometric size of \DMB/. To have a dark matter-generated track in Mica, one at least requires one encounter event or $\rho_{\rm DM}/m_{\rm DM} \, v\,A_{\rm det} \, t_{\rm exp} \sim 1$. Based on Ref.~\cite{Price:1986ky}, the experiment has $A_{\rm det} \approx 595\,\mbox{cm}^2$ and $t_{\rm exp} \approx 0.6\times 10^9\,\mbox{yr}$. For the averaged dark matter velocity with $v\sim 10^{-3}\,c$,  one has
\beqa
1 \sim \left(\frac{10^{26}\,\mbox{GeV}}{\Msol}\right) \left(\frac{A_{\rm det}}{595\,\mbox{cm}^2} \right) \left( \frac{t_{\rm exp}}{0.6\times 10^9\,\mbox{yr}}  \right) ~.
\label{eq:encounter-rate}
\eeqa
So, \DMB/ with a mass $ \lesssim 1.0 \times 10^{26}$~GeV may leave a track in Mica. 

Following the treatment in the paper by De R\'ujula and Glashow~\cite{DeRujula:1984axn}, the energy loss rate is
\beqa
\frac{dE}{dt} = - \sum_i \sigma_{{\tiny\Pcirc}A_i} \,\rho_{A_i} \,v^2 \approx  -\sigma_{{\tiny\Pcirc}A}  \, \rho_{\rm mica}\,v^2 \,.
\eeqa
Here, $\rho_{A_i}$ is the individual element energy density and $\rho_{\rm mica}\approx 2.88\,\mbox{g}/\mbox{cm}^3$. The velocity $v$ should be the speed at a depth of around $L\sim 3$\,km underground.\footnote{One may worry about the overburden effects on reducing the \DMB/ velocity when it reaches Mica. The relative change of the velocities from ground with $v_i$ to underground with $L$ and $v_f$  has $(v_f - v_i)/v_i \approx \sigma_{{\tiny\Pcirc}A}\,\rho_{\oplus}\,L/\Msol$, which is around $10^{-11}$ and negligible for $L=3$~km, $\rho_{\oplus}=2.9\,\mbox{g}/\mbox{cm}^3$ (for crust) and the benchmark model point in Eq.~\eqref{eq:cross-section-nucleus}. Beyond the \DMB/ model, we will consider the parameter region with $\sigma \lesssim 2\times 10^{-14}\,\mbox{cm}^2 \times (m_{\rm DM}/10^{16}\,\mbox{GeV})$ in order to ignore the overburden effects.} For the velocity, we simple take it to be the averaged dark matter velocity around our solar system. To leave a track in the Mica experiment~\cite{Price:1986ky}, the Mica stopping power has to be above 
\beqa
\rho_{\rm mica}^{-1}\,\frac{dE}{dx} \gtrsim 2.4~\mbox{GeV}\,\mbox{cm}^2 \,\mbox{g}^{-1} = 4.3 \times 10^{-24}\,\mbox{cm}^2 ~.
\eeqa
So, the constraint on the \DMB/ scattering cross section has
\beqa
\sigma_{{\tiny\Pcirc}A} \gtrsim v^{-2} \times 4.3 \times 10^{-24}\,\mbox{cm}^2 \approx 4.3 \times 10^{-18}\,\mbox{cm}^2  \,.
\eeqa

\subsubsection*{Xenon-1T Sensitivity} 
For the Xenon1T experiment, one could search for multi-hit signals as discussed in Ref.~\cite{Bramante:2018qbc}. Since the liquid Xenon TPC has 1 m in diameter and 1 m in height, we simply take the detector area as $A_{\rm det} \approx 10^4\,\mbox{cm}^2$. Taking the observation time of one year or $t_{\rm exp}=1\,\mbox{yr}$, the upper limit on dark matter mass is $2.8\times 10^{18}$~GeV. 

Different from the situation in a neutrino detector, one could count the numbers of hits from both the scintillation and electroluminescence of electrons that have drifted into the gas above the target liquid. The number of hits is estimated to be $N_{\rm hit} \sim  \sigma_{{\tiny\Pcirc}A}\, n_{\rm det}  \,L_{\rm det}$. The liquid Xenon has a density of $3.1\,\mbox{g}/\mbox{ml}$ with the main abundant atomic mass number of around $A=131$. Taking $L_{\rm det} \approx 1$~m and $N_{\rm hit} = 5$, the projected probing cross section has
\beqa
\sigma_{{\tiny\Pcirc}A}  \gtrsim  3.5\times 10^{-24}\,\mbox{cm}^2 ~.
\eeqa
%

\subsubsection*{BOREXINO Sensitivity} 
\label{sec:borexino}

For the neutrino experiment, BOREXINO, located in the Gran Sasso underground laboratory with the average rock cover of about 1.4 km and based on organic scintillator with a low energy trigger threshold~\cite{Bramante:2018tos}, it can also constrain some of the \DMB/ parameter space.  Using Eq.~\eqref{eq:encounter-rate} and $A_{\rm det}  = 5 \times 10^{5}\,\mbox{cm}^2$ and $t_{\rm exp} = 10\,\mbox{yr}$, the dark matter mass is required to be below around $1.4 \times 10^{21}$~GeV to have a few encounter events. 

Following the experimental paper~\cite{Bellini:2013lnn}, the trigger of BOREXINO requires at least 25 to 30 hits of PMTs or 50 to 60 keV energy threshold during a selected time window around $t_{\rm select} \approx 100$ ns. However, there is some efficiency factor $\kappa$ for the nuclear recoil energy converting to detectable photons. Following the experimental measurement~\cite{Hong:2002ec} and the semi-empirical calculation~\cite{Tretyak:2009sr}, the factor $\kappa \approx 10\%$ for the recoil energy below around 20 keV. In the experimental paper~\cite{Hong:2002ec}, the possible threshold energy around 2.8 keV for Carbon is mentioned, but was not crystal clear. The BOREXINO uses the organic scintillator pseudocumene (C$_9$H$_{12}$) with a density of $0.88\,\mbox{g}/\mbox{cm}^3$. Since Carbon occupies around 90\% of the total density and has a larger recoil energy, we mainly keep Carbon when we derive a constraint on scattering cross sections. 

Since the kinematics has $\Msol \gg m_{\rm C}$, the reduced mass of the two-body scattering system has $m_r \approx m_{\rm C}$. In terms of the scattering angle $\theta^*$ defined in the center-of-mass frame, the energy deposited in the detector is
\beqa
E_R 
= \frac{|{\bf q}|^2}{2\,m_{\rm C}}
= \frac{m_r^2 \, v^2}{m_{\rm C}} ( 1 - \cos{\theta^*}) ~.
\eeqa
For a simple Maxwellian halo and ignoring the motion of the Sun and Earth and the escaping velocity, the differential rate per recoil energy has a simple dependence on $E_R$ via~\cite{Jungman:1995df}
\beqa
T(E_R) \approx \mbox{exp}(- v^2_{\rm min} / v_0^2 )  =  \mbox{exp}[- E_R \,m_C /(2 m_r^2 v_0^2) ]  \approx \mbox{exp}[- E_R /(2 m_C v_0^2) ]  ~,
\eeqa
with $v_0=220\,\mbox{km}/\mbox{s}$.  The averaged recoil energy can be estimated to be 
\beqa
\langle E_R \rangle \approx \frac{\int d E_R \, E_R\, T(E_R)}{ \int d E_R \,  T(E_R) } \approx  2\,m_C\,v_0^2 \approx  12.1~\mbox{keV} ~.
\eeqa

For one incident \DMB/ hitting the detector, the interaction rate is $\Gamma = n_{\rm C} \, \sigma_{{\tiny\Pcirc}A} \, \overline{v}_{\rm rel}$. The constraints on the cross section after satisfying the trigger requirement is
\beqa
\label{eq:borexino-limit}
\Gamma\times t_{\rm select}\times \langle E_R \rangle \times \kappa > E_R^{\rm PMT} = 50\,\mbox{keV}\quad \Rightarrow \quad  \sigma_{{\tiny\Pcirc}A} \gtrsim  3.5\times 10^{-22}\,\mbox{cm}^2 \,.
\eeqa
For the multi-hit signal events from \DMB/ passing by, one could use the event shape to isolate them from backgrounds. Before dedicated experimental searches, we take the above equation as the potential reach from BOREXINO.

\begin{figure}[t!]
\centering
\includegraphics[width=0.7\textwidth]{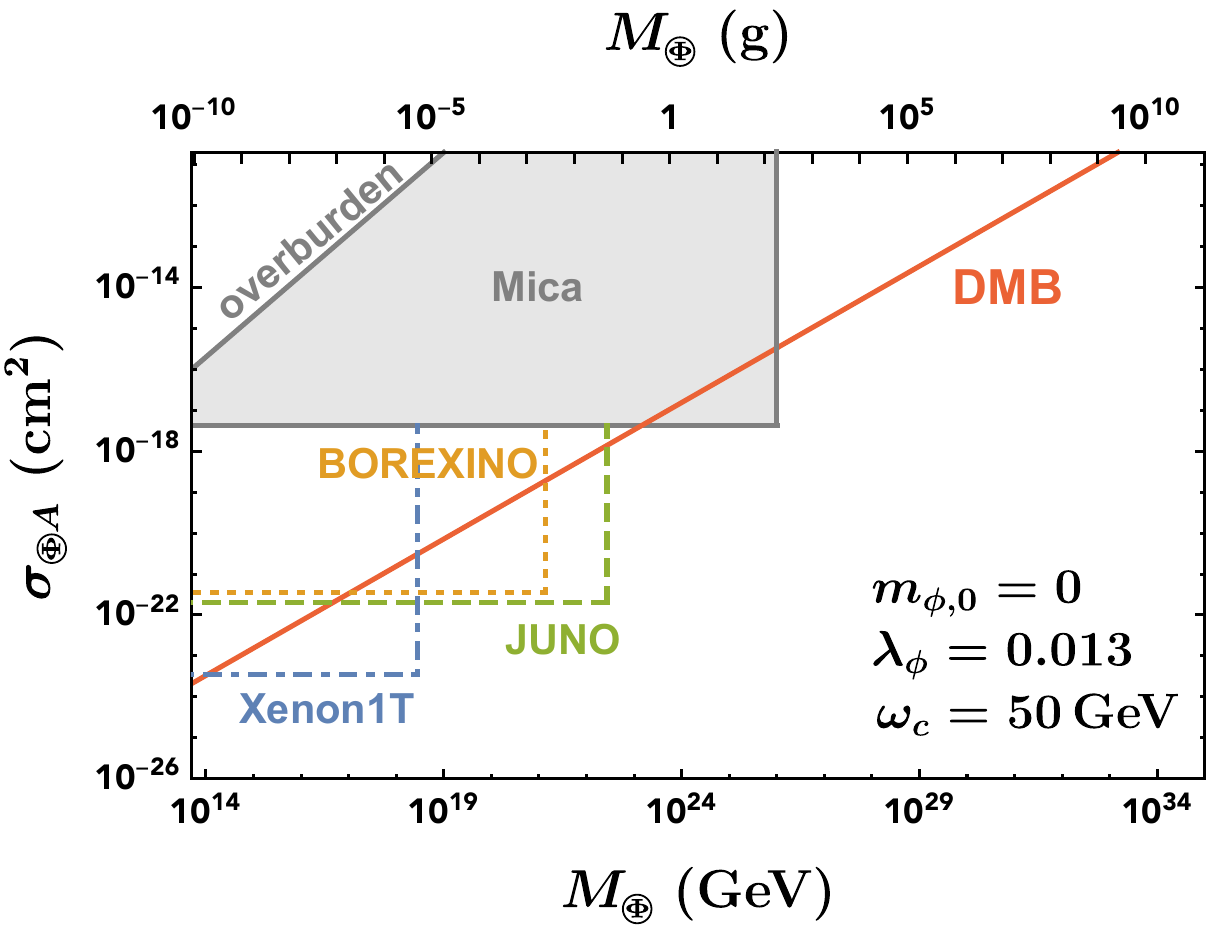}
\caption{Constraints and projected limits on \DMB/ masses and scattering cross sections off a nucleus for different experiments. The scattering cross section is taken to be a geometrical one $\sigma_{{\tiny\Pcirc}A} \approx 2\pi R_{{\tiny\Pcirc}}^2$ and proportional to $M_{\tiny\Pcirc}^{2/3}$ [see~\eqref{eq:cross-section-nucleus}]. Based on the formation from a first-order EWPT and for the range $\lambda_{\phi h} \in [2,9]$, the average \DMB/ masses vary from $1.1\times 10^{24}$~GeV to $9.2\times 10^{33}$~GeV.
}
\label{fig:sigmaexp}
\end{figure} 

\subsubsection*{JUNO Sensitivity} 
\label{sec:juno}

The JUNO neutrino experiment aiming to determine the neutrino mass hierarchy is located in Jiangmen of South China and around 700 m underground. It uses liquid scintillators with  3 g/L 2,5-diphenyloxazole as the fluor and 15 mg/L p-bis-(o-methylstyryl)-benzene as the wavelength shifter. It has the density of 0.859 g/ml with around 88\% of Carbon. The radius of the JUNO detector is $R_{\rm JUNO}=17.7$~m. So, we take the detector area to be approximately $A_{\rm det} \approx \pi R_{\rm JUNO}^2 = 9.8\times 10^6\,\mbox{cm}^2$. Taking an experimental observation time of $t_{\rm det} \approx 10\,\mbox{yr}$, the reach on the \DMB/ mass is estimated to be $2.7\times 10^{22}$~GeV. 

From the experimental studies in Refs.~\cite{An:2015jdp,JUNO-thesis}, one can have the selection time of $t_{\rm select} = 300$~ns with the trigger energy around $E_R^{\rm PMT} = 70$~keV and around 80\% trigger efficiency.  Similar to the estimation in Eq.~\eqref{eq:borexino-limit}, we have the projected limit from the JUNO experiment as
\beqa
\sigma_{{\tiny\Pcirc}A} > 2.0 \times 10^{-22} \,\mbox{cm}^2 ~.
\eeqa

In Fig.~\ref{fig:sigmaexp}, we summarize the experimental reaches from Mica, Xenon1T, BOREXINO and JUNO for fixed model parameters $\Mphi=0$, $\lambda_\phi =0.013$ and $\omega_c=50$~GeV. Varying these model parameter values does not change the signal curve much, especially in the log plot here. From this figure, one can see that a wide range of \DMBs/ with a mass below around $10^{22}$~GeV could be discovered by experiments with a large exposure. For \DMB/ with a mass above $10^{26}$~GeV or the Mica reach, there is no experimental probe at the current moment. New ideas to probe a heavy \DMB/ are worth of exploring. For instance, one could use seismic data to search for heavy \DMBs/ with a large scattering cross section~\cite{Herrin:2005kb,Cyncynates:2016rij}. 

Finally, we also comment on other bigger-size neutrino experiments. For IceCube with $A_{\rm det} \approx 10^{10}\,\mbox{cm}^2$ and $t_{\rm det} \approx 10\,\mbox{yr}$, the typical trigger energy is 100 GeV with the deep-core part as low as 10 GeV~\cite{Collaboration:2011ym}. Only relativistic incident particles can generate Cherenkov lights, which is not the case for the non-relativistic \DMB/ at hand. For the DUNE experiment with $A_{\rm det} \approx 1.0 \times 10^{8}\,\mbox{cm}^2$ and $t_{\rm det} \approx 10\,\mbox{yr}$~\cite{Acciarri:2016crz}, it potentially can probe \DMB/ mass up to $3\times 10^{23}$ GeV. However, the energy threshold is a few MeV and still too high to measure the summed recoil energy for the \DMB/ multi-hit events. Beyond the recoil energy from the \DMB/ elastic scattering, the bound state formation from \DMB/ capturing nucleons and nuclei could deposit much larger energy, 38 MeV for a nucleon and $38\times A$ MeV for a nucleus, which requires additional dedicated studies to estimate the experimental reach.

\section{Discussion and Conclusions}
\label{sec:conclusion}
The EWS-\DMB/ is a special dark matter candidate because of its close relation to the Higgs potential. Different from the collider studies for the Higgs boson properties with the Higgs boson as the quantum field around the global vacuum with $\langle h \rangle = v$, the restoration of electroweak symmetry inside a \DMB/ provides us an opportunity to probe a wide range of the Higgs VEV from 0 to $v$. When the \DMB/ interacts with the ordinary matter, the vacuum energy difference inside and outside the \DMB/ can generate an effective ``Higgs force" on nucleons or nuclei. As we discuss before, this force could also bind nucleons or nuclei inside the \DMB/. 

As discussed around Eq.~\eqref{eq:number-orbital}, the number of allowed bound states for heavy quarks inside a \DMB/ is very large. After the primordial formation of \DMBs/ from the early-universe dynamics, many of those states could be filled. As a result, the baryon number density inside a \DMB/ could be very high and even to have the QCD in the deconfined phase, when the energy per baryon is below the proton mass outside \DMB/. If this is indeed the case, the electroweak symmetry inside \DMB/ is really unbroken, not even corrected by the QCD confinement related chiral symmetry breaking. The EW sphaleron process is therefore active and can change baryons to leptons, which similar to the monopole in Grand Unified Theory that can induce catastrophic proton decays~\cite{Callan:1982ah,Rubakov:1982fp}.  Although the Earth is not dense enough to stop a \DMB/, a neutron star may have a \DMB/ stuck inside. The subsequent sphaleron-induced nucleon decays may change the properties of neutron stars and even evaporate them away.

For the \DMB/ considered in this paper, the constituent dark matter particle is a boson. A similar study can be performed for a fermionic dark matter particle. The equilibrium of the low-temperature \DMB/ state is then reached by the vacuum pressure and the degenerate fermion pressure. The situation is similar to the quark nugget in Ref.~\cite{Witten:1984rs}, except that the energy density of EWS-\DMB/ has the electroweak scale and higher than the QCD scale in the QCD quark nugget. For the early-universe formation of \DMBs/, we have simply used the first-order electroweak phase transition, which is very natural given the Higgs-portal coupling to the complex scalar dark matter particle. The stochastic gravitational waves could be another correlated signatures to cross check the scenario in this paper. 

In summary, starting from the simple Higgs-portal dark matter model, we have shown that the non-topological soliton state could be the main appearance of dark matter. The electroweak-symmetric \DMB/ is another type of macroscopical dark matter models and has its mass of $1$--$10^{10}$~g, depending on the portal quartic coupling strength and if  formed from the first-order electroweak symmetry phase transition. The radius of a spherically-symmetric \DMB/ varies from $10^{-9}$ to $10^{-6}$~cm. The energy density of \DMB/ is at the electroweak scale such that it can penetrate the Earth without stopping. The experiments with a large detector size or a long exposure time are ideal to search for \DMBs/. Indeed, the existing Xenon1T and BOREXINO or the future JUNO experiments may discover a \DMB/ with mass from $10^{-14}$~g to 0.1~g, based on multi-hit events.

\subsubsection*{Acknowledgements}
We thank Joshua Berger, Marcela Carena, Zackaria Chacko and Rog\'erio Rosenfeld for discussion.  B.J. and E.P. acknowledge support from the S\~ao Paulo Research Foundation (FAPESP), grant award numbers 2016/01343-7 and 2017/05770-0. The work of Y.B. is supported by the U. S. Department of Energy under the contract DE-SC0017647. Y.B. also thanks the host of ICTP-SAIFR when this project was initialized. This work was performed at the Aspen Center for Physics, which is supported by National Science Foundation grant PHY-1066293. This research was supported by the Munich Institute for Astro- and Particle Physics (MIAPP) of the DFG cluster of excellence ``Origin and Structure of the Universe".


\appendix
\section{Number of \DMB/ Nucleation Sites}
\label{sec:nucleation-sites}

In this appendix we derive part of the information necessary to estimate the expected \DMN/ carried by \DMBs/ produced during a first-order phase transition. Assuming that the net $Q$ charge in a Hubble volume has been set by some earlier \DMN/-genesis mechanism, the average \DMN/ for such \DMBs/ is controlled by how many \DMBs/ are produced within that Hubble patch, which we call $N^{\rm Hubble}_{\DMB/}$. Here we simply take $N^{\rm Hubble}_{\DMB/} \sim N_{\rm nucl}$, where $N_{\rm nucl}$ is the number of EWSB nucleation sites in one Hubble patch. To estimate $N_{\rm nucl}$, we follow a standard calculation (see Ref.~\cite{Enqvist:1991xw} for example).  

\begin{figure}[t] 
\centering
\includegraphics[width=0.6\textwidth]{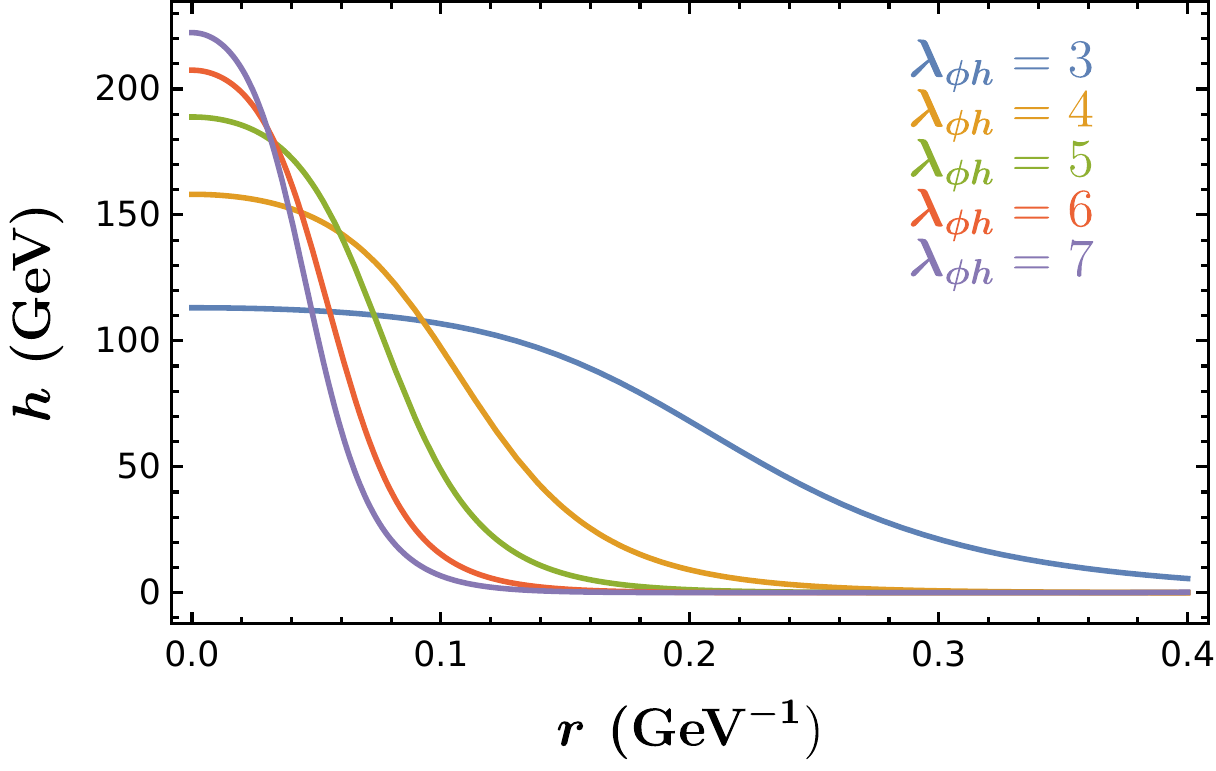} 
\caption{Bubble solution's profile functions at the nucleation temperature $T_f$ for different values of 
$\lambda_{\phi h}$ (see Table~\ref{table:S3-over-T} for numerical values).}	
\label{fig:bubble_profiles}
\end{figure}

The bubble nucleation rate is controlled by the $SO(3)$-symmetric bounce action, $S_3(T)/T$, where $S_3(T)$ is the energy of the static $SO(3)$-symmetric critical bubble,
\bea
S_3(T) &=& 4\pi \int_0^{\infty} \! dr \, r^2 \left[ \frac{1}{2} \, (h')^2 + V_{\rm eff}(h, T) \right]~,
\eea
and $V_{\rm eff}(h, T)$ is the 1-loop, finite-temperature effective potential given in Eq.~(\ref{eq:TFD}). The bounce solution can be obtained by solving the temperature-dependent Euclidean equation of motion 
\beqa
h^{\prime \prime}(r) + \frac{2}{r}\,h^\prime(r) = V^\prime_{\rm eff}\big(h(r), T\big) ~,
\eeqa
with boundary conditions $h(\infty)=0$ and $h^\prime(0)=0$. For different values of $\lambda_{\phi h}$ and in Fig.~\ref{fig:bubble_profiles}, we show the bounce profiles at $T_f$ (see Table~\ref{table:S3-over-T} for numerical numbers), approximately the temperature at the end of nucleation process. In general, the bounce profiles have a ``thick-wall" feature. As one increases the coupling $\lambda_{\phi h}$, the size of the profile decreases.

For the electroweak phase transition, we are interested in times only slightly after the critical temperature is reached. We parametrize the bounce action, in the vicinity of $T_c$, as
\beqa
\frac{S_3(T)}{T} &\approx& \frac{\overline{S}_3}{T_c} \left( 1 - \frac{T}{T_c}\right)^{-\alpha} 
~\equiv~ \frac{\overline{S}_3}{T_c} \, \eta(T)^{-\alpha}~,
\label{eq:S3oT}
\eeqa
where the two parameters $\overline{S}_3$ and $\alpha$ are functions of the model-parameter $\lambda_{\phi h}$,\footnote{The results reported in this appendix are for $\Mphi^2 = 0$. Since we are dealing with a ``one-step" phase transition, where $\phi = 0$, $\Mphi^2$ enters only at 1-loop order through the $h$-dependent $m_\phi$, while $\lambda_\phi$ enters only at 2-loop order. We have also investigated the impact of other values on $T_c$. We don't expect the estimates derived here to change significantly, as long as a strong first-order phase transition can be obtained. In particular, $\Mphi ^2$ cannot be too large, or else $\Phi$ would decouple from the plasma and one goes back to the SM result.} and we wrote the second equality in terms of the relative difference between $T_c$ and $T$: $\eta(T) \equiv (T_c - T)/T_c$. For the range $3 \lesssim \lambda_{\phi h} \lesssim 6$, we find that the index $\alpha \approx 1.7$ is relatively insensitive to $\lambda_{\phi h}$.\footnote{Ref.~\cite{Enqvist:1991xw} obtains $\alpha=2$ based on an analytic high-temperature-expansion. Here, we have used the full one-loop finite-temperature potential which leads to a slightly different index.} The coefficient $\overline{S}_3/T_c$, however, has a strong dependence on $\lambda_{\phi h}$. We show the numerical values for $\alpha$ and $\overline{S}_3/T_c$, as well as $T_c$, for a few values of $\lambda_{\phi h}$ in Table~\ref{table:S3-over-T}. Performing a numerical fit, we have a power-law dependence for $\overline{S}_3/T_c$ as a function of $\lambda_{\phi h}$
\beqa
\left(\frac{\overline{S}_3}{T_c}\right) (\lambda_{\phi h}) \approx 0.013~{\rm GeV}^{-1}\, \times \,\left(\frac{\lambda_{\phi h}}{3}\right)^{8.2} \qquad (\mbox{fit})  ~. 
\label{eq:S3bar-fit}
\eeqa
\begin{table}[t]
	\begin{center}
		\begin{tabular}{|c|c|c|c|c|c|c|}
			\hline
			\rule{0mm}{5mm} $\lambda_{\phi h}$ & $3$ & $4$ & $5$ & $6$ & $7$   \\ [0.5em]
			\hline
			\hline
			\rule{0mm}{5mm} $\alpha$ & $1.71$ & $1.72$ & $1.71$ & $1.65$ & $1.61$  \\ [0.5em]
			\hline
		         \rule{0mm}{5mm} $\overline{S}_3/T_c~({\rm GeV}^{-1})$ & $0.012$ & $0.16$ & $0.78$ & $3.59$ & $15.0$  \\ [0.5em]
			\hline
			\rule{0mm}{5mm} $T_c~({\rm GeV})$ & $134.1$ & $125.1$ & $116.5$ & $107.3$ & $96.5$  \\ [0.5em]
			\hline
			\rule{0mm}{5mm} $T_f~({\rm GeV})$ & $133.4$ & $122.3$& $110.3$ & $94.8$ & $71.1$ \\ [0.5em]
			\hline
			\rule{0mm}{5mm} $N_{\rm nucl}$ & $1.1\times 10^{13}$ & $1.5\times 10^{11}$ & $1.0\times 10^{10}$ &$7.5\times 10^{8}$ & $4.0\times 10^{7}$\\ [0.5em]
			\hline
		\end{tabular}
	\end{center}
	\label{table:S3-over-T}
	\caption{Parameters that characterize the bubble nucleation rate for several values of $\lambda_{\phi h}$ that lead to a first-order EWPT. It is assumed that the bare mass, $\Mphi^2$, can be neglected. }
\end{table}

Given $S_3(T)$, one can estimate the nucleation rate per unit volume as~\cite{Linde:1980tt}
\beqa
\gamma \approx \zeta\,T^4\, \left( \frac{S_3}{2\pi\, T}\right)^{3/2}\, e^{-S_3/T}~,
\label{gammarate}
\eeqa
where the $T$-independent $\zeta$ is expected to be of order one. Once formed, the nucleated bubble expands due to the vacuum pressure difference between the broken and unbroken phases. Due to its interaction with the particles in the plasma, a non-relativistic terminal velocity $\beta_{\rm w}$ is reached~\cite{Dorsch:2018pat}. The bubble wall is also preceded by a shock front that moves, for a strong first-order phase transition, at a speed close to the speed of sound. Here, we use the speed of sound also for the bubble wall velocity, $\beta_{\rm w} \approx 1/\sqrt{3}$, to estimate the temperature at the end of the nucleation process, and then the total number of nucleation sites produced. 

\begin{figure}[t]
\centering
\includegraphics[width=0.48\textwidth]{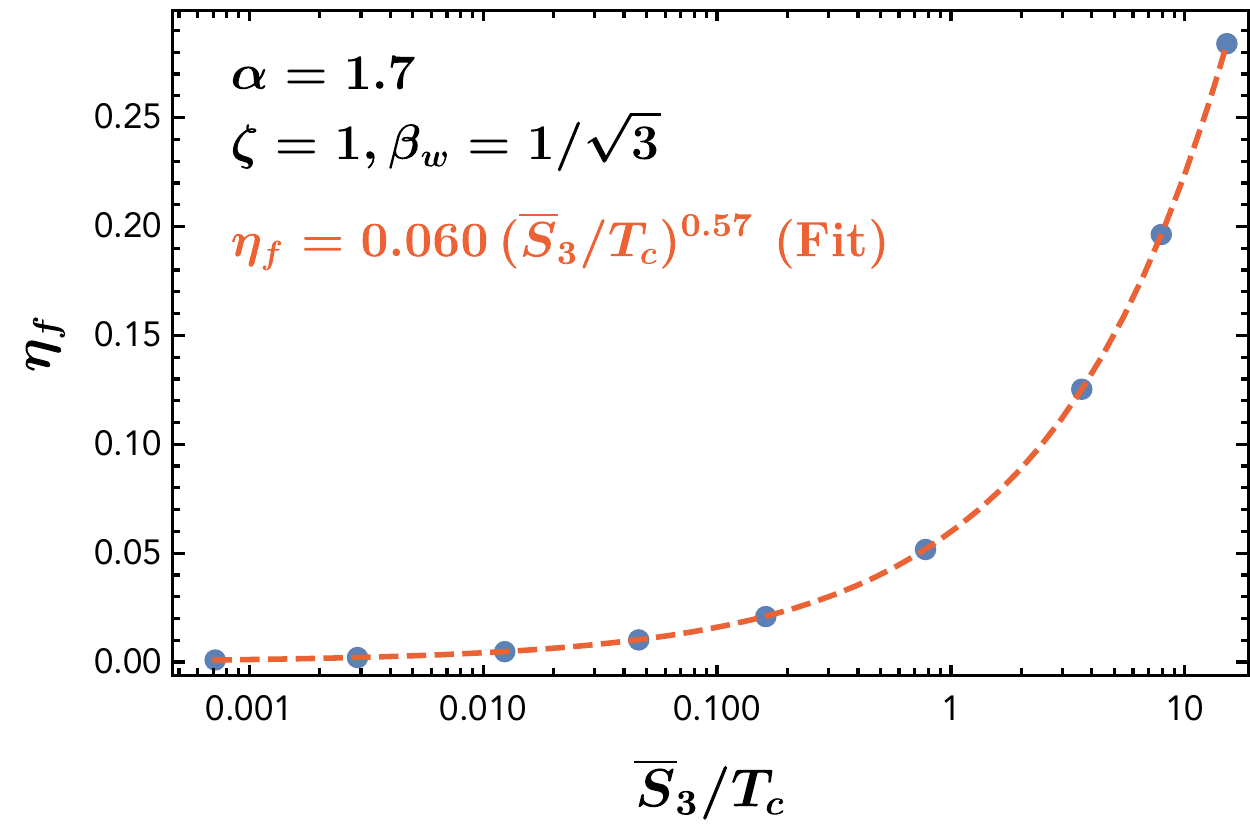}
\hspace{3mm}
\includegraphics[width=0.48\textwidth]{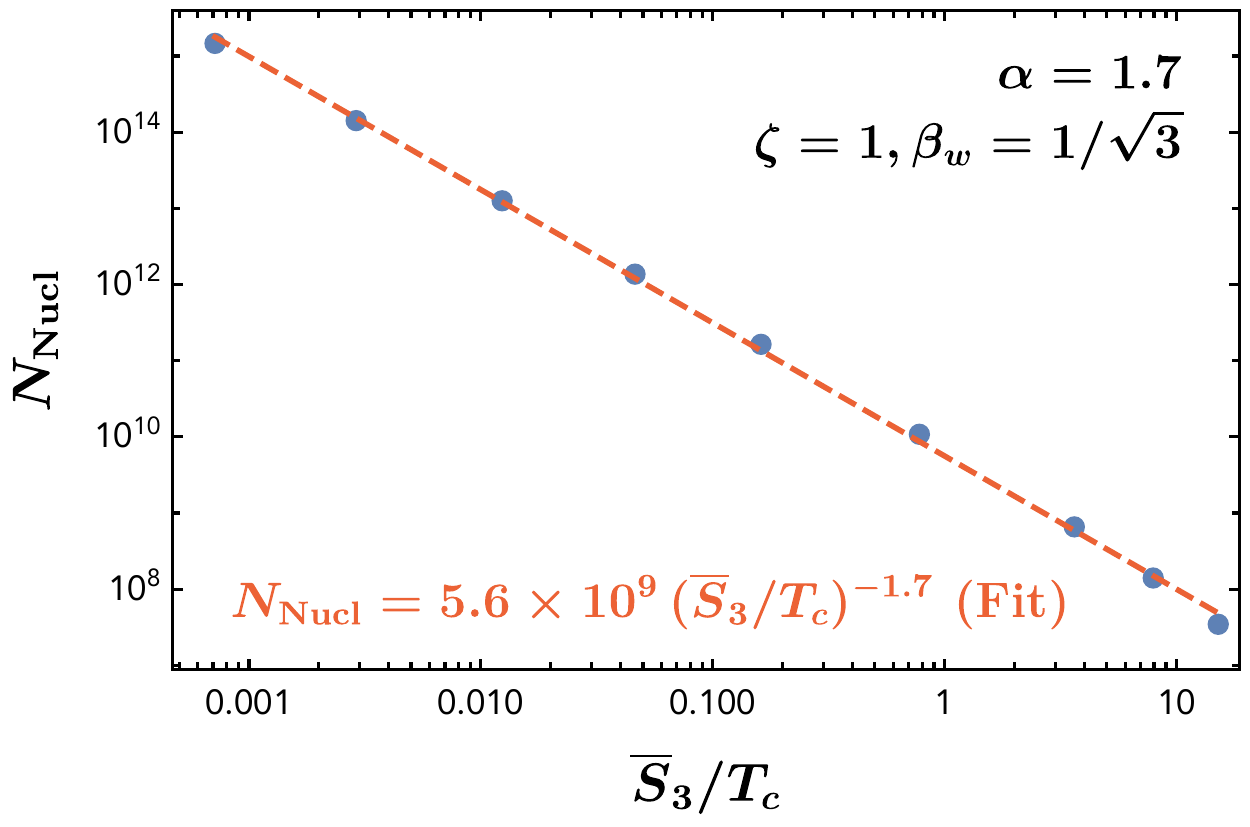}
\caption{Left panel: the end-of-nucleation temperature parameter, $\eta_f \equiv (T_c - T_f)/T_c$, as a function of the parameter $\overline{S}_3/T_c$ in Eq.~\eqref{eq:S3oT}. Right panel: the number of nucleation sites as a function of $\overline{S}_3/T_c$. 
}
\label{fig:etan-sites}
\end{figure}

Starting at time $t_c = 1.509 g_* ^{-1/2} M_{\rm P}/T_c^2$,\footnote{In the radiation dominated era. Also, close to the EW phase transition, $T \approx T_c$, the effective number of relativistic degrees of freedom, $g_*$, does not suffer any abrupt change. $M_{\rm P}$ is the reduced Planck Mass.} when the plasma temperature equals $T_c$, the fraction of space that remains in the EW unbroken phase is given by~\cite{Linde:1980tt}
\beqa
f_{\rm unbroken}(t) = \mbox{exp}\left[ - \frac{4\pi}{3} \int_{t_c}^t dt' \, \beta_{\rm w}^3\, (t - t')^3 \, \gamma(t') \right] ~,
\label{funb}
\eeqa
where the exponentiation accounts for the overlap of the bubbles~\cite{Guth:1981uk}.
Defining $T_f$ as the temperature (corresponding to time $t_f$) when the nucleation process is essentially complete by $f_{\rm unbroken}(t_f) = e^{-1}$, we  calculate $T_f$ from the bounce action as parametrized in Eq.~(\ref{eq:S3oT}) as follows. Using the steepest descent or saddle-point approximation to evaluate the integral in Eq.~(\ref{funb})~\cite{Enqvist:1991xw,Bai:2018dxf}, one gets the following approximate relation
\beqa
8\pi\,\beta_{\rm w}^3\, \gamma(\eta_f) \, \beta^{-4} \approx 1 ~, 
\label{eq:Tn}
\eeqa
where $\eta_f \equiv \eta(T_f)$, and we wrote $\gamma(T_f)$ simply as $\gamma(\eta_f)$ without changing the notaton [see Eqs.~(\ref{eq:S3oT}) and (\ref{gammarate})]. Eq.~(\ref{eq:Tn}) holds in the limit that $S_3|_{T_f} \gg 1$  and\,\footnote{For the Hubble scale we have $H = 0.331 g_*^{1/2} T^2/M_{\rm P}$, where the effective number of relativistic degrees of freedom is between $g_* = 106.75$ (the SM value) and $g_* = 108.75$, depending on the bare mass parameter $\Mphi^2$. For concreteness, we use $g_* = 108.75$.}
\bea
\beta &\equiv& \left. H \, \frac{d(S_3/T)}{d\ln T} \right|_{T_f} ~\gg~ \left(\frac{1}{2H(T_f)} - \frac{1}{2H(T_c)} \right)^{-1}~\approx~ \left. \frac{H}{\eta} \right|_{T_f}~,
\label{betaNucleation}
\eea
where we used that we are interested in times (from the phase transition temperature, $T_c$) up to $T_f$, and that $\eta(T_f) \ll 1$.
In the left panel of Fig.~\ref{fig:etan-sites}, we show $\eta_f$ as a function of $\overline{S}_3/T_c$, for $\zeta =1$ and $\alpha = 1.7$, obtained numerically from Eq.~(\ref{eq:Tn}). We see that it can be described by a simple power law: $\eta_f = 0.060 \times (\overline{S}_3/T_c)^{0.57}$ for these parameter values. Knowing $\eta_f(\overline{S}_3/T_c)$, one obtains $S_3(T_f)/T_f$ from Eq.~({\ref{eq:S3oT})}.  One finds that for the wide range $0.012 < \overline{S}_3/T_c < 15$, $S_3(T_f)$ only varies from 109 to 129 and can be fitted by $S_3(T_f) \approx 120.5 \times (\overline{S}_3/T_c)^{0.02}$, which is insensitive to the detailed information of the bounce action.

The number of nucleation sites within a Hubble patch at $t_f$ is approximately given by~\cite{Enqvist:1991xw,Bai:2018dxf} 
\beqa
N_{\rm nucl} \approx \frac{1}{8\pi\,\beta_{\rm w}^3} \, \left(\frac{\beta}{H(T_f)}\right)^3 \, \approx 5.6\times 10^{9}\,\times \,\left(\frac{\overline{S}_3/T_c}{{\rm GeV}^{-1}}\right)^{-1.7}   \approx 1.0\times 10^{13}\,\times \,\left(\frac{\lambda_{\phi h}}{3}\right)^{-14} ~,
\label{Nnucleations}
\eeqa
where in the third equality we have used the fitted relation between $\overline{S}_3/T_c$ and $\lambda_{\phi h}$ in Eq.~\eqref{eq:S3bar-fit}. This captures the dominant dependence on $\lambda_{\phi h}$. Because of the large value in the power index, the number of nucleation sites can change five orders of magnitude even for $\lambda_{\phi h}$ changes by a factor of two. We also note that taking $\alpha = 2$, the power index $-1.7$ in Eq.~(\ref{Nnucleations}) changes slightly from $-1.7$ to $-1.5$~\cite{Bai:2018dxf}. 

Finally, for our calculation of \DMBs/ charges, it is convenient to know the $\lambda_{\phi h}$ dependence of phase transition temperatures, which has 
\beqa
T_c \approx 134.5\,{\rm GeV}  - 9.3\,{\rm GeV} \,\times \, \left(\lambda_{\phi h}- 3\right) ~.
\label{eq:Tc-fit}
\eeqa
%

\section{\EDBs/ and Bound States}
\label{sec:Backreaction}

Consider adding a new (complex scalar) degree of freedom to the Higgs-$\Phi$ sector discussed in the main text. This will serve as a proxy for other ``matter" and we will denote it by $\chi$. Thus, the system to be considered in this appendix is given by
\bea
{\cal L}_{\chi} &=& {\cal L} + \partial_\mu \chi^\dagger \partial^\mu \chi - y^2 H^\dagger H \chi^\dagger \chi~,
\label{masterLFull}
\eea
where ${\cal L}$ is given by Eq.~(\ref{MasterL}). Like the SM fermions and gauge bosons, $\chi$ interacts with the scalar sector only through the Higgs field.\footnote{At loop-level, local interactions with $\Phi$ will be induced, but since our interest here is only to use $\chi$ as a proxy for the SM sector, we do not consider such terms. We also choose not to include a mass term or self-interactions for $\chi$. This means that when $\langle H \rangle=0$, as happens inside a \DMB/, $\chi$ behaves as a massless particle and relativistic effects are important. We are including here only the kinematic ones.}

We look for spherically symmetric solutions of the form
\bea
\Phi &=& \frac{1}{\sqrt{2}} \, e^{-i\omega t} \phi(r)~,
\hspace{1cm}
H ~=~ \frac{1}{\sqrt{2}} \, h(r)~,
\hspace{1cm}
\chi ~=~ e^{-i E t} \psi(r)~,
\label{wvAnsatz}
\eea
where $\phi(r)$ and $h(r)$ are real, while it is convenient to treat $\psi(r)$ as complex.\footnote{We can chose to treat $\psi$ as real, but then it is more convenient to include a factor of $1/\sqrt{2}$ in its definition to ensure canonical normalization. Treating it as complex makes the relation to the non-relativistic wavefunction more standard. However, note that here $\psi$ is not exactly the radial wavefunction as usually defined in quantum mechanical central problems. In particular, $\psi$ includes the factor $Y^0_0 = 1/\sqrt{4\pi}$, and has mass dimension one, as for relativistic canonical normalization.} They obey the boundary conditions
\bea
\frac{d\phi}{dr}(r=0) &=& 0~,
\hspace{1cm}
\phi(r=\infty) ~=~ 0~,
\\ [0.5em]
\frac{dh}{dr}(r=0) &=& 0~,
\hspace{1cm}
h(r=\infty) ~=~ v~,
\\ [0.5em]
\frac{d\psi}{dr}(r=0) &=& 0~,
\hspace{1cm}
\psi(r=\infty) ~=~ 0~.
\eea
We also assume $0 <  E < m_\chi \equiv y v/\sqrt{2}$, so that $\psi$ corresponds to a bound state in the Higgs well induced by $\phi$, as well as the analogous condition $\omega^2 < \lambda_{\phi h} v^2/2$ [see Eq.~(\ref{HillCond})], so that there is a soliton in the first place. This generalizes the case studied in the main text, and allows to discuss the issue of $\chi$ particles that get bound to the \DMB/.

The EOM are
\bea
- \frac{1}{r^2} \frac{d}{dr} \left( r^2 \frac{d\phi}{dr} \right) + \frac{1}{2} \lambda_{\phi h} h^2 \phi + V_\Phi'(\phi) &=& \omega^2 \phi~,
\label{eqphi}
\\ [0.5em]
- \frac{1}{r^2} \frac{d}{dr} \left( r^2 \frac{dh}{dr} \right) + V_H'(h) + \frac{1}{2} \lambda_{\phi h} \phi^2 h + y^2 |\psi|^2 h &=& 0~,
\label{eqh}
\\ [0.5em]
- \frac{1}{r^2} \frac{d}{dr} \left( r^2 \frac{d\psi}{dr} \right) + \frac{1}{2} y^2 h^2 \psi &=& E^2 \psi~,
\label{eqpsi}
\eea
where $V_H(h)$ is the SM Higgs potential and $V_\Phi(\phi)$ was defined in Eq.~(\ref{Vphi}).
We see that, at the level of the EOM, the distinction between $\phi$ and $\psi$ is mainly in $\lambda_{\phi h}$ vs $y^2$ (and potentially in their self-interactions, which play a secondary role in the present discussion). We make a notational distinction as a reminder that the $y$ interaction is meant to mimic a Yukawa interaction. The important physical difference between $\phi$ and $\chi$ arises instead from the normalizations we impose. For $\phi$ we impose that the charge $Q$ as defined by Eq.~(\ref{QCharge}) be much larger than one:
\bea
Q &=& \omega \int \! d^3x \, \phi^2 ~\gg~ 1~.
\eea
For $\psi$, we impose instead the relativistic normalization for ``one particle".
\bea
2E \int \! d^3x \, \psi^\dagger \psi &=& 1~.
\label{relnorm}
\eea
These normalizations do not affect directly the EOM for $\phi$ or $\psi$, but they affect the Higgs EOM, saying that, even if $\lambda_{\phi h} \sim y^2$, the Higgs well is sustained mainly by $\phi$ (there are many more particles in $\phi$ than in $\psi$). Then $\psi$ can be thought as a bound state in this well. 

We are interested in the total energy of the system, denoted by
\bea
\Msol[\psi] &=& 4\pi \int_0^\infty \! dr \, r^2 \left\{ \frac{1}{2} \omega^2 \phi^2 + \frac{1}{2} (\phi')^2 + \frac{1}{2} (h')^2 + V_H(h) + \frac{1}{4} \lambda_{\phi h} h^2 \phi^2 \right.
\nonumber \\ [0.5em]
&& \hspace{2cm} \left.
\mbox{} +
E^2 |\psi|^2 + |\psi'|^2 + \frac{1}{2} y^2 h^2 |\psi|^2 \right\}~,
\label{MPhiBound}
\eea
and, in particular, on how it compares to the energy of an ``empty" \DMB/, $\Msol^{(0)}\equiv \Msol[\psi=0]$, wih the same charge:
\bea
\Delta \Msol &=& \left. \rule{0mm}{5mm} \Msol[\psi] - \Msol^{(0)}\right|_{Q \textrm{ fixed}}~.
\eea
We want to compare two solutions with the same $Q$: one where the soliton and a ``free" $\chi$ are far away, so that the soliton mass is given by $\Msol^{(0)}$, and the solution where $\chi$ is bound to the soliton. 
Let us call $\phi_0$, $h_0$ the solutions to the EOM with $\psi = 0$, i.e.~the \sol/ configurations we considered in Section~\ref{sec:soliton}. Since we want to keep 
\bea
Q &=& \frac{\partial \Msol}{\partial \omega} ~=~ \omega \int \! d^3x \, \phi^2
\label{Q}
\eea
fixed, we can consider the Legendre transform
\bea
F[\psi] &=& \Msol[\psi] - \omega Q~,
\hspace{1cm}
\omega ~=~ - \frac{\partial F}{\partial Q}~.
\eea
The EOM for $\phi$ and $h$, Eqs.~(\ref{eqphi}) and (\ref{eqh}) can be written as
\bea
\frac{\delta F}{\delta \phi} = 0~,
\hspace{1cm}
\frac{\delta F}{\delta h} = 0~.
\label{FEOM}
\eea
A bound state, i.e.~$\psi \neq 0$, induces a perturbed $\phi$-ball solution $\phi = \phi_0 + \delta \phi$, $h = h_0 + \delta h$, and therefore
\bea
\Delta \Msol &=& \delta \omega\, Q + \frac{\delta F}{\delta \phi} \delta \phi + \frac{\delta F}{\delta h} \delta h
+ \int \! d^3x \, \psi^\dagger \left\{ E^2 \psi - \frac{1}{r^2} \frac{d}{dr} \left( r^2 \frac{d\psi}{dr} \right) + \frac{1}{2} y^2 h^2 \psi  \right\}
\nonumber \\ [0.5em]
&=& \delta \omega\,Q + 2 \int \! d^3x \, E^2 |\psi|^2~,
\eea
where the EOM (\ref{FEOM}) and (\ref{eqpsi}) have been used. We also have from Eq.~(\ref{Q})
\bea
\delta \omega &=& - \frac{2\, \omega^2_0}{Q} \int \! d^3 x \, \phi_0\, \delta \phi~,
\eea
so we can finally write
\bea
\Delta \Msol &=& 2 \int \! d^3x \, \left\{ E^2 |\psi|^2 - \omega^2_0\, \phi_0\, \delta \phi \right\}
\nonumber \\
&=& E - 2\, \omega^2_0 \int \! d^3x \, \phi_0 \,\delta \phi~,
\label{DeltaMExact}
\eea
where we used Eq.~(\ref{relnorm}). The second term corresponds to the backreaction of $\chi$ on the \sol/ when it gets bounded to it.
One can get an estimate for $\delta\phi$ by considering the linearized EOM satisfied by the perturbations, but in essence one expects the backreaction of a single particle that gets bound to the $Q \gg1$ particle system to be small. Thus,
\bea
\Delta \Msol &\approx& E~.
\label{DeltaMApprox}
\eea
We see that the change in the total mass is positive and approximately set by the energy of the bound state, as determined by the eigenvalue problem (\ref{eqpsi}) in the \textit{unperturbed} background for $h$. Note that to obtain the binding energy, we must compare the bound state mass to $\Msol^{(0)} + m_\chi$, i.e.~the total energy of the empty \sol/ plus a $\chi$ particle at rest at infinity (where $h=v$). This gives
\bea
E_B &\approx& m_\chi - E~.
\eea

We can find the solution to the full EOM (\ref{eqphi})-(\ref{eqpsi}) numerically by proceeding iteratively as follows. We start from the 0-th order solutions to Eqs.~(\ref{eqphi}) and (\ref{eqh}) (with $y=0$), that we are calling here $\phi_0$ and $h_0$. We then solve Eq.~(\ref{eqpsi}) in the fixed background $h_0$ using the ``shooting method" to obtain a bound state wavefunction, $\psi^{(n)}_0$, obeying the desired boundary conditions, and corresponding to the $n$-th bound state.
We then go back to Eqs.~(\ref{eqphi})-(\ref{eqh}), inserting (the properly normalized) $\psi^{(n)}_0$ as a fixed background, and solving the two-variable system as described in Section~\ref{ClassicalEOM}. Here we need to readjust $\omega = \omega_0 + \delta \omega$, where $\omega_0$ corresponds to the $y=0$ soliton, so as to keep $Q$ fixed. This produces new solutions $\phi_1 = \phi_0 + \delta \phi$ and $h_1 = h_0 + \delta h$. The procedure is iterated to obtain the corrected $n$-th bound state in the $h_1$ background, which is then used to obtain $(\phi_2, h_2)$, etc. We find that this procedure typically converges fast.

\begin{figure}[t]
\centering
\includegraphics[width=0.6\textwidth]{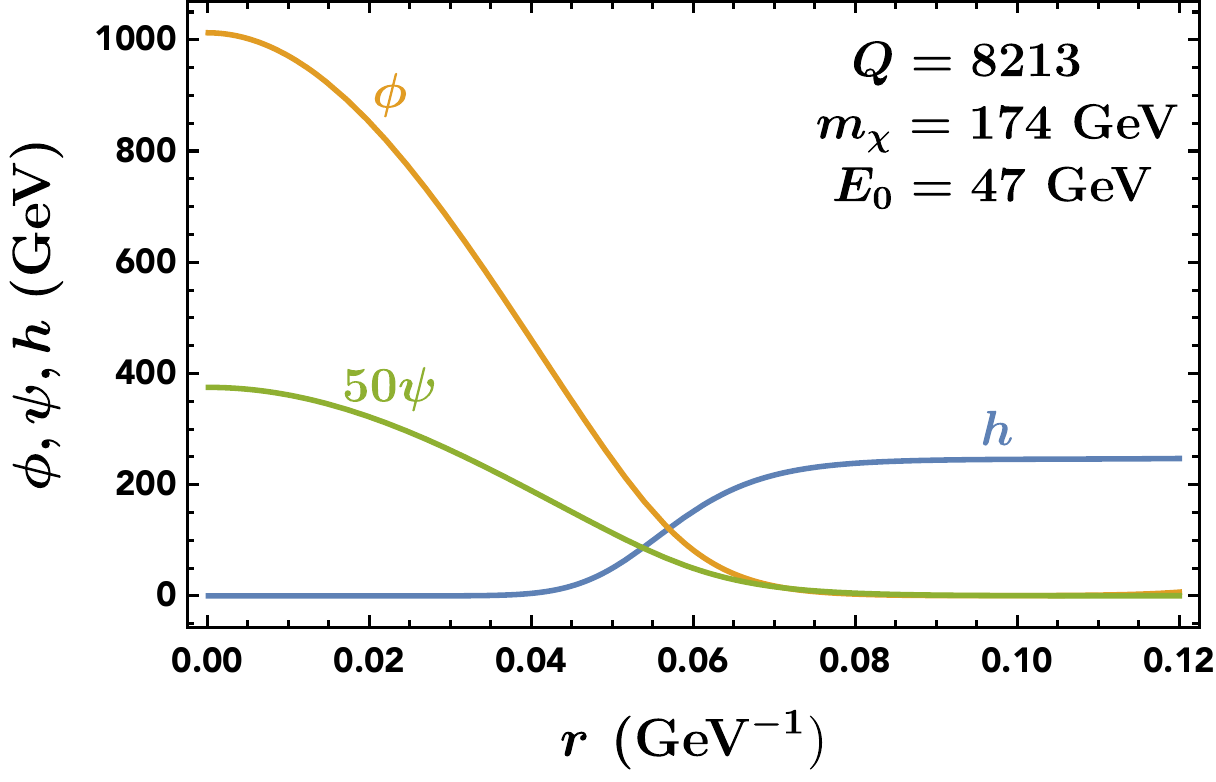}
\caption{Self-consistent solution with $Q \approx 8213$ and containing a single bound particle of ``vacuum mass" $m_\chi = 174~{\rm GeV}$. The total energy is $49~{\rm GeV}$ larger than the energy in the absence of the bound state (but with the same charge $Q$). The bound state energy (which corresponds to the ground state) is $E_0 = 47$~GeV. The small mismatch of 2~GeV is due to the backreaction. Note that the bound state wavefunction has been multiplied by 50. The model parameters are $\lambda_{\phi h} =1$, $\Mphi^2 = \lambda_\phi = 0$.}
\label{fig:backreaction}
\end{figure}
As an example, we consider the model defined by $\lambda_{\phi h} =1$, $\Mphi^2 = \lambda_\phi = 0$, and the 0-th order soliton with $\omega = 50~{\rm GeV}$, which has $Q \approx 8213$ and a radius of about $R = 0.06~{\rm GeV}^{-1}$ (this is an example of a \DMB/). Taking a particle with vacuum mass $m_{\chi} = 174~{\rm GeV}$ (corresponding to $y = 1/\sqrt{2}$) one finds a ground state with energy $E_0 \approx 47~{\rm GeV}$, while the total mass $\Msol[\psi]$ is about 49~GeV heavier than for the empty \DMB/. We see that the backreaction amounts to about $2$~GeV. $\Delta \Msol = 49~{\rm GeV}$ is the available energy if $\chi$ ``disappeared". The corresponding profiles are shown in Fig.~\ref{fig:backreaction}. 

Consider a second, lighter particle with $m_{\chi'} \approx 49~{\rm GeV}$ ($y = 0.2$). One finds $\Delta \Msol = 35.93~{\rm GeV}$ and $E_0 = 32.48~{\rm GeV}$ (not as deeply bound as $\chi$ above). The decay $\chi \to 2\chi'$ is allowed outside the \DMB/.  But if initially $\chi$ is bound inside the \DMB/ in the ground state, and if we assume that a state with two bound $\chi'$ particles has $\Delta \Msol = 2 \times 35.93 \approx 72~{\rm GeV}$ (i.e.~that the effects are approximately additive), we see that energy conservation would not allow the decay to proceed inside such a soliton.

\section{Bound States in a Higgs Potential Well}
\label{sec:HiggsWell}

If one neglects the backreaction effects discussed in Appendix~\ref{sec:Backreaction}, we can get a handle on the spectrum of bound states in a \sol/ by solving Eq.~(\ref{eqpsi}) in the fixed $h$ background of the unperturbed \sol/. With some abuse of notation, it will be useful to state the starting point here as follows:
\bea
{\cal L}_\chi &=& \partial_\mu \chi^\dagger \partial^\mu \chi - y^2 H^\dagger H \chi^\dagger \chi - J \chi^\dagger \chi~,
\label{masterL}
\eea
where the source term $J$ is added for convenience in taking various limits below. Furthermore, we will assume that the $H$ background corresponds to a (spherically symmetric) 3D potential well of size $R$:
\bea
H(r) &=& \frac{1}{\sqrt{2}} \, v \, \Theta(r-R)~.
\eea
We shall treat the EOM that follows from ${\cal L}_\chi$ above as a single particle wave equation. When looking for solutions of the form specified in Eq.~(\ref{wvAnsatz}), it takes the form
\bea
\left( -\nabla^2 + y^2 |H|^2 + J \right) \psi &=& E^2 \psi~,
\label{KGEigenvalueEq}
\eea
where $\psi$ is regarded as complex.
The structure of this equation is the same as the corresponding eigenvalue Schr\"odinger problem, using a non-relativistic Hamiltonian with a potential, $V$, defined by $2MV = y^2 |H|^2 + J$, and taking $E^2/2M \rightarrow E_{\rm NR}$, where $E_{\rm NR}$ stands for the non-relativistic energy eigenvalue, and $M$ would correspond to the mass of the particle in the corresponding non-relativistic problem. For potentials which are piecewise constant, the solutions can be expressed in terms of spherical Bessel functions. We are interested in bound states, which means that $E^2$ must be below $m^2 + J$ and above $J$ (for particles, as opposed to antiparticles, we also take $E>0$). Here we defined $m \equiv y v/\sqrt{2}$, the mass of the $\chi$ field in the normal EWSB vacuum, when $J = 0$.

For the $s$-wave case, we have $\nabla^2 \psi = r^{-2} \frac{d}{dr}(r^2 \psi')$ and~\footnote{We use $K$ for the wavevector in the inner region to reserve the symbol $k$ for the wavevector in the outside region for scattering states with $E>0$. In the non-relativistic limit this means $E = k^2/2m$ and $k = i\kappa$ when $E<0$ and the wavefunction is in the classically forbidden region [as in the second line of Eq.~(\ref{swavesol})].}
\bea
\psi(r) &=& 
\left\{ 
\begin{array}{ccc}
\frac{A}{r} \, \sin(K r) & \textrm{   for} &   r < R   \\ [0.5em]
\frac{B}{r} \, e^{-\kappa r}  & \textrm{   for} &   r > R   
\end{array}
\right.~,
\label{swavesol}
\eea
where
\bea
K &=& \sqrt{E^2 - J}~,
\label{Kelectron}
\\ [0.5em]
\kappa &=& \sqrt{m^2 + J - E^2}
\label{kappaelectron}
\eea
are both real. 
Matching the functions and their first derivatives at $r=R$ and dividing the two relations leads to the equation for the spectrum
\bea
- \cot(K R) = \frac{\kappa}{K}~.
\label{spec1}
\eea
Setting $J=0$ leads to Eq.~(\ref{relativisticspectrumSection}), describing the $s$-wave spectrum of a particle that gets its mass only from the background $H$. A similar analysis can be applied to higher partial waves. 

Let us establish the condition for the existence of at least one bound state. Writing $E \equiv m + \Delta E$, the first possible bound state appears when $\Delta E < 0$ with $|\Delta E| \to 0$, i.e.~just at threshold. In this case, Eq.~(\ref{relativisticspectrumSection}) reads
\bea
- \cot\left(m R \right) &\approx& \sqrt{\frac{-2\Delta E}{m}}~.
\eea
If $mR = \frac{\pi}{2}+\epsilon$ for $0 < \epsilon \ll 1$, the cotangent is negative and there is a solution for $\Delta E \approx - (\epsilon^2/2) \, m$, consistent with the approximation $|\Delta E| \ll m$ made above. Thus, we see that the (minimum) threshold radius for the existence of a bound state is $R_{\rm th} = \pi/(2 m)$. Note that this problem is not exactly the same as a non-relativistic 3D well, since we are taking into account here the fact that the mass vanishes inside $R$ (recall that $m$ above simply stands for $yv/\sqrt{2}$).\footnote{\label{RthNR}We note that the threshold $R$ is mapped into the nonrelativistic result $R_{\rm th} = \frac{1}{\sqrt{2mV_0}} \, \frac{\pi}{2}$ with the identification $V_0 = m/2$, as expected from the discussion after Eq.~(\ref{KGEigenvalueEq}), taking $M = m$.} So, even though the $\chi$ particle inside the well is ``massless", it can be bound. For larger $R$, more bound states appear with an asymptotic spacing
\bea
|E_{n+1} - E_n| &\sim& \frac{\pi}{2\,R}~.
\eea
In the large $R$ limit, the ground state has energy of order $E \approx 0$, which corresponds to a binding energy of order $-\Delta E \approx m$.

Let us now consider nucleons, whose mass is essentially independent of $v$. The relevant terms in the (physical) nucleon Lagrangian read
\bea
{\cal L}_N &\supset& - m_N \overline{N} N - y_{hNN} h \overline{N} N 
\nonumber \\
&=& - m_N \overline{N} N - y_{hNN} (\sqrt{2} H-v) \overline{N} N~.
\nonumber
\eea
For a ``scalar nucleon" we can then set $J = m_N^2$ and $|H|^2 \rightarrow 2 |H|^2 - v^2$ in Eq.~(\ref{masterL}), so that~\footnote{Alternatively, one can make $H \rightarrow \sqrt{2} H - v$, adding an explicit minus sign in front of the $y_{hNN}^2 (\sqrt{2}H-v)^2$ term in Eq.~(\ref{masterL}) to model a well instead of a barrier.}
\bea
K &=& \sqrt{E^2 - (-y_{hNN}^2 v^2 + m_N^2)}~,
\label{KNucleon}
\\ [0.5em]
\kappa &=& \sqrt{m_N^2 - E^2}~,
\label{kappaNucleon}
\eea
where we also relabelled $y \to y_{hNN}$. This leads to Eq.~(\ref{NucleonspectrumSection}) in the $s$-wave case.

Setting $E \equiv m_N + \Delta E$, one can see that the first bound state with $\Delta E \approx 0$ appears for~\footnote{The threshold $R$ coincides with the one found in the nonrelativistic analysis of a particle of mass $m_N$ in a potential well of depth $2m_N V_0 = (y_{hNN} v)^2$. See footnote~\ref{RthNR}.} 
\bea
R_{\rm th} &=& \frac{1}{y_{hNN} v} \, \frac{\pi}{2} ~\approx~ \frac{4}{m_N} \, \frac{\pi}{2}~,
\label{nonrelRNucleon}
\eea
where the second equation is specific to a nucleon. If we considered a nucleus with $A$ nucleons one gets an additional factor of $1/A$. Again, as $R$ increases one gets additional bound states, and in the large $R$ limit, their ($s$-wave) spacing is of order $\pi/(2R)$.

\section{Simple Example of Scattering Against a Heavy Object}
\label{sec:heavysource}

In this appendix we illustrate the method used in the main text to compute scattering cross sections with a simple toy model. We take the interaction terms
\bea
{\cal L}_{\rm int} &=& \mu\, \Phi^\dagger \Phi\, h + g\,h \,\overline{\chi} \chi~,
\eea
where $\Phi$ and $h$ are scalars (complex and real, respectively), and $\chi$ is a Dirac fermion. We assume that $\Phi$ is much heavier than $\chi$. In this situation, the approach is to first find the $h$ field induced by $\Phi$, then consider the scattering of $\chi$ in such an $h$ background. 

Consider an ``elementary" $\Phi$ configuration in its rest frame given by
\bea
\Phi(\vec{x}, t) &=& \frac{1}{\sqrt{2}} \, \phi_0 \, e^{-i M_\Phi t}~,
\label{ParticleField}
\eea
where $M_\Phi$ is the $\Phi$ mass. In the $h$ EOM this enters as a source
\bea
(\Box + m_h^2) \, h &=& \frac{1}{2} \, \mu\, \phi_0^2 ~=~ \int \! d^3x \, \frac{1}{2} \, \mu \,\phi_0^2 \, \delta(\vec{x})~,
\eea
which can be interpreted as the superposition of point-like sources
\bea
(\Box + m_h^2) \, h &=& \frac{1}{2} \, \mu\, \phi_0^2 \, \Delta V \,\delta(\vec{x})~.
\label{hEOMtemp}
\eea
The $\Phi$ charge contained in volume $\Delta V$ in the configuration of Eq.~(\ref{ParticleField}) is
\bea
Q &=& M_\Phi \,\phi_0^2 \, \Delta V~,
\eea
so we can write Eq.~(\ref{hEOMtemp}) as
\bea
(\Box + m_h^2) \, h &=& \frac{1}{2} \, \frac{\mu}{M_\Phi} Q \, \delta(\vec{x})~.
\label{hEOMFinal}
\eea
The static and spherically symmetric $h$ field induced by the point-like source is
\bea
h(r) &=& - \frac{1}{4\pi} \left( \frac{1}{2} \, \frac{\mu}{M_\Phi} Q \right) \, \frac{1}{r} \, e^{-m_h r}~.
\eea
We can now turn to the fermion, which sees a potential $V(r) = g\, h(r)$. In the Born approximation, the differential scattering cross section is given by 
\bea
f(E,\theta) &=& - \frac{2m_\chi}{q} \, \int_0^\infty \! dr \, r \sin(qr) V(r)
\nonumber \\
&=& \frac{g}{4\pi} \, \frac{m_\chi}{M_\Phi} \, \frac{\mu \, Q}{\vec{q}^2 + m_h^2} ~,
\label{fAmpl}
\eea
where $q \equiv |\vec{q}|$. The total cross section for $\vec{q} = 0$ is
\bea
\sigma &=& \frac{g^2 \mu^2 Q^2}{4\pi m_h^4} \, \left( \frac{m_\chi}{M_\Phi} \right)^2~.
\label{XSBackground}
\eea
Let us now repeat the computation following a textbook quantum field theory approach. The tree-level invariant amplitude for $\Phi \chi \to \Phi \chi$ scattering is
\bea
{\cal M} &=& (i \mu) \frac{i}{q^2 - m_h^2} (ig) (\overline{u} u)~,
\label{invAmpl}
\eea
where for elastic scattering $q^2 = q_\mu q^\mu = - |\vec{q}|^2$. In the heavy $M_\Phi$ limit, i.e.~taking the Mandelstam $s \approx M_\Phi^2$, we have
\bea
\frac{d\sigma}{d\Omega} &\approx& \frac{1}{64\pi^2 M_\Phi^2} \, |{\cal M}|^2~.
\label{diffXS}
\eea
Using $\overline{u} u = 2 m_\chi$, and taking $\vec{q} = 0$, we get 
\bea
\sigma &\approx& \frac{m_\chi^2}{4\pi M_\Phi^2} \, \frac{g^2\mu^2}{m_h^4}~,
\eea
which reproduces Eq.~(\ref{XSBackground}) with $Q = 1$. This corresponds to the fact that in the QFT computation we are scattering a single $\Phi$ particle, and establishes how to implement this concept in the classical language, or generalize it for aggregates of particles.
Note also that the only real assumption we have made is that $M_\Phi \gg m_\chi$. While we assumed that $\vec{q} = 0$ to compute the total cross section, this was done only for simplicity. Indeed, one can see from Eqs.~(\ref{diffXS}) and (\ref{invAmpl}) that
\bea
\frac{d\sigma}{d\Omega} &\approx& |f(E,\theta)|^2~,
\eea
with $f(E,\theta)$ given by Eq.~(\ref{fAmpl}), for any $\vec{q}$. Although clearly the QFT approach is significantly more efficient in this example, the ``fixed background" approach is better suited for the scattering against heavy extended objects, which is more naturally discussed in configuration space.

\bibliographystyle{JHEP}
\bibliography{HiggsPortal}

\providecommand{\href}[2]{#2}\begingroup\raggedright\begin{thebibliography}{10}

\bibitem{Aprile:2018dbl}
{\scshape XENON} collaboration, \emph{{Dark Matter Search Results from a One
  Ton-Year Exposure of XENON1T}},
  \href{https://doi.org/10.1103/PhysRevLett.121.111302}{\emph{Phys. Rev. Lett.}
  {\bfseries 121} (2018) 111302}
  [\href{https://arxiv.org/abs/1805.12562}{{\ttfamily 1805.12562}}].

\bibitem{Hawking:1971ei}
S.~Hawking, \emph{{Gravitationally collapsed objects of very low mass}},
  {\emph{Mon. Not. Roy. Astron. Soc.} {\bfseries 152} (1971) 75}.

\bibitem{Witten:1984rs}
E.~Witten, \emph{{Cosmic Separation of Phases}},
  \href{https://doi.org/10.1103/PhysRevD.30.272}{\emph{Phys. Rev.} {\bfseries
  D30} (1984) 272}.

\bibitem{RosenSoliton}
G.~{Rosen}, \emph{{Particlelike Solutions to Nonlinear Complex Scalar Field
  Theories with Positive-Definite Energy Densities}},
  \href{https://doi.org/10.1063/1.1664693}{\emph{Journal of Mathematical
  Physics} {\bfseries 9} (1968) 996}.

\bibitem{Friedberg:1976me}
R.~Friedberg, T.~D. Lee and A.~Sirlin, \emph{{A Class of Scalar-Field Soliton
  Solutions in Three Space Dimensions}},
  \href{https://doi.org/10.1103/PhysRevD.13.2739}{\emph{Phys. Rev.} {\bfseries
  D13} (1976) 2739}.

\bibitem{Coleman:1985ki}
S.~R. Coleman, \emph{{Q Balls}},
  \href{https://doi.org/10.1016/0550-3213(85)90286-X,
  10.1016/0550-3213(86)90520-1}{\emph{Nucl. Phys.} {\bfseries B262} (1985)
  263}.

\bibitem{Lee:1991ax}
T.~D. Lee and Y.~Pang, \emph{{Nontopological solitons}},
  \href{https://doi.org/10.1016/0370-1573(92)90064-7}{\emph{Phys. Rept.}
  {\bfseries 221} (1992) 251}.

\bibitem{Frieman:1988ut}
J.~A. Frieman, G.~B. Gelmini, M.~Gleiser and E.~W. Kolb, \emph{{Solitogenesis:
  Primordial Origin of Nontopological Solitons}},
  \href{https://doi.org/10.1103/PhysRevLett.60.2101}{\emph{Phys. Rev. Lett.}
  {\bfseries 60} (1988) 2101}.

\bibitem{Griest:1989bq}
K.~Griest and E.~W. Kolb, \emph{{Solitosynthesis: Cosmological Evolution of
  Nontopological Solitons}},
  \href{https://doi.org/10.1103/PhysRevD.40.3231}{\emph{Phys. Rev.} {\bfseries
  D40} (1989) 3231}.

\bibitem{Frieman:1989bx}
J.~A. Frieman, A.~V. Olinto, M.~Gleiser and C.~Alcock, \emph{{Cosmic Evolution
  of Nontopological Solitons. 1.}},
  \href{https://doi.org/10.1103/PhysRevD.40.3241}{\emph{Phys. Rev.} {\bfseries
  D40} (1989) 3241}.

\bibitem{Kusenko:1997si}
A.~Kusenko and M.~E. Shaposhnikov, \emph{{Supersymmetric Q balls as dark
  matter}}, \href{https://doi.org/10.1016/S0370-2693(97)01375-0}{\emph{Phys.
  Lett.} {\bfseries B418} (1998) 46}
  [\href{https://arxiv.org/abs/hep-ph/9709492}{{\ttfamily hep-ph/9709492}}].

\bibitem{Dvali:1997qv}
G.~R. Dvali, A.~Kusenko and M.~E. Shaposhnikov, \emph{{New physics in a
  nutshell, or Q ball as a power plant}},
  \href{https://doi.org/10.1016/S0370-2693(97)01378-6}{\emph{Phys. Lett.}
  {\bfseries B417} (1998) 99}
  [\href{https://arxiv.org/abs/hep-ph/9707423}{{\ttfamily hep-ph/9707423}}].

\bibitem{Kusenko:1997vp}
A.~Kusenko, V.~Kuzmin, M.~E. Shaposhnikov and P.~G. Tinyakov,
  \emph{{Experimental signatures of supersymmetric dark matter Q balls}},
  \href{https://doi.org/10.1103/PhysRevLett.80.3185}{\emph{Phys. Rev. Lett.}
  {\bfseries 80} (1998) 3185}
  [\href{https://arxiv.org/abs/hep-ph/9712212}{{\ttfamily hep-ph/9712212}}].

\bibitem{Witten:1983tx}
E.~Witten, \emph{{Current Algebra, Baryons, and Quark Confinement}},
  \href{https://doi.org/10.1016/0550-3213(83)90064-0}{\emph{Nucl. Phys.}
  {\bfseries B223} (1983) 433}.

\bibitem{Aad:2015zhl}
{\scshape ATLAS, CMS} collaboration, \emph{{Combined Measurement of the Higgs
  Boson Mass in $pp$ Collisions at $\sqrt{s}=7$ and 8 TeV with the ATLAS and
  CMS Experiments}},
  \href{https://doi.org/10.1103/PhysRevLett.114.191803}{\emph{Phys. Rev. Lett.}
  {\bfseries 114} (2015) 191803}
  [\href{https://arxiv.org/abs/1503.07589}{{\ttfamily 1503.07589}}].

\bibitem{Lubeck:1986if}
E.~G. Lubeck, M.~C. Birse, E.~M. Henley and L.~Wilets, \emph{{Momentum
  Projection and Relativistic Boost of Solitons: Coherent States and
  Projection}}, \href{https://doi.org/10.1103/PhysRevD.33.234}{\emph{Phys.
  Rev.} {\bfseries D33} (1986) 234}.

\bibitem{Gross:1980br}
D.~J. Gross, R.~D. Pisarski and L.~G. Yaffe, \emph{{QCD and Instantons at
  Finite Temperature}},
  \href{https://doi.org/10.1103/RevModPhys.53.43}{\emph{Rev. Mod. Phys.}
  {\bfseries 53} (1981) 43}.

\bibitem{Parwani:1991gq}
R.~R. Parwani, \emph{{Resummation in a hot scalar field theory}},
  \href{https://doi.org/10.1103/PhysRevD.45.4695,
  10.1103/PhysRevD.48.5965.2}{\emph{Phys. Rev.} {\bfseries D45} (1992) 4695}
  [\href{https://arxiv.org/abs/hep-ph/9204216}{{\ttfamily hep-ph/9204216}}].

\bibitem{Arnold:1992rz}
P.~B. Arnold and O.~Espinosa, \emph{{The Effective potential and first order
  phase transitions: Beyond leading-order}},
  \href{https://doi.org/10.1103/physrevd.50.6662.2,
  10.1103/PhysRevD.47.3546}{\emph{Phys. Rev.} {\bfseries D47} (1993) 3546}
  [\href{https://arxiv.org/abs/hep-ph/9212235}{{\ttfamily hep-ph/9212235}}].

\bibitem{Carrington:1991hz}
M.~E. Carrington, \emph{{The Effective potential at finite temperature in the
  Standard Model}}, \href{https://doi.org/10.1103/PhysRevD.45.2933}{\emph{Phys.
  Rev.} {\bfseries D45} (1992) 2933}.

\bibitem{Delaunay:2007wb}
C.~Delaunay, C.~Grojean and J.~D. Wells, \emph{{Dynamics of Non-renormalizable
  Electroweak Symmetry Breaking}},
  \href{https://doi.org/10.1088/1126-6708/2008/04/029}{\emph{JHEP} {\bfseries
  04} (2008) 029} [\href{https://arxiv.org/abs/0711.2511}{{\ttfamily
  0711.2511}}].

\bibitem{Carena:2008vj}
M.~Carena, G.~Nardini, M.~Quiros and C.~E.~M. Wagner, \emph{{The Baryogenesis
  Window in the MSSM}},
  \href{https://doi.org/10.1016/j.nuclphysb.2008.12.014}{\emph{Nucl. Phys.}
  {\bfseries B812} (2009) 243}
  [\href{https://arxiv.org/abs/0809.3760}{{\ttfamily 0809.3760}}].

\bibitem{Katz:2014bha}
A.~Katz and M.~Perelstein, \emph{{Higgs Couplings and Electroweak Phase
  Transition}}, \href{https://doi.org/10.1007/JHEP07(2014)108}{\emph{JHEP}
  {\bfseries 07} (2014) 108} [\href{https://arxiv.org/abs/1401.1827}{{\ttfamily
  1401.1827}}].

\bibitem{Curtin:2016urg}
D.~Curtin, P.~Meade and H.~Ramani, \emph{{Thermal Resummation and Phase
  Transitions}},  \href{https://arxiv.org/abs/1612.00466}{{\ttfamily
  1612.00466}}.

\bibitem{Jain:2017sqm}
B.~Jain, S.~J. Lee and M.~Son, \emph{{Validity of the effective potential and
  the precision of Higgs field self-couplings}},
  \href{https://doi.org/10.1103/PhysRevD.98.075002}{\emph{Phys. Rev.}
  {\bfseries D98} (2018) 075002}
  [\href{https://arxiv.org/abs/1709.03232}{{\ttfamily 1709.03232}}].

\bibitem{Coleman:1973jx}
S.~R. Coleman and E.~J. Weinberg, \emph{{Radiative Corrections as the Origin of
  Spontaneous Symmetry Breaking}},
  \href{https://doi.org/10.1103/PhysRevD.7.1888}{\emph{Phys. Rev.} {\bfseries
  D7} (1973) 1888}.

\bibitem{Chung:2012vg}
D.~J.~H. Chung, A.~J. Long and L.-T. Wang, \emph{{125 GeV Higgs boson and
  electroweak phase transition model classes}},
  \href{https://doi.org/10.1103/PhysRevD.87.023509}{\emph{Phys. Rev.}
  {\bfseries D87} (2013) 023509}
  [\href{https://arxiv.org/abs/1209.1819}{{\ttfamily 1209.1819}}].

\bibitem{Patel:2011th}
H.~H. Patel and M.~J. Ramsey-Musolf, \emph{{Baryon Washout, Electroweak Phase
  Transition, and Perturbation Theory}},
  \href{https://doi.org/10.1007/JHEP07(2011)029}{\emph{JHEP} {\bfseries 07}
  (2011) 029} [\href{https://arxiv.org/abs/1101.4665}{{\ttfamily 1101.4665}}].

\bibitem{Aghanim:2018eyx}
{\scshape Planck} collaboration, \emph{{Planck 2018 results. VI. Cosmological
  parameters}},  \href{https://arxiv.org/abs/1807.06209}{{\ttfamily
  1807.06209}}.

\bibitem{Patrignani:2016xqp}
{\scshape Particle Data Group} collaboration, \emph{{Review of Particle
  Physics}}, \href{https://doi.org/10.1088/1674-1137/40/10/100001}{\emph{Chin.
  Phys.} {\bfseries C40} (2016) 100001}.

\bibitem{Cheng:2012qr}
H.-Y. Cheng and C.-W. Chiang, \emph{{Revisiting Scalar and Pseudoscalar
  Couplings with Nucleons}},
  \href{https://doi.org/10.1007/JHEP07(2012)009}{\emph{JHEP} {\bfseries 07}
  (2012) 009} [\href{https://arxiv.org/abs/1202.1292}{{\ttfamily 1202.1292}}].

\bibitem{Sirunyan:2018owy}
{\scshape CMS} collaboration, \emph{{Search for invisible decays of a Higgs
  boson produced through vector boson fusion in proton-proton collisions at
  $\sqrt{s} =$ 13 TeV}},  \href{https://arxiv.org/abs/1809.05937}{{\ttfamily
  1809.05937}}.

\bibitem{Aaboud:2017phn}
{\scshape ATLAS} collaboration, \emph{{Search for dark matter and other new
  phenomena in events with an energetic jet and large missing transverse
  momentum using the ATLAS detector}},
  \href{https://doi.org/10.1007/JHEP01(2018)126}{\emph{JHEP} {\bfseries 01}
  (2018) 126} [\href{https://arxiv.org/abs/1711.03301}{{\ttfamily
  1711.03301}}].

\bibitem{Price:1986ky}
P.~B. Price and M.~H. Salamon, \emph{{Search for Supermassive Magnetic
  Monopoles Using Mica Crystals}},
  \href{https://doi.org/10.1103/PhysRevLett.56.1226}{\emph{Phys. Rev. Lett.}
  {\bfseries 56} (1986) 1226}.

\bibitem{DeRujula:1984axn}
A.~De~Rujula and S.~L. Glashow, \emph{{Nuclearites: A Novel Form of Cosmic
  Radiation}}, \href{https://doi.org/10.1038/312734a0}{\emph{Nature} {\bfseries
  312} (1984) 734}.

\bibitem{Bramante:2018qbc}
J.~Bramante, B.~Broerman, R.~F. Lang and N.~Raj, \emph{{Saturated Overburden
  Scattering and the Multiscatter Frontier: Discovering Dark Matter at the
  Planck Mass and Beyond}},
  \href{https://doi.org/10.1103/PhysRevD.98.083516}{\emph{Phys. Rev.}
  {\bfseries D98} (2018) 083516}
  [\href{https://arxiv.org/abs/1803.08044}{{\ttfamily 1803.08044}}].

\bibitem{Bramante:2018tos}
J.~Bramante, B.~Broerman, J.~Kumar, R.~F. Lang, M.~Pospelov and N.~Raj,
  \emph{{Foraging for dark matter in large volume liquid scintillator neutrino
  detectors with multiscatter events}},
  \href{https://doi.org/10.1103/PhysRevD.99.083010}{\emph{Phys. Rev.}
  {\bfseries D99} (2019) 083010}
  [\href{https://arxiv.org/abs/1812.09325}{{\ttfamily 1812.09325}}].

\bibitem{Bellini:2013lnn}
{\scshape Borexino} collaboration, \emph{{Final results of Borexino Phase-I on
  low energy solar neutrino spectroscopy}},
  \href{https://doi.org/10.1103/PhysRevD.89.112007}{\emph{Phys. Rev.}
  {\bfseries D89} (2014) 112007}
  [\href{https://arxiv.org/abs/1308.0443}{{\ttfamily 1308.0443}}].

\bibitem{Hong:2002ec}
J.~Hong, W.~W. Craig, P.~Graham, C.~J. Hailey, N.~J.~C. Spooner and D.~R.
  Tovey, \emph{{The scintillation efficiency of carbon and hydrogen recoils in
  an organic liquid scintillator for dark matter searches}},
  \href{https://doi.org/10.1016/S0927-6505(01)00114-1}{\emph{Astropart. Phys.}
  {\bfseries 16} (2002) 333}.

\bibitem{Tretyak:2009sr}
V.~I. Tretyak, \emph{{Semi-empirical calculation of quenching factors for ions
  in scintillators}},
  \href{https://doi.org/10.1016/j.astropartphys.2009.11.002}{\emph{Astropart.
  Phys.} {\bfseries 33} (2010) 40}
  [\href{https://arxiv.org/abs/0911.3041}{{\ttfamily 0911.3041}}].

\bibitem{Jungman:1995df}
G.~Jungman, M.~Kamionkowski and K.~Griest, \emph{{Supersymmetric dark matter}},
  \href{https://doi.org/10.1016/0370-1573(95)00058-5}{\emph{Phys. Rept.}
  {\bfseries 267} (1996) 195}
  [\href{https://arxiv.org/abs/hep-ph/9506380}{{\ttfamily hep-ph/9506380}}].

\bibitem{An:2015jdp}
{\scshape JUNO} collaboration, \emph{{Neutrino Physics with JUNO}},
  \href{https://doi.org/10.1088/0954-3899/43/3/030401}{\emph{J. Phys.}
  {\bfseries G43} (2016) 030401}
  [\href{https://arxiv.org/abs/1507.05613}{{\ttfamily 1507.05613}}].

\bibitem{JUNO-thesis}
von Rikhav~Shah, ``{Studies on the trigger configuration for the JUNO
  experiment}.''
  {\footnotesize\url{http://collaborations.fz-juelich.de/ikp/neutrino/group_mem/documents/Rikhav_Master_Thesis.pdf}}.

\bibitem{Herrin:2005kb}
E.~T. Herrin, D.~C. Rosenbaum and V.~L. Teplitz, \emph{{Seismic search for
  strange quark nuggets}},
  \href{https://doi.org/10.1103/PhysRevD.73.043511}{\emph{Phys. Rev.}
  {\bfseries D73} (2006) 043511}
  [\href{https://arxiv.org/abs/astro-ph/0505584}{{\ttfamily
  astro-ph/0505584}}].

\bibitem{Cyncynates:2016rij}
D.~Cyncynates, J.~Chiel, J.~Sidhu and G.~D. Starkman, \emph{{Reconsidering
  seismological constraints on the available parameter space of macroscopic
  dark matter}}, \href{https://doi.org/10.1103/PhysRevD.95.063006,
  10.1103/PhysRevD.95.129903}{\emph{Phys. Rev.} {\bfseries D95} (2017) 063006}
  [\href{https://arxiv.org/abs/1610.09680}{{\ttfamily 1610.09680}}].

\bibitem{Collaboration:2011ym}
{\scshape IceCube} collaboration, \emph{{The Design and Performance of IceCube
  DeepCore}},
  \href{https://doi.org/10.1016/j.astropartphys.2012.01.004}{\emph{Astropart.
  Phys.} {\bfseries 35} (2012) 615}
  [\href{https://arxiv.org/abs/1109.6096}{{\ttfamily 1109.6096}}].

\bibitem{Acciarri:2016crz}
{\scshape DUNE} collaboration, \emph{{Long-Baseline Neutrino Facility (LBNF)
  and Deep Underground Neutrino Experiment (DUNE)}},
  \href{https://arxiv.org/abs/1601.05471}{{\ttfamily 1601.05471}}.

\bibitem{Callan:1982ah}
C.~G. Callan, Jr., \emph{{Disappearing Dyons}},
  \href{https://doi.org/10.1103/PhysRevD.25.2141}{\emph{Phys. Rev.} {\bfseries
  D25} (1982) 2141}.

\bibitem{Rubakov:1982fp}
V.~A. Rubakov, \emph{{Adler-Bell-Jackiw Anomaly and Fermion Number Breaking in
  the Presence of a Magnetic Monopole}},
  \href{https://doi.org/10.1016/0550-3213(82)90034-7}{\emph{Nucl. Phys.}
  {\bfseries B203} (1982) 311}.

\bibitem{Enqvist:1991xw}
K.~Enqvist, J.~Ignatius, K.~Kajantie and K.~Rummukainen, \emph{{Nucleation and
  bubble growth in a first order cosmological electroweak phase transition}},
  \href{https://doi.org/10.1103/PhysRevD.45.3415}{\emph{Phys. Rev.} {\bfseries
  D45} (1992) 3415}.

\bibitem{Linde:1980tt}
A.~D. Linde, \emph{{Fate of the False Vacuum at Finite Temperature: Theory and
  Applications}},
  \href{https://doi.org/10.1016/0370-2693(81)90281-1}{\emph{Phys. Lett.}
  {\bfseries 100B} (1981) 37}.

\bibitem{Dorsch:2018pat}
G.~C. Dorsch, S.~J. Huber and T.~Konstandin, \emph{{Bubble wall velocities in
  the Standard Model and beyond}},
  \href{https://doi.org/10.1088/1475-7516/2018/12/034}{\emph{JCAP} {\bfseries
  1812} (2018) 034} [\href{https://arxiv.org/abs/1809.04907}{{\ttfamily
  1809.04907}}].

\bibitem{Guth:1981uk}
A.~H. Guth and E.~J. Weinberg, \emph{{Cosmological Consequences of a First
  Order Phase Transition in the SU(5) Grand Unified Model}},
  \href{https://doi.org/10.1103/PhysRevD.23.876}{\emph{Phys. Rev.} {\bfseries
  D23} (1981) 876}.

\bibitem{Bai:2018dxf}
Y.~Bai, A.~J. Long and S.~Lu, \emph{{Dark Quark Nuggets}},
  \href{https://doi.org/10.1103/PhysRevD.99.055047}{\emph{Phys. Rev.}
  {\bfseries D99} (2019) 055047}
  [\href{https://arxiv.org/abs/1810.04360}{{\ttfamily 1810.04360}}].

\end{thebibliography}\endgroup

\end{document}